\documentclass{article}

\usepackage{PRIMEarxiv}

\usepackage[utf8]{inputenc} 
\usepackage[T1]{fontenc}    
\usepackage{hyperref}       
\usepackage{url}            
\usepackage{booktabs}       
\usepackage{amsmath,amsthm,amssymb}
\usepackage{amsfonts}       
\usepackage{nicefrac}       
\usepackage{microtype}      
\usepackage{lipsum}
\usepackage{fancyhdr}       
\usepackage{graphicx}       
\graphicspath{{media/}}     

\usepackage[numbers]{natbib}

\title{Sharing with Frictions:\\ Limited Transfers and Costly Inspections
}

\author{
  Federico Bobbio, Randall A. Berry, Michael L. Honig \\
  Department of Electrical and Computer Engineering \\
  Northwestern University \\
  Evanston, IL, USA \\
  \texttt{\{federico.bobbio, rberry, mhonig\}@northwestern.edu} \\
  \And
  Thanh Nguyen \\
  Daniels School of Business \\
  Purdue University \\
  West Lafayette, IN, USA \\
  \texttt{nguye161@purdue.edu} \\
  \And
  Vijay G. Subramanian \\
  Department of Electrical Engineering and Computer Science \\
  University of Michigan \\
  Ann Arbor, MI, USA \\
  \texttt{vgsubram@umich.edu} \\
  \And
  Rakesh V. Vohra \\
  Department of Economics \\
  University of Pennsylvania \\
  Philadelphia, PA, USA \\
  \texttt{rvohra@seas.upenn.edu} \\
}

\usepackage{our_commands}

\allowdisplaybreaks

\newcommand{\Xomit}[1]{}

\begin{document}
\maketitle

\begin{abstract}
The radio spectrum suitable for commercial wireless services  is limited.  A portion of the radio spectrum has been reserved for institutions using it for non-commercial purposes such as federal agencies, defense, public safety bodies and scientific institutions. In order to operate efficiently, these incumbents need clean spectrum access. However, commercial users also want access, and granting them access may materially interfere with the existing activity of the incumbents. 
Conventional market based mechanisms for allocating scarce resources in this context are problematic. Allowing direct monetary transfers to and from  public or scientific institutions risks distorting their non-commercial mission. Moreover, often only the incumbent knows the exact value of the interference it experiences, and, likewise, only commercial users can predict accurately the expected monetary outcome from sharing the resource. Thus, our problem is to determine the efficient allocation of resources in the presence of private information without the use of direct monetary transfers. The problem is not unique to spectrum. Other resources that governments hold in trust share the same feature. We propose a novel mechanism design formulation of the problem, characterize the optimal mechanism and describe some of its qualitative properties. 
\end{abstract}

\keywords{Mechanism Design \and Costly Inspection \and Spectrum Sharing}

\section{Introduction}\label{sec:introduction}

Resources intended for non-commercial use often attract commercial users seeking access. However, once commercial users are allowed to share the resource, their involvement may erode its value for the original incumbents.  
To illustrate,  consider a given band of wireless spectrum allocated exclusively to radio astronomers for the purposes of scientific exploration. Scientific uses of the spectrum rely on
passive sensing which requires access to \lq\lq clean\rq\rq\, spectrum, meaning there are no other man-made transmitters operating in the band or in neighboring bands detectable by the scientific instrument. 
Indeed, that same spectrum bandwidth can also be used for commercial purposes such as broadcasting or broadband access via terrestrial or satellite networks.  
Risks and opportunities arise with the possibility to allow commercial users access to the band reserved for scientific purposes. On one side,   
passive receivers that search for and measure astronomical phenomena, for instance detecting gravitational waves or measuring cosmic radiation, look for trace signals well below the noise floor for commercial users. The adoption of wide guard-bands or the enforcement of spatial \lq\lq quiet zones\rq\rq\, to protect the bands allocated for scientific purposes may leave significant bandwidth unoccupied and unused. On the other side, sharing with a commercial user (e.g., satellites) may impose a negative externality (e.g, interference) upon the astronomers as it may interfere (even faintly) with the measurements they make. Is it possible to manage bandwidth sharing while keeping interference under control, thereby enabling  effective coexistence among rival applications?


A classical approach would involve a direct monetary transfer from the commercial user to the scientific user to compensate for the interference. 
However, this may not be feasible for several reasons. First, there may be absence of market-based signals about the monetary value of uncontaminated scientific use (see \cite{WSRD2015sharing}). 
Second, even if a monetary value can be assigned to the non-commercial use of the resource (for example, based on public and private funding received by the incumbents), compensating them for the negative externality would still be problematic. In fact, such a monetary transfer can appear as researchers profiting from resources held in trust by the government for scientific purposes, making it politically infeasible. What if, instead, the monetary transfer went directly to a government agency? In that case, it is unclear whether the agency would fully account for the benefits of scientific activity when deciding how much commercial access to allow on bands allocated for scientific use; plus, anti-corruption laws may forbid such transfers at all.

The problem is not unique to spectrum. Other resources that governments hold in trust, such as water in national parks or the preservation of natural habitats, share similar features. The benefits of non-commercial use are diffuse and not directly priced in any market. The entity assigned to receive compensation for the negative externality may not fully internalize the benefits of non-commercial activity. For example, river water could be used for both drinking and leisure activities. A firm using the river to dispose of waste may be able to estimate the negative externality on consumption but not leisure activities.
How, then, should one determine efficient allocations in these contexts?

We approach this as a mechanism design problem. We consider three agents: an incumbent currently using a resource (e.g., a spectrum band, a national park, or a marine habitat), a commercial user seeking access,\footnote{For spectrum, access may involve co-channel or adjacent-channel use; our analysis applies to both. For a national park or marine habitat, sharing can include regulated hunting, water rights, or related activities.}\footnote{Note the commercial user may simply be described as an \emph{entrant}. For instance, an entrant can be another federal agency that seeks access for purposes that could also be profitable.} and a regulator with the authority to grant such access.\footnote{In the spectrum case, the regulator could be the Federal Communications Commission, the incumbent a radio telescope, and the commercial user a satellite constellation. In the national reservoir example, the regulator might be the Department of Commerce, the commercial user a fishery, and the incumbent ecological groups advocating against overharvesting.} The  regulator grants access when it sustains its objective to optimize the social welfare. When the resource is shared, the incumbent incurs a negative externality whose magnitude is their private information. In contrast, the resource’s value under the incumbent’s exclusive access is publicly known in monetary terms\footnote{For example, the monetary value of scientific research conducted with telescopes could be approximated by the total annual expenditures for building, operating, and maintaining the instruments and for the salaries of the researchers.} and positive. The commercial user also holds private information about its own potential gains from shared use of the resource. 

While our regulator can use monetary transfers to incentivize the commercial user to disclose their private information, the regulator cannot do so with the incumbent for the reasons discussed above. 
Nonetheless, the regulator can often rely on costly inspections to verify the incumbent's claims about the extent of the negative externality they would suffer.\footnote{In the spectrum example such negative externality is the interference.} To illustrate through our spectrum example, interference measurements can potentially help to increase efficiency by informing the agents when and where coexistence may be problematic. We assume interference measurements to be costly, such as hiring a third party firm to replicate experiments that measure the interference. Certainly there is an opportunity to exploit interference measurements not only in bands dedicated to passive sensing, but also in those allocated to other applications and services, for which passive sensing might complement measurements in dedicated bands. 

Our regulator is further constrained in that it cannot inject subsidies to cover for the cost of inspection or run a surplus from the money raised from the commercial user. Both constraints are driven by political concerns. The first is motivated by the fact that the commercial entrant generating negative externality should be paying for the costly inspection; while in the second, a regulator focused on surplus maximization who does not internalize the non-commercial benefits of the resource will simply sell the resource to the commercial user. 

To summarize, the regulator can incentivize truthful reporting by the commercial user through monetary transfers.\footnote{The commercial user does not have to be a commercial entity as long as it has the ability to make transfers.} Truthful reporting by the incumbent can be incentivized using costly inspection. To ensure that the regulator makes no subsidies or enjoys a surplus, the transfer from the commercial user is used to pay for inspection. The planner's goal is to design an incentive compatible mechanism to maximize efficiency subject to these constraints.\footnote{Our methods extend to variants of this set up, which are discussed in the body of the paper.}

\subsection*{Contributions} 
We study a three-agent mechanism design problem with an incumbent, an entrant (commercial user), and a regulator. 
The model applies to a general setting in which a regulator allocates a contested resource between an incumbent and a commercial user (CU) when (i) the incumbent’s value is possibly non-monetary, (ii) direct compensation monetary transfers from the CU are infeasible and can neither be absorbed by the regulator nor the incumbent, and (iii) inspections are costly.
The regulator maximizes expected welfare subject to incentive compatibility, enforced via costly inspections and a fee used to cover the inspection. We denote the monetary value of the good when it is exclusively granted to the incumbent as $v$, the cost of inspection as $K$, the incumbent's private information about the interference as $\alpha$ (in $[0, \Bar{\alpha}]$), and the commercial user's private information about the expected utility if access is granted as $u$ (in $[0, \Bar{u}]$). Our contributions are:

    \paragraph{Model.} We characterize the regulator’s optimal \emph{direct} mechanism under two interference technologies: (i) \emph{independent} interference, where the externality on the incumbent does not vary with the entrant’s activity; and (ii) \emph{power} interference, where the externality scales with the entrant’s economic activity. The choice of the term \emph{power} comes from the intuition of using the commercial utility as a proxy for the intensity of the commercial activity over the good. In the case of a wireless entrant, transmitting at a higher power could yield more commercial value but it also causes a greater negative externality over scientific users. The choice of the term \emph{independent} comes from the intuition that the visual interference of the trajectory of satellites within the field of view of a telescopes, does not depend on the commercial activity of the satellites. 
    \paragraph{Solution structure.} In both settings we allow the decision variables to be randomized. We find that the optimal mechanisms are deterministic and characterized by two threshold functions: one threshold function determines sharing vs.\ exclusivity, the other threshold function determines whether inspection takes place or not. As the optimal rule is threshold-based and transparent, the reported pairs of private information are mapped into the following decision areas:

\begin{itemize}

\item[]\emph{Exclusivity without inspection} when CU value \(u\) is too low to justify funding an audit.

\item[]\emph{Sharing:}  if \(u>v\), sharing is optimal. Interestingly, for \(v>>\Bar{u}\), there still exists a possibly negligible \(\alpha^\dagger>0\) such that for some profiles \((\alpha,u)\) with \(\alpha<\alpha^\dagger\), sharing remains optimal because expected harm is very low. For instance, in the spectrum sharing example, this may be interpreted as scientific experiments having different silence requirements in the neighboring band. Note however that as $v$ increases, the region of sharing vanishes, thus making sharing viable only for commercial utilities comparable to the market price of the good. 

\item[]\emph{Targeted inspection} for intermediate \(u\): audit if the reported interference crosses the inspection cutoff; implement exclusivity only if verified.

\end{itemize}
    \paragraph{Distortions.}  Costly inspections and incentive constraints generate inefficiencies, at low, intermediate, and high entrant values \(u\). Such inefficiencies distort the structure of the optimal policy with respect to the optimal policy when there is truthful revelation for both the entrant and the incumbent. Low \(u\) is associated with an increased likelihood of exclusive access without running inspection; for intermediate \(u\), exclusivity is granted subject to inspection, and sharing can be both first-best and second-best in order to reduce inspection costs; high \(u\) and high inspection cost results in more sharing. Inefficiencies are also bounded to the values of $v$ and $K$. As the value $v$ increases, the region of sharing diminishes accordingly, vanishing in the limit. As the inspection cost \(K\) rises, the inspection region shrinks and eventually vanishes; the allocation frontier collapses to a single sharing/exclusivity boundary. 
The nature of the intermediate trade-off depends on the nature of the interference taken in consideration:
\begin{itemize}
  \item[]\emph{Independent interference:} inspection takes place when the declared interference (plus some inefficiency)  is greater than CU’s value; sharing occurs only when reported harms are sufficiently low relative to \(u\).
  \item[]\emph{Power interference:} interference scales with \(u\); the inspection cutoff depends solely on the incumbent’s technical parameter \(\alpha\), simplifying screening.
\end{itemize}
    \paragraph{Computation.} The threshold functions are characterized analytically in terms of two parameters $u_{\text{bot}}$ and $u_{\text{top}}$. Thanks to such a characterization, we can reduce the mathematical optimization formulation of the problem as a continuous knapsack problem with an implicit greedy ordering by benefit–to-cost ratio. This makes calibration and policy counterfactuals straightforward.

Since the optimal inspection is deterministic, the regulator either inspects or not with certainty. It is sub-optimal to do a partial inspection or a random allocation, meaning that if an audit is worth doing at all, it is worth doing fully. In practice, this may be  true for resources that are particularly valuable and whose integrity must be preserved, such as water, natural conservation areas and spectrum for national security or critical scientific experiments.  

Regarding implementation, the two allocation and inspection thresholds are computed from prior distributions via a continuous knapsack ordering (benefit–to-cost ratio), which suggests the following implementation guidelines:
\begin{itemize}
  \item[]\emph{Safe harbors:} the regulator may publish the allocation and inspection cutoffs; above the allocation threshold permits sharing; below it, grants exclusivity; the regulator audits only when the inspection threshold is crossed.
  \item[]\emph{Audit targeting:} prioritizing inspections with the highest expected harm-reduction per audit dollar; updating thresholds as data arrive.

  \item[]\emph{Participation guidance:} since the inspection cost \(K\) is public and the thresholds move monotonically in \(K\), a CU can \emph{only} pre-classify its outcome \(u\) into two certainty regions: (i) below the lower cutoff, implying default \emph{exclusivity}; (ii) above the inspection band, implying default \emph{sharing}. Inside the intermediate band, the outcome depends on the incumbent’s private information (which can vary over time), so entry is profitable only under beliefs about that information that yield nonnegative expected surplus.
\end{itemize}

\paragraph{\textbf{Paper structure.}}
Relevant prior literature is summarized next. Section \ref{sec:model} introduces the model followed by Section \ref{sec:properties} detailing key properties of the optimal mechanism. Section~\ref{sec:inefficiency} analyses the inefficiencies for both settings. We
conclude in Section~\ref{sec:conclusions}.  All the relevant proofs can be found in the appendices~\ref{sec:appendix} and \ref{sec:solution}.

\subsection{Related Work}\label{sec:literature}


Our work is related to the literature on mechanism design with inspection that begins with \cite{gl1986}. Since then, there has been a great deal of interest in understanding how the design of mechanisms is affected when the planner can verify (partially or perfectly) the report of an agent's type.\footnote{This is distinct from the agent being able to prove or provide evidence of their type. See, for example, \cite{bull2007hard}.} This verification, often called inspection, is treated as costly~\cite{townsend1979optimal}. Inspection is a substitute to monetary transfers when these are infeasible, for instance as in~\cite{bull2007hard,bpdl2014}. The most closely related papers to our work are those that employ the simultaneous use of both inspection {\em and} monetary transfers. In these papers {\em all} agents can make transfers, in our case there is one agent who {\em cannot} receive or make transfers.

In \cite{patel2022costly}, the planner bears the cost of inspection but agents can also be assessed a transfer which does not make its way to the planner. In their setting, the transfer is interpreted as money-burning. Thus, transfers cannot be used to cover the cost of inspection. In \cite{belloni}, all agents can make and receive transfers, but inspection takes place {\em after} an allocation has been made. In our case, inspection takes place {\em before} an allocation is selected. In~\cite{li2020mechanism,mylovanov2017optimal} if the agent who received the allocation is found to have lied, will incur in a penalty, in our case there is always one agent who incurs in a \lq\lq penalty\rq\rq\,: if the incumbent was truthful, the CU does not have access and has a zero utility, otherwise the incumbent suffers the interference as a punishment. In \cite{https://doi.org/10.3982/TE3907}, the author uses transfers to cover the planner's cost of inspection. However, agent types are two dimensional consisting of value and budget. Inspection reveals the budget component of the type but not the valuation. In our case, the type is one dimensional and only one of the agents is subject to inspection. When types are scalar (single-parameter), truthful implementability reduces to monotone allocations with appropriate threshold payments, enabling truthful mechanisms that run in polynomial time whenever monotonicity can be enforced~\cite{archer2001truthful}. In settings that admit linear programming formulations with suitable structure, polynomial-time truthful mechanisms can also be obtained~\cite{bikhchandani2001linear}. Our environment is equipped with single-parameter types, yet it departs from these linear templates (our underlying formulation is nonlinear); we show that a polynomial-time and truthful mechanism is still attainable.

Our paper is also related to the literature on the use of mechanism design to allocate property rights. Examples are \cite{10.1257/pol.20200426} and \cite{RePEc:fme:wpaper:86}. Both papers are motivated by the same consideration as motivates this paper. Specifically, because of technological changes, it would be efficient to reallocate an initial grant of a property right. These papers argue that because of asymmetric information, a market for the exchange of the property right may not yield an efficient allocation. \cite{10.1257/pol.20200426} focuses on the design of a mechanism to re-allocate the property right ex-post so as to maximize efficiency. \cite{RePEc:fme:wpaper:86}, on the other hand, consider the ex-ante case of designing the initial property right so as to facilitate efficient re-allocation in the future. Our paper focuses on the ex-post case  as in \cite{10.1257/pol.20200426}, but is constrained by the fact that the holder of the property right (the incumbent) cannot accept or make transfers.  Moreover, a parallel algorithmic contract-theory literature treats inspection design itself as a computational problem: a principal chooses payments and allocates a limited inspection budget to induce costly, desirable actions. Optimal deterministic and randomized inspection schemes can be computed under a natural structure (e.g., submodular inspection costs), and inspection probabilities co-move in predictable ways with cost parameters~\cite{ezra2024contracts, fallah2024contract}. Although framed as contracts, these techniques translate to auction-style private-value environments in which the designer can spot-check claims at a cost.

Finally, our paper is inspired by \cite{bobbio2025costly}, which also studies a regulator–incumbent–CU triad, though with one-sided private information: a linear interference borne by the incumbent under sharing. We depart in two respects. First, we allow two-sided private information (the CU’s value and the incumbent’s harm). Second, as the regulator covers the cost of inspection, they place inspection costs in the objective, which changes both the objective’s form and the audit design (and thus the induced inspection probabilities). Closest in application to spectrum markets,~\cite{jia2009revenue} designs truthful auctions for dynamic spectrum access that maximize seller revenue under interference constraints, combining a Myerson-style optimal auction with a polynomial-time truthful approximation; while not using inspection, their allocation/payment design informs auction-style private-value settings with complex feasibility constraints.

\section{Model}\label{sec:model}

In our model, three agents interact: the \emph{incumbent} (\textit{I}), the \emph{commercial user} (\textit{CU}), and a \emph{regulator}. The incumbent, representing the scientific user in our motivating spectrum example, may be required to share a resource with the commercial user. If sharing occurs, the incumbent suffers a negative externality.
\footnote{The analysis extends easily to the case of multiple commercial users who impose an identical negative externality. For economy of exposition only, we restrict attention to a single commercial user.}

The utility enjoyed by \textit{CU} from sharing the resource is $u \in [0, \Bar{u}] $ with $\Bar{u}>0$, and is their private information. Agent \textit{I} derives value $v > 0$ from sole use of the resource which is common knowledge but bears a cost $t(\alpha,u)$ from having to share it with \textit{CU}, where $\alpha \in [0, \Bar{\alpha}]$  is a parameter that summarizes \textit{I}'s private information. 
Assume that $\alpha$ and $u$ are independent draws from  distributions with density functions $f$ over $[0, \Bar{\alpha}]$ and $g$ over $[0, \Bar{u}]$, respectively. Denote the corresponding cumulative distribution functions by $F$ and $G$, respectively. Densities are assumed to be positive and continuous over their support. We focus on a class of density functions $g$ that has a decreasing survival mass, specifically we require that $r(u):=1-G(u) -u g(u) = (1-G(u))(1-u h(u))$ is decreasing in $u$, where $h(u)= \frac{g(u)}{1-G(u)}$ is the hazard rate.\footnote{Another alternative condition could be that the cost of inspection $K$ is sufficiently large so that $K> \sup_{u>v} r(u)$. These conditions are usual in mechanism design, see for example~\cite{myerson1981optimal}.}

We examine two specifications for $t(\alpha, u).$  The first is when $t(\alpha, u) = \alpha$ (so that $t(\alpha, u)\in[0,\Bar{\alpha}]$). Hence, the burden borne by  \textit{I} is independent of \textit{CU}'s intensity of activity. The second specification is $t(\alpha, u)= \alpha u$ (here, $t(\alpha, u) \in [0,\Bar{\alpha}\Bar{u}]$).  This can be interpreted as the externality scaling with the `volume' of \textit{CU}'s activities if one uses $u$ as a proxy for `volume.' 
%
In both cases \textit{I}'s net payoff from sharing is $\max \{v- t(\alpha,u), 0\}$; in words, if $t(\alpha,u) \geq v$, \textit{I} may cease using the resource. In the context of our motivating example, we are assuming that \textit{I} {\em internalizes} all the benefits from deploying the resource. We can easily accommodate, at the expense of additional notation, the scenario where \textit{I} internalizes a fraction of $v$ only. For example, \textit{I} cares only about the immediate scientific payoff and not the positive spillovers generated by discovery. 

The regulator must decide whether \textit{I} can use the resource exclusively or share it with \textit{CU} with the goal to maximize the sum of \textit{I} and \textit{CU}'s payoffs. By the revelation principle, the regulator can employ a direct mechanism. The first component of the mechanism we describe is the allocation rule. If \textit{I} reports $\alpha$ and \textit{CU} reports $u$, let $a(\alpha,u)\in [0,1]$ be the probability with which the regulator assigns exclusive use to \textit{I}. Thus, $1-a(\alpha,u)$ is the probability with which the regulator allows sharing.
The regulator's objective function is social welfare, which is given by

\begin{equation}\label{expression:objective_function}
    v a(\alpha,u) + (1-a(\alpha,u))(\max(v-t(\alpha,u),0) +u).
\end{equation} 

For a cost $K$, the regulator can inspect the veracity of any report of \textit{I}. Inspection is assumed to be perfect and reveals the magnitude of the negative externality. If it is discovered that \textit{I} has lied, it is without loss to suppose that \textit{I} will be denied exclusive use. So, the second component of the mechanism is the inspection rule $c(\alpha,u)\in [0,1]$, where $c(\alpha,u)$ denotes the probability with which \textit{I}'s report will be inspected given \textit{I} reported $\alpha$, and \textit{CU} reported $u$ . As it makes no sense to inspect \textit{I} if one decides to allow \textit{CU} to share, it follows that $c(\alpha,u) \leq a(\alpha,u) .$ If the incumbent is found to be lying, the regulator punishes with sharing, otherwise the incumbent is prized with exclusive access.\footnote{Note that the allocation of exclusive access \emph{or} sharing follows after the inspection.} 

The regulator chooses the allocation-inspection policy $(a(\cdot, \cdot), c(\cdot, \cdot))$ to give \textit{I} an incentive to truthfully report $\alpha$. Hence, for every report $u$ of \textit{CU}, if \textit{I} is of type $\alpha$ and reports type $s$ instead, we need: 
\begin{equation}\label{eq:IC_SU}
\begin{split}
    v a(\alpha,u) + (1 - a(\alpha,u)) \max(v - t(\alpha,u), 0) \geq \, c(s,u) \max(v - t(\alpha,u), 0) + \\
    + (1 - c(s,u)) \big(v a(s,u) + (1 - a(s,u)) \max(v - t(\alpha,u), 0)\big), \,\, \forall s \neq \alpha.
\end{split}
\end{equation}
The left-hand side of (\ref{eq:IC_SU})  is \textit{I}'s expected payoff when truthfully reporting their type, while the right-hand side is the expected payoff when reporting type $s$ instead. 
Rearranging terms we get

\begin{align*}
&\bigl[v - \max (v-t(\alpha,u),0)\bigr]a(\alpha,u) + \max (v-t(\alpha,u),0) \ge va(s,u)
  +\bigl(1 - a(s,u)\bigr)\max (v-t(\alpha,u),0)  
  \\
&\quad \quad\quad\quad\quad\quad\quad\quad\quad\quad +\Bigl[\max (v-t(\alpha,u),0) - v a(s,u) -\bigl(1 - a(s,u)\bigr) \max (v-t(\alpha,u),0)\Bigr] 
    c(s,u)
  \\[6pt]
&\quad \Rightarrow 
  \min (v, t(\alpha,u)) a(\alpha,u) + \max (v-t(\alpha,u),0)
  \ge \\
  & \quad\quad\quad\quad\quad -\min (v, t(\alpha,u)) a(s,u) c(s,u) 
  + \min (v, t(\alpha,u)) a(s,u)
  +\max (v-t(\alpha,u),0) 
  \\[6pt]
&\quad \Rightarrow 
  \min (v, t(\alpha,u)) a(\alpha,u) 
  \ge
  - \min (v, t(\alpha,u)) a(s,u) c(s,u) 
  + \min (v, t(\alpha,u)) a(s,u).
\end{align*}
When \(t(\alpha,u)\) is non-zero, the incumbent's incentive compatibility constraint becomes
\begin{equation}\label{formula:IC_incumbent}
    a(s,u) \;-\; a(s,u)\,c(s,u) \;\le\; a(\alpha,u).
\end{equation}
We  assume that $a(s,u) \geq a(0,u)$ for all $s$ and call this property $\alpha$-monotonicity. It can be derived as a consequence of incentive compatibility and optimality, but the proof is tedious and uninstructive.\footnote{A similar derivation made in a different setting is detailed in~\cite{bobbio2025costly}.} Thus we choose to impose it, as it is a reasonable condition. As constraint (\ref{formula:IC_incumbent}) must hold for all $\alpha$, it must hold with $\alpha =0$, i.e.  
\begin{equation}\label{formula:IC_incumbent_0}
    a(s,u) \;-\; a(s,u)\,c(s,u) \;\le\; a(0,u).
\end{equation}
Note that $\alpha$-monotonicity and (\ref{formula:IC_incumbent_0}) imply (\ref{formula:IC_incumbent}).\footnote{Note that (\ref{formula:IC_incumbent}) is equivalent to $\alpha$-monotonicity and (\ref{formula:IC_incumbent_0}).} Thus, we replace it with constraint~\eqref{formula:IC_incumbent_0} along with the $\alpha$-monotonicity constraints.

To incentivize truthful reporting of $u$ by \textit{CU}, the regulator can use monetary transfers. One can interpret this as as an entry fee that allows participation. Hence, the third component of the mechanism is a rule $p(\alpha,u)$ that specifies the payment \textit{CU} must make when \textit{I} reports $\alpha$ and \textit{CU} reports $u.$ If we impose dominant strategy incentive compatibility for \textit{CU} as well, standard arguments tell us that $a(\alpha,u)$ is decreasing in $u$ for each fixed $\alpha$ (see \cite{vohra2011mechanism})  and 
\begin{equation}
 \label{assumption:two} p(\alpha,u) = u(1-a(\alpha,u)) - \int_0^u \big(1-a(\alpha,x)\big) dx = \int_0^ua(\alpha,x)dx - ua(\alpha,u) = - \int_0^u x \frac{\partial a(\alpha,x)}{\partial x}dx.
\end{equation}

The payment collected is used to cover the cost of inspection, so that there is no transfer to or from agent \textit{I}. The {\em expected} cost of inspection is, 
$$K \int_0^{\Bar{u}} g(u) \int_0^{\Bar{\alpha}} f(\alpha) c(\alpha,u)d\alpha du.$$
The following ex-ante budget balance constraint ensures that the expected cost of inspection is covered by the transfer from \textit{CU}:
$$\int_0^{\Bar{u}}g(u)\int_0^{\Bar{\alpha}}f(\alpha)p(\alpha,u) d\alpha du = K \int_0^{\Bar{u}}g(u)\int_0^{\Bar{\alpha}} f(\alpha)c(\alpha,u) d\alpha du,$$
which becomes,
$$\int_0^{\Bar{u}}g(u)\int_0^{\Bar{\alpha}}f(\alpha)\Big[  \int_0^ua(\alpha,x)dx - ua(\alpha,u) \Big]d\alpha du = K \int_0^{\Bar{u}}g(u)\int_0^{\Bar{\alpha}} f(\alpha)c(\alpha,u) d\alpha du.$$

Recall that $r(u):= \frac{1-G(u)}{g(u)}-u$, captures a hazard rate, which we assume to be is decreasing in $u$. Thanks to Fubini, we can rearrange the left-hand side as follows, details can be found in the proof of Lemma~\ref{lemma:sufficient_small_value_K} (in Appendix~\ref{sec:appendix}).

$$\int_0^{\Bar{\alpha}}f(\alpha) \int_0^{\Bar{u}}g(u) r(u) a(\alpha,u) du d\alpha = K \int_0^{\Bar{u}}g(u)\int_0^{\Bar{\alpha}} f(\alpha)c(\alpha,u) d\alpha du.$$

Before we present the complete mechanism design problem, we rearrange the terms of the regulator's objective function (\ref{expression:objective_function}) as follows:
$$\big(\min(v,t(\alpha,u))-u\big)a(\alpha,u) + \max(v- t(\alpha,u),0) +u.$$
This allows us to ignore the constant terms $\max(v- t(\alpha,u),0) +u.$ 
The regulator's optimization problem therefore takes the following form.

\begin{subequations}\label{formulation:main_problem}
\begin{alignat}{2}
\max_{a,c} \quad & \int_0^{\Bar{u}} g(u) \int_0^{\Bar{\alpha}} f(\alpha) \big( \min(v , t(\alpha,u) ) - u \big) \cdot a(\alpha,u) \, d\alpha \, du \label{formulation:main_problem_reduced_obj}\\
\text{ s.t.   }\quad & \notag\\
&  \int_0^{\Bar{\alpha}}f(\alpha) \int_0^{\Bar{u}}g(u) r(u) a(\alpha,u) du d\alpha  = K \int_0^{\Bar{u}} g(u) \int_0^{\Bar{\alpha}} f(\alpha) c(\alpha,u)  d\alpha  du \label{formulation:main_problem_reduced_budget} \\
&   a(\alpha,u) - a(\alpha,u)c(\alpha,u) \leq a(0,u) \quad\quad\quad \forall \alpha,\, \forall u \label{formulation:main_problem_reduced_IC} \\
 &  c(\alpha,u) \leq a(\alpha,u) \quad\quad\,\,\quad\quad\quad\quad\quad\quad\,\,\,\,\,\, \forall \alpha,\, \forall u  \label{formulation:main_problem_reduced_inspection}\\
&  a(\alpha,u) \leq a(\alpha,u') \quad\quad\,\,\quad\quad\quad\quad\quad\quad\,\,\, \forall \alpha,\, \forall u \geq u'  \label{formulation:main_problem_reduced_umono}\\
&   a(0,u) \leq a(\alpha,u) \quad\quad\,\,\quad\quad\quad\quad\quad\quad\,\,\, \forall u,\, \forall  \alpha\label{formulation:main_problem_reduced_tmono}\\
&  0 \leq a(\alpha,u) \leq 1 \quad\quad\,\,\quad\quad\quad\quad\,\quad\quad\quad \forall \alpha,u, \\
&  0 \leq c(\alpha,u) \leq 1 \quad\quad\,\,\quad\quad\quad\quad\,\,\quad\quad\quad \forall \alpha,u.
\end{alignat}
\end{subequations}

Constraint~\eqref{formulation:main_problem_reduced_budget} is the budget constraint, making sure that all the budget raised from the entry-fee imposed on the CU will be used for inspection purposes. Constraints~\eqref{formulation:main_problem_reduced_IC} are the incentive compatibility constraints for \textit{I}. Constraints~\eqref{formulation:main_problem_reduced_inspection} allow for inspection only when \textit{I} has positive probability of receiving exclusive access, while constraints~\eqref{formulation:main_problem_reduced_umono} and~\eqref{formulation:main_problem_reduced_tmono} are monotonicity requirements over $u$ and $\alpha$, respectively.

In this paper we characterize the optimal solution for the case when $K$ is sufficiently large, consistent with our motivating example. See Lemma~\ref{lemma:sufficient_small_value_K} in Appendix~\ref{sec:appendix} for the precise threshold $K_{\text{low}}$ that depends only on $v,g$ and $f$. The case of small $K$ raises some interesting conceptual issues. To see why, suppose $K=0$ and $t(\alpha, u) =\alpha$. 
In this case the regulator can costlessly determine the true value of $\alpha$. However, $K=0$ and the inability to run a surplus, mean the regulator {\em cannot} assess \textit{CU} a transfer. Thus, the regulator is unable to determine any information about $u$ beyond what is contained in the prior. Therefore, the decision to allow sharing or not reduces to a comparison of $v$ with $\max\{v-\alpha, 0\} + \mathbb{E}(u).$ If $v \geq \alpha$, sharing is always allowed. If $v < \alpha$, sharing is allowed only if $v < \mathbb{E}(u)$. Hence, for small $K$, the inability to run a surplus severely restricts the regulators ability to identify an efficient allocation.

One could propose an alternative specification where the regulator is allowed to run a surplus, but the surplus must be `burnt' to ensure incentive compatibility for both \textit{I} and \textit{CU}.
To model this, it is sufficient to relax the budget constraint into an inequality to guarantee that the entry-fee on the CU is sufficient to cover for the cost of inspection. The slack variable for this constraint would correspond to the amount burnt and appear with a coefficient of -1 in the objective function. Mathematically, the problem is similar to the one we are solving and our methods would extend to it. However, as we consider money burning in our motivating applications to be implausible, we do not pursue it. Nevertheless, we think it important to highlight how constraining the no surplus requirement is when $K$ is small. For a qualitative analysis of the optimal solution please refer to Appendix~\ref{sec:appendix:small_K}. 

\section{Properties of the Optimal Mechanism}\label{sec:properties}

Denote the optimal solution  to Problem \eqref{formulation:main_problem} by $(a^\circ(\alpha,u), c^\circ(\alpha,u))$. The next result shows that it must be deterministic. Thus, allocations are deterministic and there is no
random inspection. 

\begin{theorem}\label{theo:binary_optimal_solution_mainbody}
  Let $(a^\circ, c^\circ)$ be an optimal solution of Formulation~\ref{formulation:main_problem}.
  Then, for every $(\alpha,u)$ in the domain, both $a^\circ(\alpha,u)$ and $c^\circ(\alpha,u)$ are binary almost everywhere. 
\end{theorem}


Interestingly, Lemma~\ref{lemma:alpha_full_monotonicity} in Appendix~\ref{sec:solution} shows that $a^0(\alpha,u)$ is non-decreasing in $\alpha$. As $a^\circ(\alpha,u)$ is non-decreasing in $\alpha$ and \emph{non-increasing} in $u$, the domain over which the optimal policy imposes sharing must be connected. Specifically, for each fixed $\alpha$, the function $u \mapsto a^\circ(\alpha,u)$ takes on binary values (either $1$ or $0$) and is monotonic in $u$. Consequently, it can only \emph{switch} from $1$ to $0$ at most once. Hence, for each $\alpha$, there is a unique threshold in $u$ that determines whether sharing is imposed. 
Denote it by $\phi(\alpha,u,K,v)$, where $K$ and $v$, recall, are common knowledge; hence, we simply write $\phi(\alpha,u)$. Formally, the optimal allocation rule can be expressed as follows:
\[
  a^\circ(\alpha,u) \;=\;
  \begin{cases}
    1 & \text{if } u \leq \phi(\alpha,u),\\[6pt]
    0  & \text{if } u > \phi(\alpha,u).
  \end{cases}
\]
Because $a^\circ(\alpha,u)$ is also non-decreasing in $\alpha$, the threshold function $\phi(\alpha,u)$ itself inherits the non-decreasing monotonic dependence on $\alpha$. In fact, $\phi(\alpha,u)$ can be described using just two cutoffs as formalized in the next two sections. 

\subsection{Optimal Policy for Independent Interference}

In this section we consider  $t(\alpha,u)=\alpha$, and provide the optimal threshold for both the allocation and inspection policy.

\begin{theorem}\label{thm:threshold_mainbody}
Given $K,v\geq 0$, there exist two constants $u_{\mathrm{bot}}$ and $u_{\mathrm{top}}$ satisfying
  \[
    0 \;\le\; u_{\mathrm{bot}} 
          \;\le\; u_{\mathrm{top}}
          \;\le\; v,
  \]
  such that the optimal allocation policy $\phi(\alpha,u)$ takes the following form:
  \[
    \phi(\alpha,u) \;=\;
    \begin{cases}
      u_{\mathrm{top}}, & \text{if } \alpha \,\ge\, v,\\[6pt]
      \alpha- (v-u_{\mathrm{top}}),                & \text{if } u_{\mathrm{bot}} + v-u_{\mathrm{top}} \,<\, \alpha \,<\, v,\\[6pt]
      u_{\mathrm{bot}}, & \text{if } \alpha \,\le\, u_{\mathrm{bot}}+v-u_{\mathrm{top}}.
    \end{cases}
  \]
\end{theorem}


Note that in this case, $\phi(\alpha,u)$ only depends on $\alpha$. Moreover, the two constants $u_{\mathrm{bot}}$ and $u_{\mathrm{top}}$ are necessary for an analytical description of the allocation policy as they separate different policy regions over the type space $(\alpha,u)$.  As it turns out, they are also crucial for the description of the optimal inspection policy.

Define a threshold function $\psi(\alpha,u,K,v, u_{\mathrm{bot}}, u_{\mathrm{top}})$ for the choice of inspection. The optimal inspection policy can be written as

\[
  c^\circ(\alpha,u) \;=\;
  \begin{cases}
     1 & \text{if } u_{\mathrm{bot}} < u < \psi(\alpha,u,K,v, u_{\mathrm{bot}}, u_{\mathrm{top}}),\\[6pt]
    0  & \text{otherwise}. 
  \end{cases}
\]

As for the threshold function for the optimal allocation policy, we can characterize the threshold for the optimal inspection policy.

\begin{corollary}
    Given $\alpha$, $K$, $u_{\mathrm{bot}}$, and $u_{\mathrm{top}}$  the threshold function $\psi(\alpha,K,v, u_{\mathrm{bot}}, u_{\mathrm{top}})$ takes the following form:
  \[
    \psi(\alpha,u,K,v, u_{\mathrm{bot}}, u_{\mathrm{top}}) \;=\;
    \begin{cases}
      u_{\mathrm{top}}, & \text{if } \alpha \,\ge\, v,\\[6pt]
      \alpha - (v-u_{\mathrm{top}}),                & \text{if } u_{\mathrm{bot}} + v-u_{\mathrm{top}} \,<\, \alpha \,<\, v,\\[6pt]
      0, & \text{if } \alpha \,\le\, u_{\mathrm{bot}} + v-u_{\mathrm{top}}.
    \end{cases}
  \]
\end{corollary}

\begin{figure}
    \centering
    \includegraphics[width=0.65\linewidth]{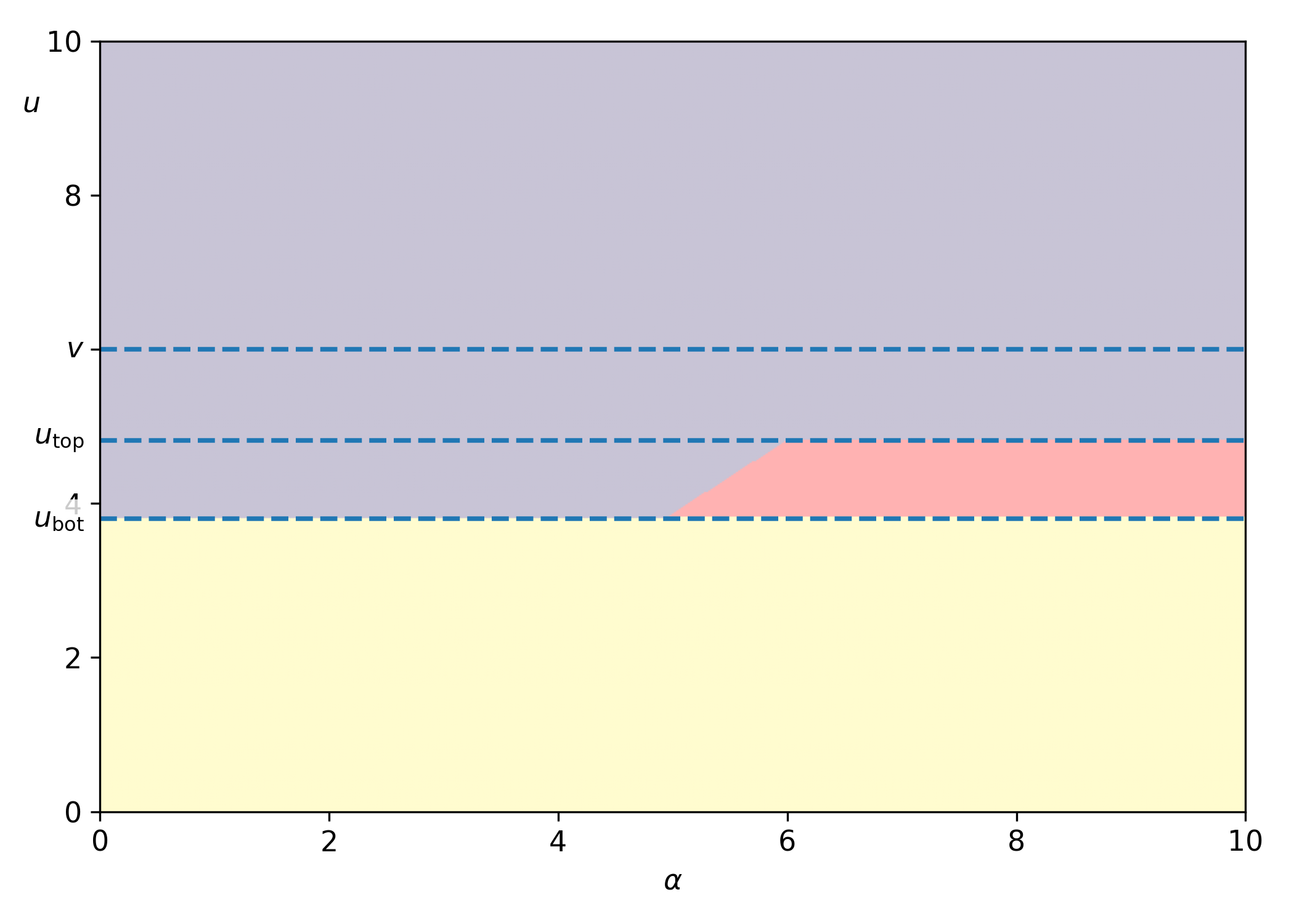}
    \caption{Optimal allocation and inspection under independent interference. Gray = sharing (no inspection). Yellow = default exclusivity (no inspection). Red = exclusivity only if an audit verifies the interference; otherwise sharing.}
    \label{fig:fig1_indep}
\end{figure}
 
We consider Figure~\ref{fig:fig1_indep} as an illustrative example. Unless noted otherwise, numerical examples use Gaussian $f$ and $g$ on $[0,10]$ and $[0,10]$ with $(\mu=5,\sigma=7)$, $v=6$ and $K=20$. In Figure~\ref{fig:fig1_indep}, we also label the threshold values of $u_{\mathrm{bot}}$ and $u_{\mathrm{top}}$. 

The threshold function partitions the type space $(\alpha,u)$ into three regions, that we color differently. 
Light yellow indicates \emph{exclusivity without inspection}; gray indicates \emph{sharing} (where inspection is not needed); red indicates the \emph{inspection band}, where an audit is run and exclusivity is implemented only if the reported interference is verified, otherwise sharing applies; by incentive compatibility, in equilibrium the inspection verifies truthful reporting. Let us now analyze these regions from the perspective of the magnitude of $u$.

For small values of $u$ ($u\leq u_{\mathrm{bot}}$) the regulator assigns by default exclusive access to the incumbent, this happens when the budget raised from $u$ is not sufficient to cover the cost of inspection and the contribution of \textit{CU} to the social welfare is not large enough to justify sharing.  

The other edge scenario is when $u$ is large enough ($u\geq u_{\mathrm{top}}$); in this case, the contribution of \textit{CU} to the social welfare is sufficiently large to justify sharing independently of the value of the interference incurred by the incumbent. 

Finally, for values of $u$ between $u_{\mathrm{bot}}$ and $u_{\mathrm{top}}$, the regulator allocates exclusive access when the value of the declared interference is higher than $u$ plus an inefficiency (caused by the elevated cost of inspection).

%

Thus, when the the commercial utility $u $, is `small' or `large' the regulator decides whether to mandate sharing or not by comparing $u$ with the  fixed thresholds $ u_{\mathrm{bot}}$ and $ u_{\mathrm{top}}$. Only when $u$ is in an intermediate range, does the regulator employ inspection and decides based on a direct comparison of $u$ with $\alpha- (v-u_{\mathrm{top}})$. 

In the optimal mechanism, an incumbent who reports a low $\alpha$ can still be allocated exclusive use with positive probability, since the regulator does not find it worthwhile to pay for inspection when the budget raised from $u$ is insufficient to cover the cost.

The threshold function is characterized by the values of \(u_{\mathrm{top}}\) and \(u_{\mathrm{bot}}\), which depend on the parameters $K$ and $v$. Next, we show how to determine the specific values of \(u_{\mathrm{top}}\) and \(u_{\mathrm{bot}}\) for the optimal mechanism, given $v$ and $K$.

\begin{theorem}\label{thm:knapsack_independent_main} 
  Let \(K,v \geq 0\), the solution to the \emph{Knapsack problem} given in Formulation~\ref{formulation:knapsack_main} characterizes a threshold function $\phi,$ for the allocation rule. 
  \begin{subequations}\label{formulation:knapsack_main}
  \begin{alignat}{1}
  \max_{u_{\mathrm{top}},\,u_{\mathrm{bot}}} 
  \,  \quad & \int_{0}^{u_{\mathrm{bot}}} g(u) \int_{0}^{\Bar{\alpha}} f(\alpha) (\min(v,\alpha) -u ) \mathrm{d}\alpha  \mathrm{d}u  \, + \notag \\
     &    \quad\quad\quad\quad\quad\quad\quad\quad\quad\quad\quad\quad\quad + 
     \int_{u_{\mathrm{bot}}}^{u_{\mathrm{top}}} g(u) \int_{u+ v - u_{\mathrm{top}}}^{\Bar{\alpha}} f(\alpha)  (\min(v,\alpha) -u )  \mathrm{d}\alpha        
             \mathrm{d}u
         \label{formulation:knapsack_obj_main}\\
  \text{s.t.}\quad 
& \int^{u_{\mathrm{bot}}}_{0} \int_0^{\Bar{\alpha}} f(\alpha)  \left( 1-G(u)-ug(u) \right) \mathrm{d}\alpha \mathrm{d}u \, + 
   \notag \\
      &   
 + \int_{u_{\mathrm{bot}}}^{u_{\mathrm{top}}} \int^{\Bar{\alpha}}_{u+v
-u_{\mathrm{top}} } f(\alpha)\left( 1-G(u)-ug(u) \right) \mathrm{d}\alpha \mathrm{d} u
       \geq   K \int_{u_{\mathrm{bot}}}^{u_{\mathrm{top}}} g(u) \int_{u+v-u_{\mathrm{top}} }^{\Bar{\alpha}} f(\alpha) \mathrm{d}\alpha \mathrm{d}u, 
      \label{formulation:knapsack_budget_main}\\
  & 0 \;\le\; u_{\mathrm{bot}} \;\le\;  u_{\mathrm{top}} \;\le\; v.
  \end{alignat}
  \end{subequations}
\end{theorem}

Note that Formulation~\ref{formulation:knapsack_main} is a classical continuous \emph{Knapsack problem}. Formally, let
$
V_1(u_{\mathrm{bot}}) 
:= \int_{0}^{u_{\mathrm{bot}}} g(u)\int_{0}^{\bar{\alpha}} f(\alpha)\bigl(\min\{v,\alpha\}-u\bigr)\,\mathrm d\alpha\,\mathrm du$ and 
$ V_2(u_{\mathrm{bot}},u_{\mathrm{top}})
:= \int_{u_{\mathrm{bot}}}^{u_{\mathrm{top}}} g(u)\int_{u+v-u_{\mathrm{top}}}^{\bar{\alpha}} f(\alpha)\bigl(\min\{v,\alpha\}-u\bigr)\,\mathrm d\alpha\,\mathrm du$,  
and rewrite the budget constraint by moving the right–hand side to the left as 
$W_1(u_{\mathrm{bot}}) + W_2(u_{\mathrm{bot}},u_{\mathrm{top}}) \;\ge 0,$
where
$W_1(u_{\mathrm{bot}})
:= \int_{0}^{u_{\mathrm{bot}}}\int_{0}^{\bar{\alpha}} f(\alpha)\bigl(1-G(u)-u g(u)\bigr)\,\mathrm d\alpha\,\mathrm du$ and 
$W_2(u_{\mathrm{bot}},u_{\mathrm{top}})
:= \int_{u_{\mathrm{bot}}}^{u_{\mathrm{top}}}\int_{u+v-u_{\mathrm{top}}}^{\bar{\alpha}} 
f(\alpha)\bigl(1-G(u)-u g(u)-K g(u)\bigr)\,\mathrm d\alpha\,\mathrm du$.
The problem can then be written as
\[
\max_{0\le u_{\mathrm{bot}}\le u_{\mathrm{top}}\le v} \; V_1(u_{\mathrm{bot}})+V_2(u_{\mathrm{bot}},u_{\mathrm{top}})
\, \text{ subject to }\,
W_1(u_{\mathrm{bot}})+W_2(u_{\mathrm{bot}},u_{\mathrm{top}})\ge 0.
\]
Thus \(u_{\mathrm{bot}}\) and \(u_{\mathrm{top}}\) continuously choose the “sizes’’ of two blocks (the two integrals), each with a value \(V_i\) and a net budget contribution \(W_i\), subject to a single linear constraint. The second block has  non-positive marginal contribution to the constraint because of the \(-K g(u)f(\alpha)\) term in \(W_2\) and since we are assuming $K\geq K_{\text{low}}$ (see Lemma~\ref{lemma:sufficient_small_value_K});  so expanding it tightens the budget while increasing the objective, exactly as in a continuous knapsack trade-off.

Moreover, in the continuous setting the optimization over $(u_{\mathrm{bot}},u_{\mathrm{top}})$ can be carried out in polynomial time in the desired accuracy as follows. Fix $u_{\mathrm{bot}}$ and write
\[
\Psi(u_{\mathrm{bot}},u_{\mathrm{top}}):=W_1(u_{\mathrm{bot}})+W_2(u_{\mathrm{bot}},u_{\mathrm{top}}).
\]
Under $K\ge K_{\text{low}}$, the map $z\mapsto\Psi(u_{\mathrm{bot}},z)$ is continuous and strictly decreasing on $[u_{\mathrm{bot}},v]$, so the optimal
\[
u_{\mathrm{top}}^*(u_{\mathrm{bot}}):=\sup\{z\in[u_{\mathrm{bot}},v]:\Psi(u_{\mathrm{bot}},z)\ge 0\}
\]
is the unique root where the constraint binds and can be found by bisection in $O(\log(1/\varepsilon))$ evaluations of $\Psi$. Since $V_2(u_{\mathrm{bot}}, \cdot)$ is increasing in the range $[u_{\mathrm{bot}},u^*_{\mathrm{top}}(u_{\mathrm{bot}})]$, then $V_2(u_{\mathrm{bot}}, u^*_{\mathrm{top}}(u_{\mathrm{bot}}))$ is the optimal value. 
Substituting $u_{\mathrm{top}}^*(u_{\mathrm{bot}})$  reduces the problem to the one–dimensional maximization
\[
\max_{0\le u_{\mathrm{bot}}\le v} \Phi(u_{\mathrm{bot}}):=V_1(u_{\mathrm{bot}})+V_2\bigl(u_{\mathrm{bot}},u_{\mathrm{top}}^*(u_{\mathrm{bot}})\bigr),
\]
which, under the same regularity assumptions, can again be solved up to precision $\varepsilon$ by a one–dimensional search (e.g.\ bisection or ternary search) in $O(\log(1/\varepsilon))$ evaluations of $\Phi$. Hence the continuous knapsack problem can be solved to $\varepsilon$–accuracy in overall time $O(\log^2(1/\varepsilon))$, up to the cost of evaluating the primitives $F,G,f,g$ and the associated integrals.

\subsection{Optimal Policy for Power Interference}

In this section, we turn our focus on the analytical description of the  allocation-inspection optimal policy for the power interference $t(\alpha,u)= \alpha\cdot u$. As for the independent interference setting, two threshold values $ u_{\mathrm{bot}}$ and $ u_{\mathrm{top}}$ conveniently help determine the optimal policies. The main difference stands in the fact that for intermediate values of $u$, the slope of the threshold is constant in $u$. The intuition is that the interference already takes into account the intensity of the CU's activity, hence, when considering intermediate values of $u$, the determinant factor deciding whether sharing is allowed or not, is the private value of $\alpha$.

Before we formalize this intuition, let us define

\[  
C_0(u_{\mathrm{bot}}) \;:=\; \Bigl(\int_{0}^{\bar{\alpha}} f(\alpha)\,d\alpha\Bigr)\Bigl(\int_{0}^{u_{\mathrm{bot}}} g(u)\,r(u)\,du\Bigr), \text{ and }
B(u_{\mathrm{bot}},u_{\mathrm{top}}) \;:=\; \int_{u_{\mathrm{bot}}}^{u_{\mathrm{top}}} g(u)\,\bigl(r(u)-K\bigr)\,du,
\]

where $B<0$ for $K>K_{\text{low}}$ by the decreasing monotonicity condition on $r(u)$. Moreover, we can re-write the budget raised starting at $\alpha=x$ as
\[
C(x,u_{\mathrm{bot}},u_{\mathrm{top}})=C_0(u_{\mathrm{bot}})+\int_{u_{\mathrm{bot}}}^{u_{\mathrm{top}}}\!g(u)\!\int_{x}^{\bar{\alpha}} f(\alpha)\,\bigl(r(u)-K\bigr)\,d\alpha\,du
\;=\; C_0(u_{\mathrm{bot}}) + B(u_{\mathrm{bot}},u_{\mathrm{top}})\bigl(1-F(x)\bigr).
\]

\begin{figure}
    \centering
    \includegraphics[width=0.65\linewidth]{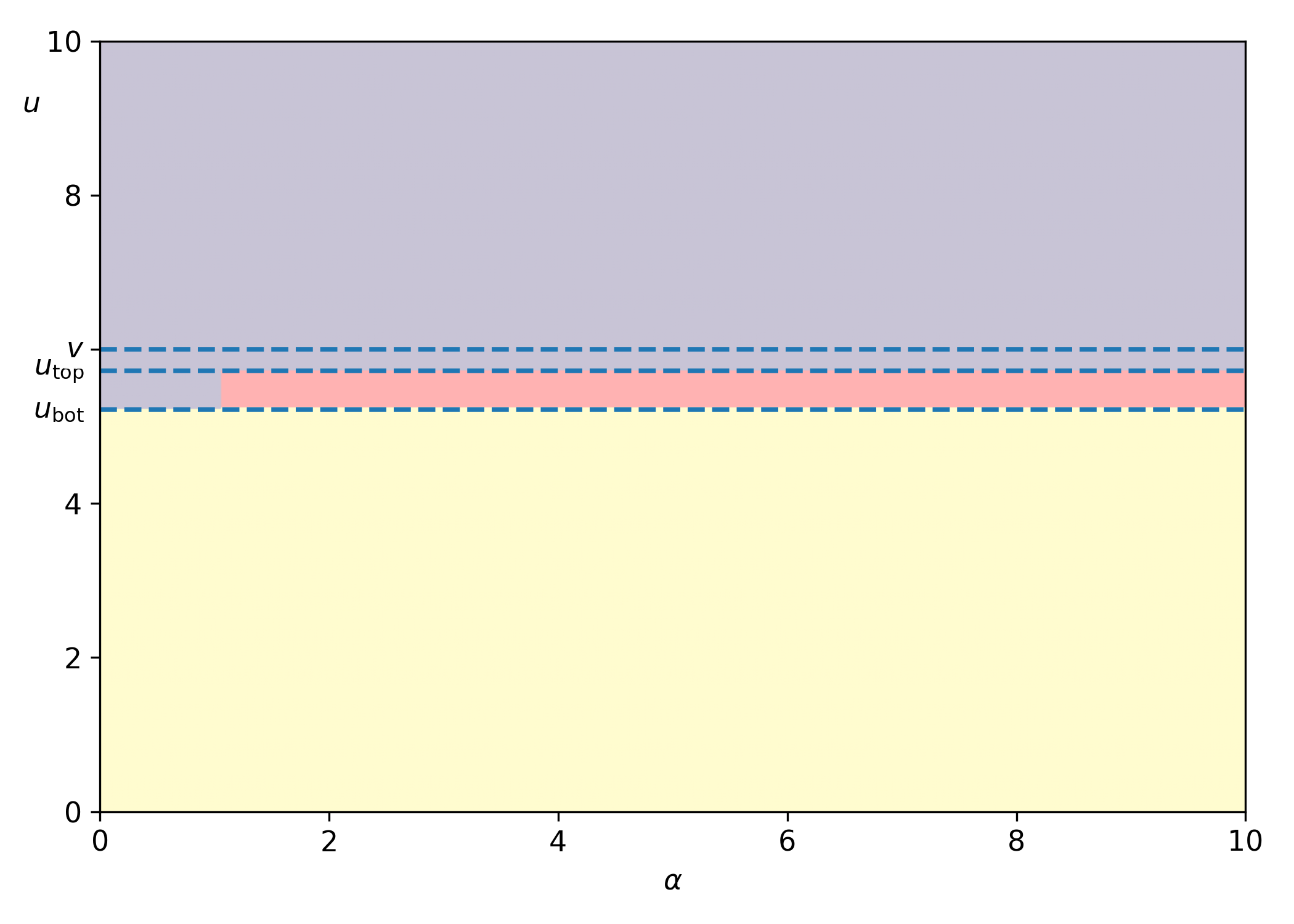}
    \caption{Optimal allocation under power interference. Gray = sharing. Yellow = default exclusivity (no inspection). Red = exclusivity only if an audit verifies the interference; otherwise sharing.}
    \label{fig:fig2_power}
\end{figure}

For easier reference, once $u_{\mathrm{bot}},u_{\mathrm{top}}$ are fixed, we denote $C(x,u_{\mathrm{bot}},u_{\mathrm{top}})$ by $C(x)$, and $C_0(u_{\mathrm{bot}})$ by $C_0$. 
Throughout this section, we assume $F$ is absolutely continuous with a positive density function on $[1,\bar{\alpha}]$, hence we also assume $\Bar{\alpha}>1$. 

\begin{theorem}\label{thm:main_threshold_power}
  Let $K,v \geq 0$. There exist two constants $u_{\mathrm{bot}}$ and $u_{\mathrm{top}}$ satisfying
  \[
    0 \;\le\; u_{\mathrm{bot}} 
          \;\le\; u_{\mathrm{top}}
          \;\le\; v,
  \]
  such that the threshold allocation function $\phi(\alpha,u)$ takes the following piecewise form.
  \[
\phi(\alpha,u)=
\begin{cases}
u_{\mathrm{top}}, & \text{if }\, \dfrac{\alpha}{\alpha^{\mathrm{opt}}} \cdot u\ge u_{\mathrm{top}},\\[6pt]
\dfrac{\alpha}{\alpha^{\mathrm{opt}}} \cdot u, & \text{if }\, u_{\mathrm{bot}}< \dfrac{\alpha}{\alpha^{\mathrm{opt}}} \cdot u< u_{\mathrm{top}},\\[10pt]
               u_{\mathrm{bot}}, & \text{if }\, \dfrac{\alpha}{\alpha^{\mathrm{opt}}} \cdot u \le u_{\mathrm{bot}}.
\end{cases}
\]

Moreover, the threshold function for the inspection policy $
c^\circ(\alpha,u)
\,=\,
\mathbf{1}\!\left\{\,u_{\mathrm{bot}} < u < \psi(\alpha,u,K,v,u_{\mathrm{bot}},u_{\mathrm{top}})\,\right\}$ 
 takes the following form.

  \[
    \psi(\alpha,u,K,v, u_{\mathrm{bot}}, u_{\mathrm{top}}) \;=\;
    \begin{cases}
u_{\mathrm{top}}, & \text{if }\, \dfrac{\alpha}{\alpha^{\mathrm{opt}}} \cdot u\ge u_{\mathrm{top}},\\[6pt]
\dfrac{\alpha}{\alpha^{\mathrm{opt}}} \cdot u, & \text{if }\, u_{\mathrm{bot}}< \dfrac{\alpha}{\alpha^{\mathrm{opt}}} \cdot u< u_{\mathrm{top}},\\[10pt]
               u_{\mathrm{bot}}, & \text{if }\, \dfrac{\alpha}{\alpha^{\mathrm{opt}}} \cdot u \le u_{\mathrm{bot}}.
\end{cases}
  \]

  where \[
\alpha^{\mathrm{opt}}=
\begin{cases}
1, & \text{if } C(1)\ge 0,\\[6pt]
F^{-1}\!\Bigl(F(\bar{\alpha})+\dfrac{C_0(u_{\mathrm{bot}})}{B(u_{\mathrm{bot}},u_{\mathrm{top}})}\Bigr), & \text{if } C(1)<0.
\end{cases}
\]
\end{theorem}

Note that if $C(1)\ge 0$ then $F(\bar{\alpha})+\tfrac{C_0}{B}\le F(1)$ and we can choose
$\alpha^{\mathrm{opt}}=1$; however, if $C(1)<0$ the optimal threshold $\alpha^{\mathrm{opt}}$ lies in $(F(1),F(\bar{\alpha})]$ and
$\alpha^{\mathrm{opt}}=F^{-1}\!\big(F(\bar{\alpha})+\tfrac{C_0}{B}\big)\in(1,\bar{\alpha}]$. 

Let us consider Figure~\ref{fig:fig2_power} as a reference. Theorem~\ref{thm:main_threshold_power} shows that the threshold is flat at $u_{\mathrm{bot}}$ for small interference values $t(\alpha , u)=\alpha \cdot u$, prescribing exclusive access for every $u$ sufficiently low to make inspection infeasible (yellow region); then, holding $\alpha$ fixed, the threshold scales linearly in $u=t(\alpha , u)/\alpha^{\mathrm{opt}}$ which lies in the intermediate region, necessitating an inspection to verify the incumbent's claim of interference (red region); finally, it saturates at $u_{\mathrm{top}}$ for large signals (gray region above $u_{\mathrm{top}}$). The cutoff $\alpha^{\mathrm{opt}}$ equals $1$ when the budget constraint is slack at $x=1$, and otherwise is the unique quantile solving $F(\alpha^{\mathrm{opt}})=F(\bar{\alpha})+\frac{C_0(u_{\mathrm{bot}})}{B(u_{\mathrm{bot}},u_{\mathrm{top}})}\leq 1$ (binding-constraint case). 

The threshold function is expressed as a function of $u_{\mathrm{top}},\,u_{\mathrm{bot}}$, and the value of  $\alpha^{\mathrm{opt}}$ is determined in polynomial time given $u_{\mathrm{top}},\,u_{\mathrm{bot}}$. Next,  we show there is a polynomial time way to retrieve the values to these parameters. 

\begin{theorem}\label{thm:main_knapsack_power_interference}
Let $K,v\ge 0$ and $t(\alpha,u)=\alpha u$. The solution to
Formulation~\ref{formulation:main_knapsack_power} is characterized by a
threshold allocation rule $a(\alpha,u)=\mathbf{1}\{\,u<\phi(\alpha,u)\,\}$
with two cutoffs $0\le u_{\mathrm{bot}}\le u_{\mathrm{top}}\le v$ and an
$\alpha$–cutoff $\alpha^{\mathrm{opt}}\in[1,\bar{\alpha}]$ solving
\begin{subequations}\label{formulation:main_knapsack_power}
\begin{alignat}{2}
\max_{\,u_{\mathrm{top}},\,u_{\mathrm{bot}}}\quad
& \int_{0}^{u_{\mathrm{bot}}} g(u)\!\int_{0}^{\bar{\alpha}} f(\alpha)\,\bigl(\min\{v,\alpha u\}-u\bigr)\,d\alpha\,du
\;+\;
\int_{u_{\mathrm{bot}}}^{u_{\mathrm{top}}} g(u)\!\int_{\alpha^{\mathrm{opt}}}^{\bar{\alpha}} f(\alpha)\,\bigl(\min\{v,\alpha u\}-u\bigr)\,d\alpha\,du
\label{formulation:main_knapsack_obj_power}\\[2pt]
\text{s.t.}\quad
& \int_{0}^{u_{\mathrm{bot}}} g(u)\!\int_{0}^{\bar{\alpha}} f(\alpha)\,r(u)\,d\alpha\,du
\;+\;
\int_{u_{\mathrm{bot}}}^{u_{\mathrm{top}}} g(u)\!\int_{\alpha^{\mathrm{opt}}}^{\bar{\alpha}} f(\alpha)\,\bigl(r(u)-K\bigr)\,d\alpha\,du
\;\ge\;0,
\label{formulation:main_knapsack_budget_power}\\[4pt]
& \alpha^{\mathrm{opt}} \;=\; \max\left(1,
F^{-1}\!\left(1+\frac{C_0(u_{\mathrm{bot}})}{B(u_{\mathrm{bot}},u_{\mathrm{top}})}\right)
\right),
\label{eq:main_alpha_opt_operational}\\[2pt]
& 0\le u_{\mathrm{bot}}\le u_{\mathrm{top}}\le v,
\end{alignat}
\end{subequations}
The induced threshold for $u$ is
\[
\phi(\alpha,u)=\min\Bigl\{\,u_{\mathrm{top}},\ \max\Bigl\{\,u_{\mathrm{bot}},\ \frac{t(\alpha,u)}{\alpha^{\mathrm{opt}}}\Bigr\}\Bigr\}.
\]
\end{theorem}

Formulation~\ref{formulation:main_knapsack_power} is a continuous knapsack
problem: the two integrals in the objective correspond to two continuous
blocks whose “sizes’’ are controlled by $(u_{\mathrm{bot}},u_{\mathrm{top}})$,
and the single budget constraint
\eqref{formulation:main_knapsack_budget_power} trades off their
contributions. Under $K\ge K_{\text{low}}$, the integrand in the second
budget term is non–positive, so for each fixed $u_{\mathrm{bot}}$ the
objective is increasing and the left–hand side of
\eqref{formulation:main_knapsack_budget_power} is decreasing in
$u_{\mathrm{top}}$, and the optimal $u_{\mathrm{top}}^*(u_{\mathrm{bot}})$ is
the unique value at which the constraint binds and can be found by bisection
in $O(\log(1/\varepsilon))$ time. Substituting $u_{\mathrm{top}}^*(u_{\mathrm{bot}})$
and the closed-form expression \eqref{eq:main_alpha_opt_operational} for
$\alpha^{\mathrm{opt}}$ then reduces the problem to a one–dimensional search
over $u_{\mathrm{bot}}$, again solvable by bisection in
$O(\log(1/\varepsilon))$ evaluations. Hence, the continuous knapsack can be
solved to accuracy $\varepsilon$ in overall time $O(\log^2(1/\varepsilon))$,
up to the cost of evaluating the primitives and the integrals.

\section{Qualitative Results and Inefficiency}\label{sec:inefficiency}

In this section we study how the change in parameters $v,\, K$ affects the optimal allocation policy, and the inefficiency imposed by the cost of inspection. 

\subsection{Variations in the Optimal Policy}

Our running assumptions are that $v>0$, i.e., the good is valued positively by the incumbent, and that $K>K_{\text{low}}$, i.e., the inspection is costly and thus the budget constraint is always binding.  

\begin{figure}[htbp]
  \centering
  \begin{subfigure}[t]{0.48\textwidth}
    \centering
    \includegraphics[width=\linewidth]{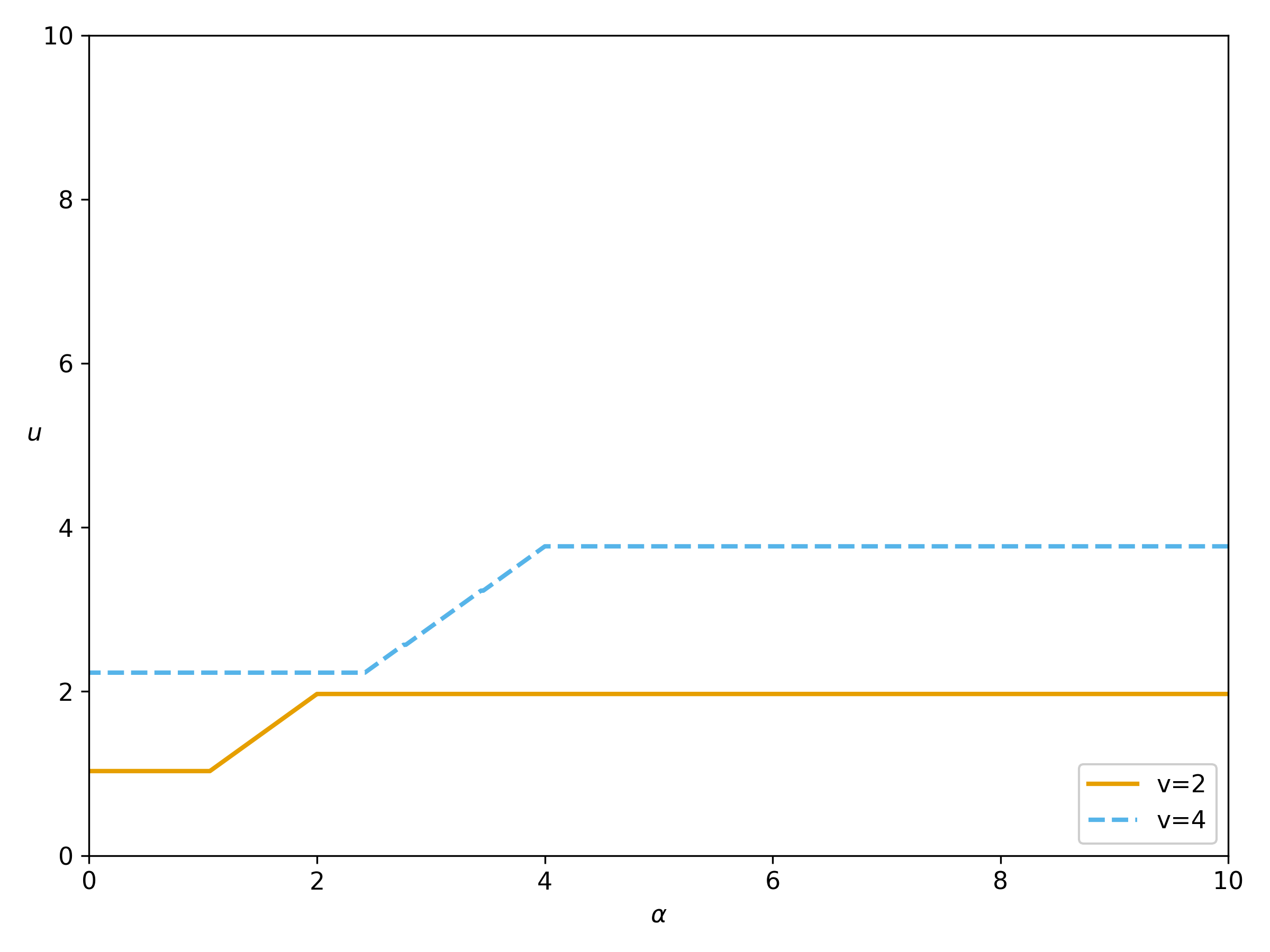}
  \end{subfigure}\hfill
  \begin{subfigure}[t]{0.48\textwidth}
    \centering
    \includegraphics[width=\linewidth]{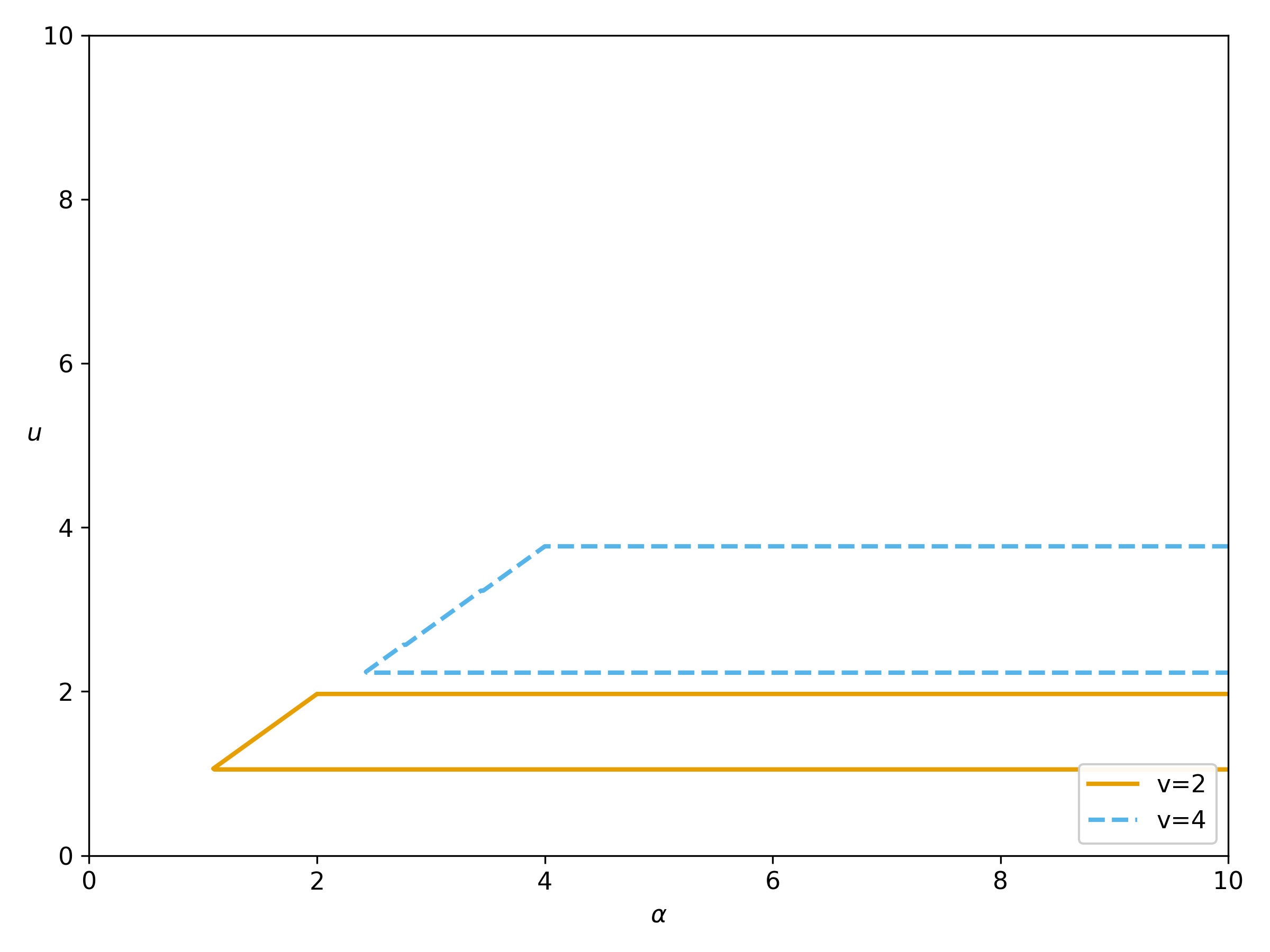}
  \end{subfigure}

  \vspace{0.5em}

  \begin{subfigure}[t]{0.48\textwidth}
    \centering
    \includegraphics[width=\linewidth]{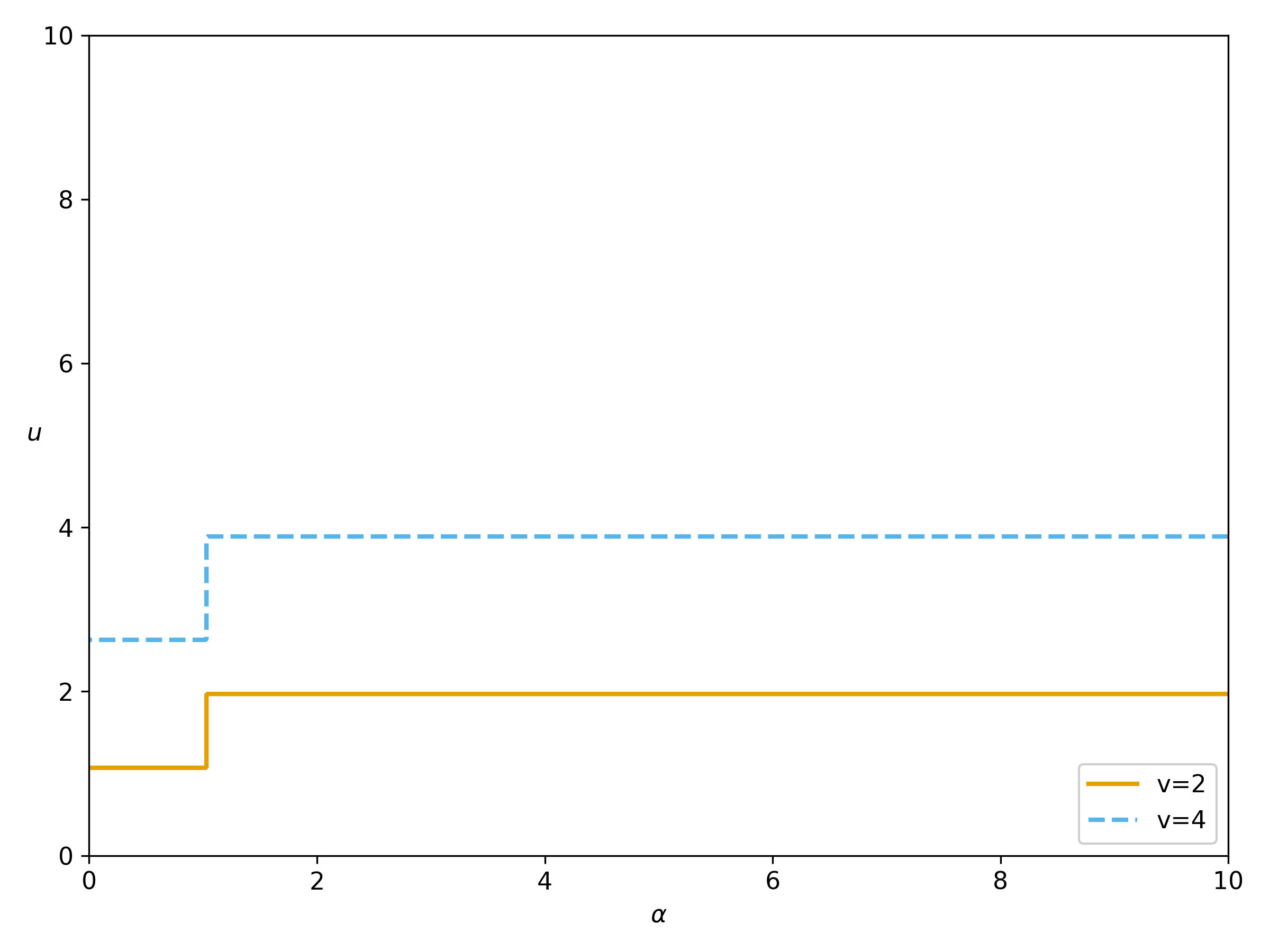}
  \end{subfigure}\hfill
  \begin{subfigure}[t]{0.48\textwidth}
    \centering
    \includegraphics[width=\linewidth]{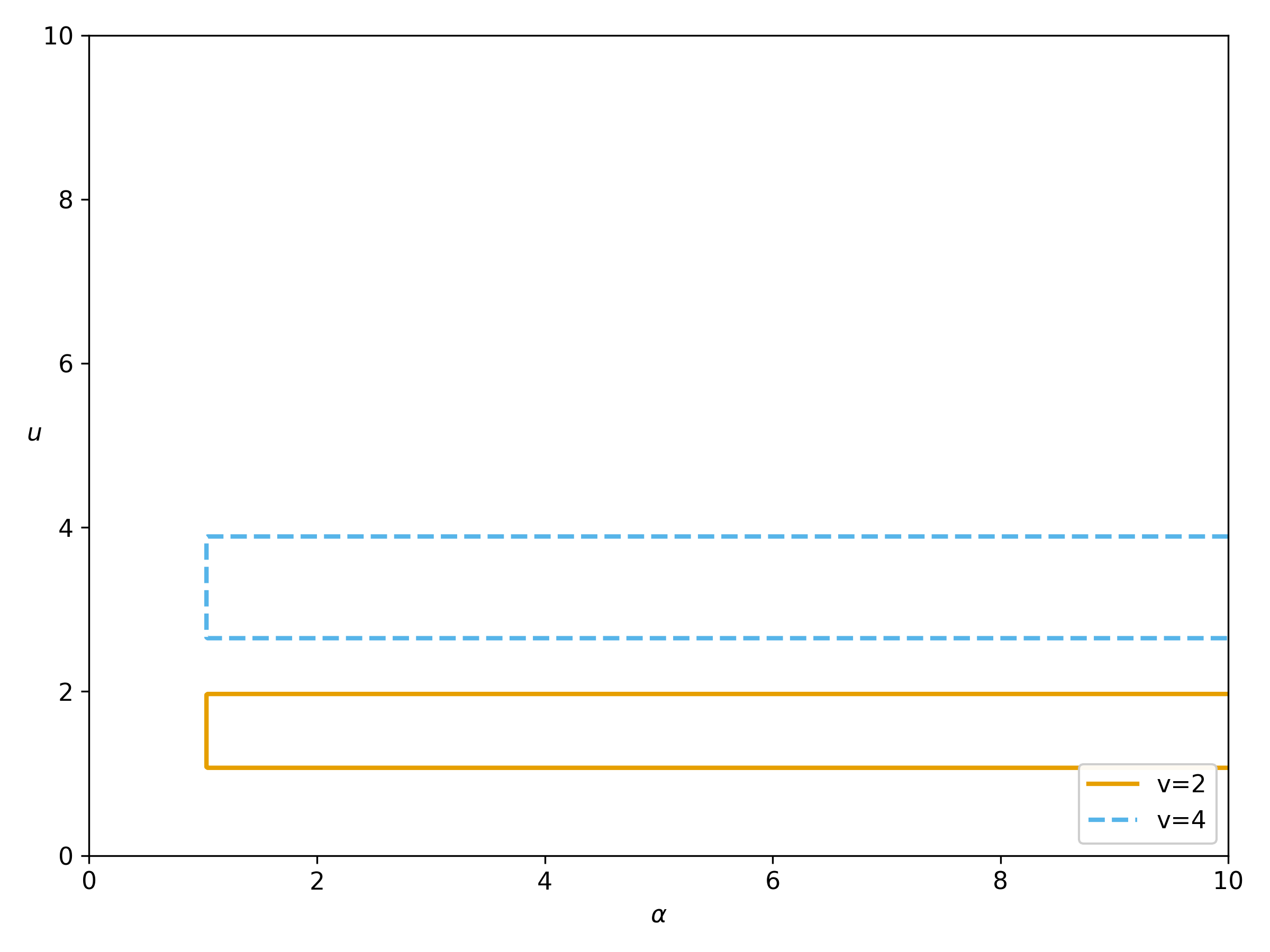}
  \end{subfigure}

  \caption{Optimal thresholds varying the value $v$. The top (bottom) row uses independent (power) interference. The left (right) column shows allocation (inspection) thresholds.}
  \label{fig:variation_v}
\end{figure}

\paragraph{\textbf{Varying the value $v$.} } 
Figure~\ref{fig:variation_v} reports allocation and inspection thresholds for $v\in\{2, 4\}$. Both $f$ and $g$ are truncated Gaussian on $[0,10]$ and $[0,10]$ with $(\mu=5,\sigma=7)$, and $K=20$. The top (bottom) row uses independent (power) interference. The left (right) column shows allocation (inspection) thresholds. Above the allocation threshold, sharing occurs. Inside the inspection contour, inspection occurs.

The main pattern is clear: as $v$ rises, the exclusive–access region expands along the optimal threshold until its upper edge reaches $u_{\mathrm{top}}=\bar u$. Yet for any $v>0$, a sharing zone persists at low interference (though negligible).
From the allocation thresholds, both cutoffs move up with $v$: $u_{\mathrm{bot}}(v)$ and $u_{\mathrm{top}}(v)$ are increasing. As $v$ grows, the sharing band narrows and can vanish once $u_{\mathrm{top}}$ hits $\bar u$. This shrinkage is more pronounced under power interference because the effective conflict scales with $\alpha\cdot u$.
Inspection mirrors this logic. Higher $v$ shifts the inspection region to larger $u$, and the inspected set contracts, eventually becoming negligible.

For policymaking, when the good’s market value is on the same order as the commercial user’s utility, sharing is feasible. As the good’s value rises well above that utility, the sharing region contracts and optimal allocations tilt toward exclusivity.

Concerning the independent interference, we observe that not only the region of inspection shrinks in size, but also the region of inspection is translated towards the right. For the power interference, instead, there is not horizontal translation.  

\begin{figure}[htbp]
  \centering
  \begin{subfigure}[t]{0.48\textwidth}
    \centering
    \includegraphics[width=\linewidth]{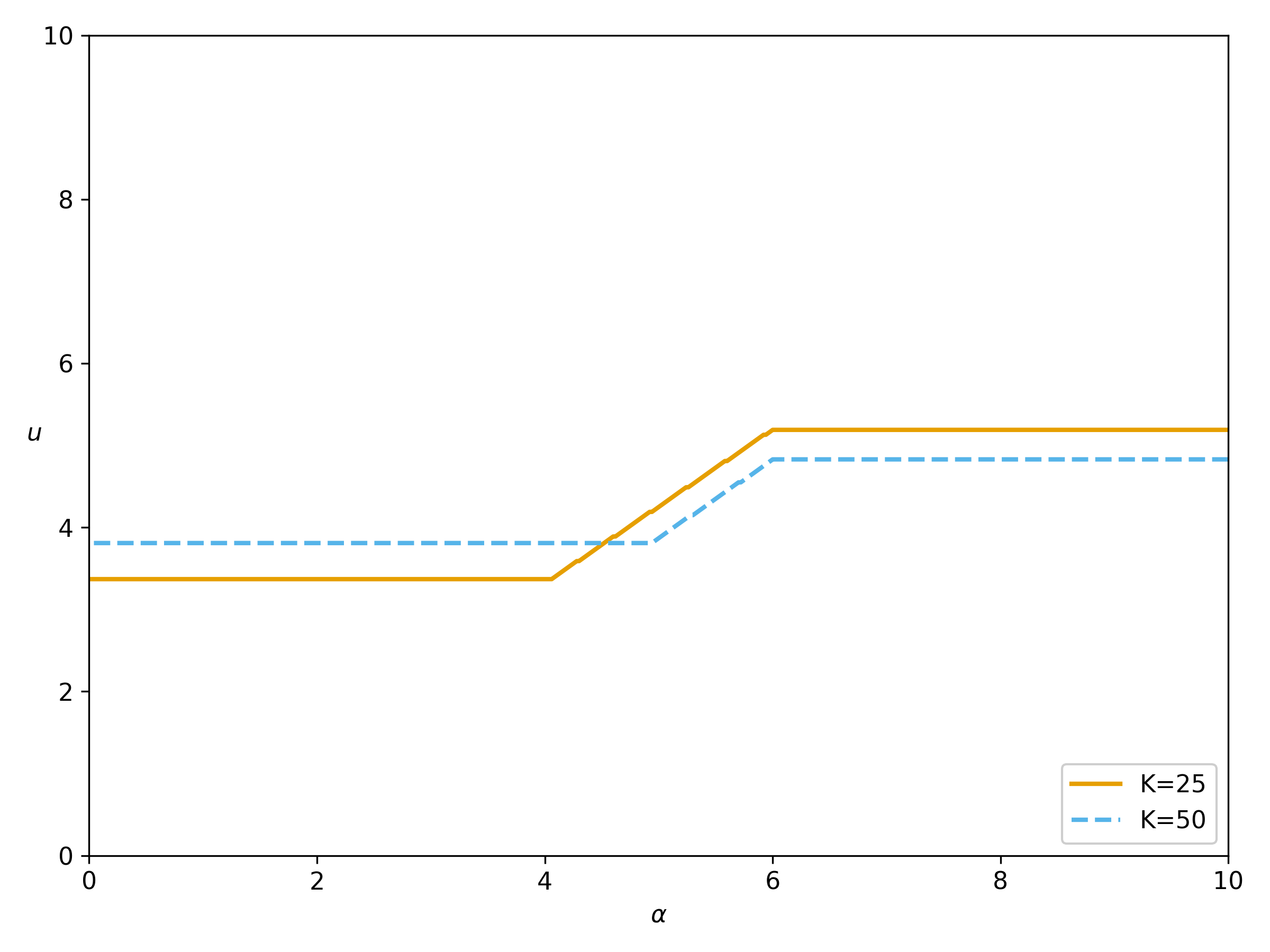}
  \end{subfigure}\hfill
  \begin{subfigure}[t]{0.48\textwidth}
    \centering
    \includegraphics[width=\linewidth]{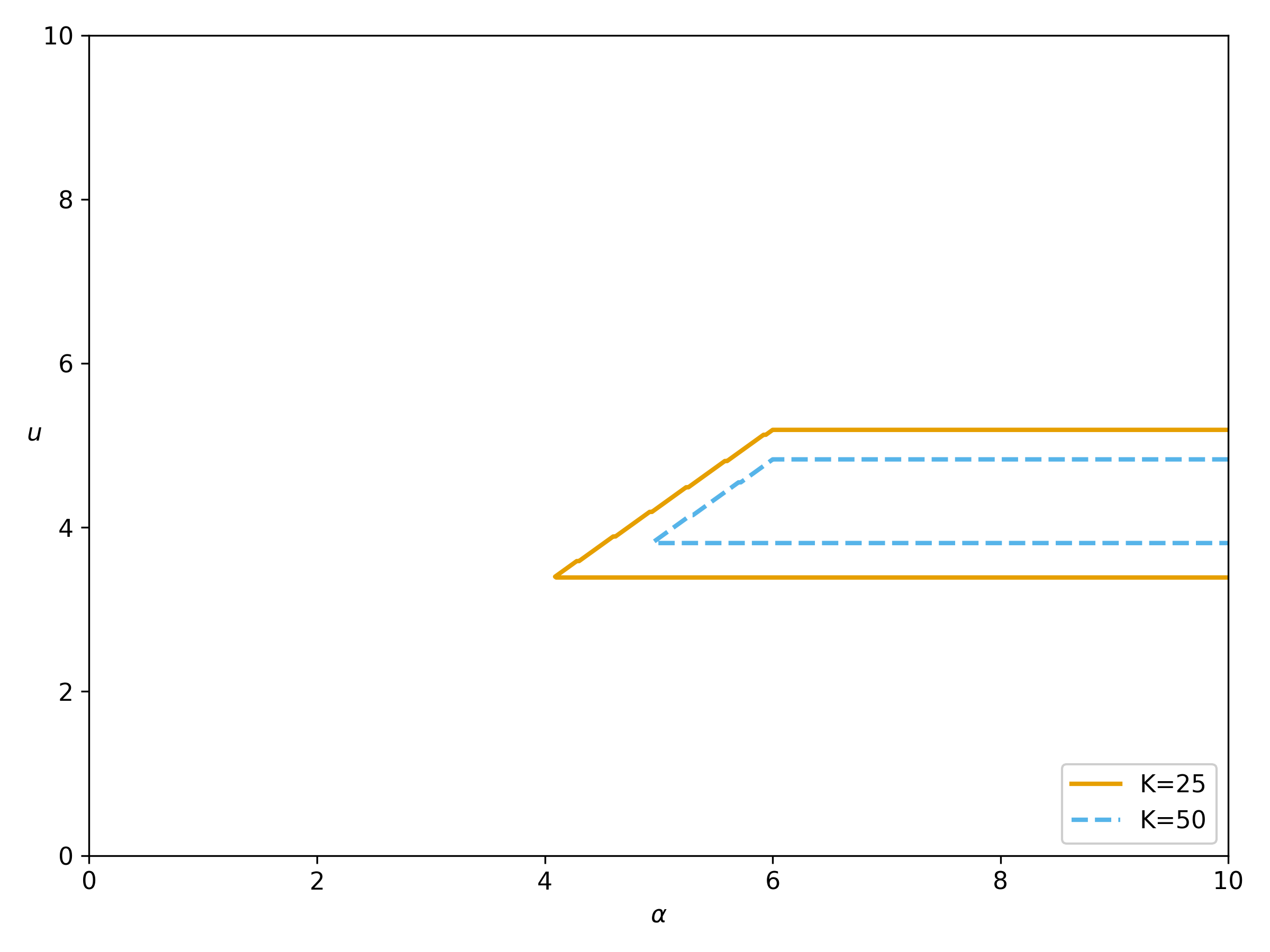}
  \end{subfigure}

  \vspace{0.5em}

  \begin{subfigure}[t]{0.48\textwidth}
    \centering
    \includegraphics[width=\linewidth]{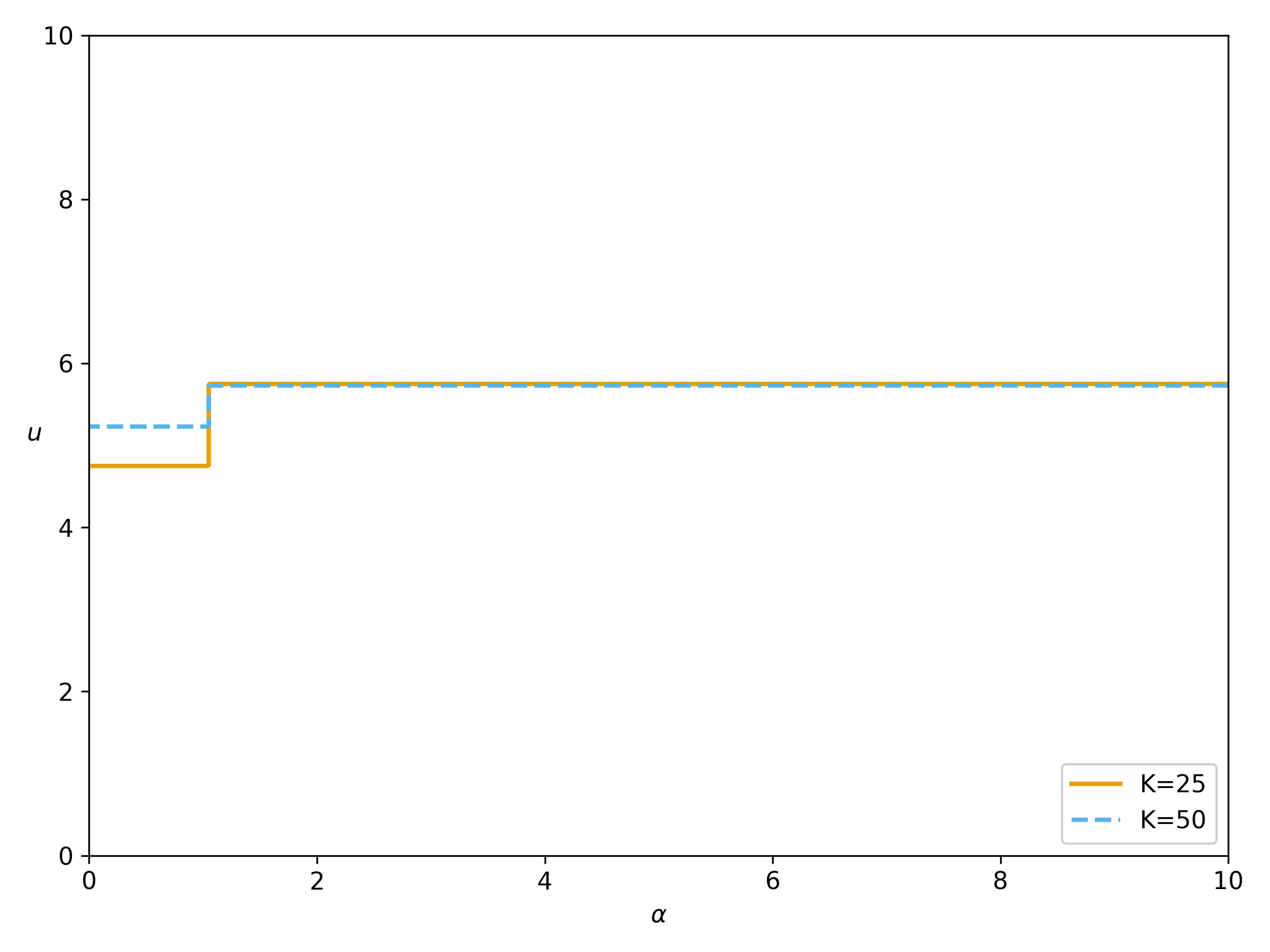}
  \end{subfigure}\hfill
  \begin{subfigure}[t]{0.48\textwidth}
    \centering
    \includegraphics[width=\linewidth]{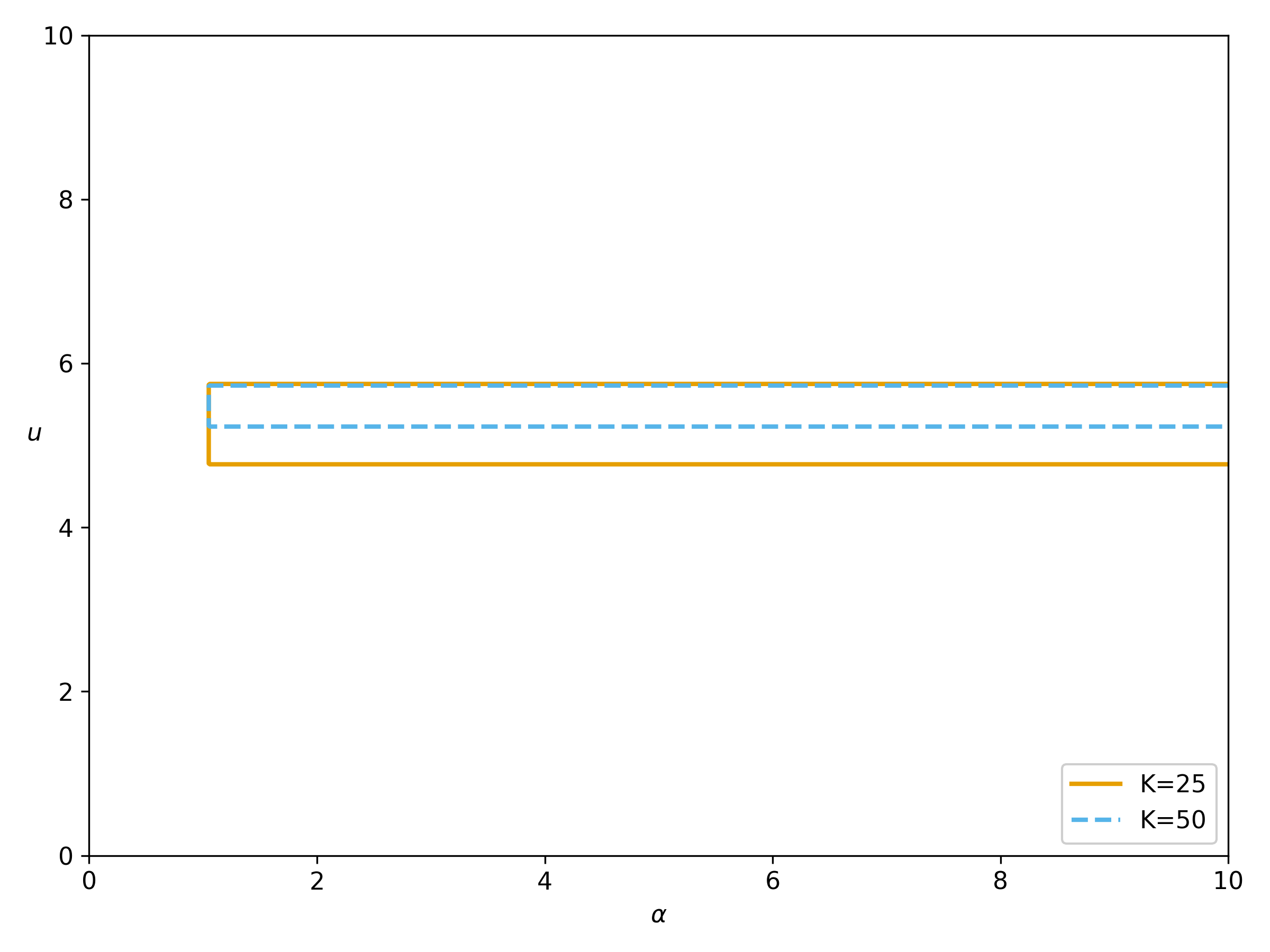}
  \end{subfigure}

  \caption{Optimal thresholds increasing the cost of inspection $K$. The top (bottom) row uses independent (power) interference. The left (right) column shows allocation (inspection) thresholds.}
  \label{fig:variation_K}
\end{figure}

\paragraph{\textbf{Increasing the inspection cost $K$.}}
Figure~\ref{fig:variation_K} displays allocation and inspection thresholds for $K\in\{25, 50\}$, with $f$ and $g$ Gaussian on $[0,10]$ $(\mu=5,\sigma=7)$ and $v=6$. The top (bottom) row uses independent (power) interference; the left (right) column shows allocation (inspection). Above the allocation cutoff there is sharing; inside the inspection contour there is inspection.

As $K$ rises, the inspection band $\,[u_{\mathrm{bot}}(K),\,u_{\mathrm{top}}(K)]\,$ contracts. In the limit $K\to\infty$, inspection vanishes and the band collapses to a point:
\[
u_{\mathrm{top}}(K)\downarrow u_\infty,\qquad
u_{\mathrm{bot}}(K)\uparrow u_\infty,\qquad
u_\infty\ge 0,
\]
consistent with Lemma~\ref{lemma:K_infinite_threshold}.

Under \emph{independent} interference, the band not only shrinks but also shifts on the right to higher values of $\alpha$. Under \emph{power} interference, the contraction is primarily vertical: the wedge tightens with no horizontal displacement, and the shrinkage is most pronounced at larger $K$.

A policy interpretation from this is that making inspection more expensive prunes the inspected states and increases the regions of default allocation.

\subsection{Inefficiency}

In this section, we turn to the inefficiency introduced by costly inspection: even though inspection preserves incentive compatibility, it inevitably distorts the allocation and reduces the objective value.\footnote{Clearly, without any tool to enforce incentive compatibility, the regulator cannot rely on the information from the incumbent and commercial user to decide the allocation.} To measure this distortion, we define the \emph{inefficiency gap} as the relative loss in expected welfare compared to the first best:
\begin{equation*}\label{eq:gap_def}
  \mathrm{Gap}(u;K,v)
  \;=\;
  \frac{\; \mathbb{E}_{\alpha}\!\left[ W^{\mathrm{FB}}(\alpha,u) \right]
        - \mathbb{E}_{\alpha}\!\left[ W^{\star}_{K,v}(\alpha,u) \right]\;}
       {\mathbb{E}_{\alpha}\!\left[ W^{\mathrm{FB}}(\alpha,u) \right]},
\end{equation*}

where $W^{\mathrm{FB}}(\alpha,u)$ is the optimal value of the first best solution at $(\alpha,u)$,\footnote{The first best solution assumes that information is revealed truthfully. The optimal allocation policy for the first best solution is $a(\alpha, u)=1$ if $ [\min(v,t(\alpha,u))-u]>0$ and $a(\alpha, u)=0$ otherwise. } and $W^{\star}_{K,v}(\alpha,u)$ is the optimal value of the constrained solution with cost of inspection $K$ and incumbent's valuation $v$ at $(\alpha,u)$.

\begin{figure}[htbp]
  \centering
  \begin{subfigure}[t]{0.48\textwidth}
    \centering
    \includegraphics[width=\linewidth]{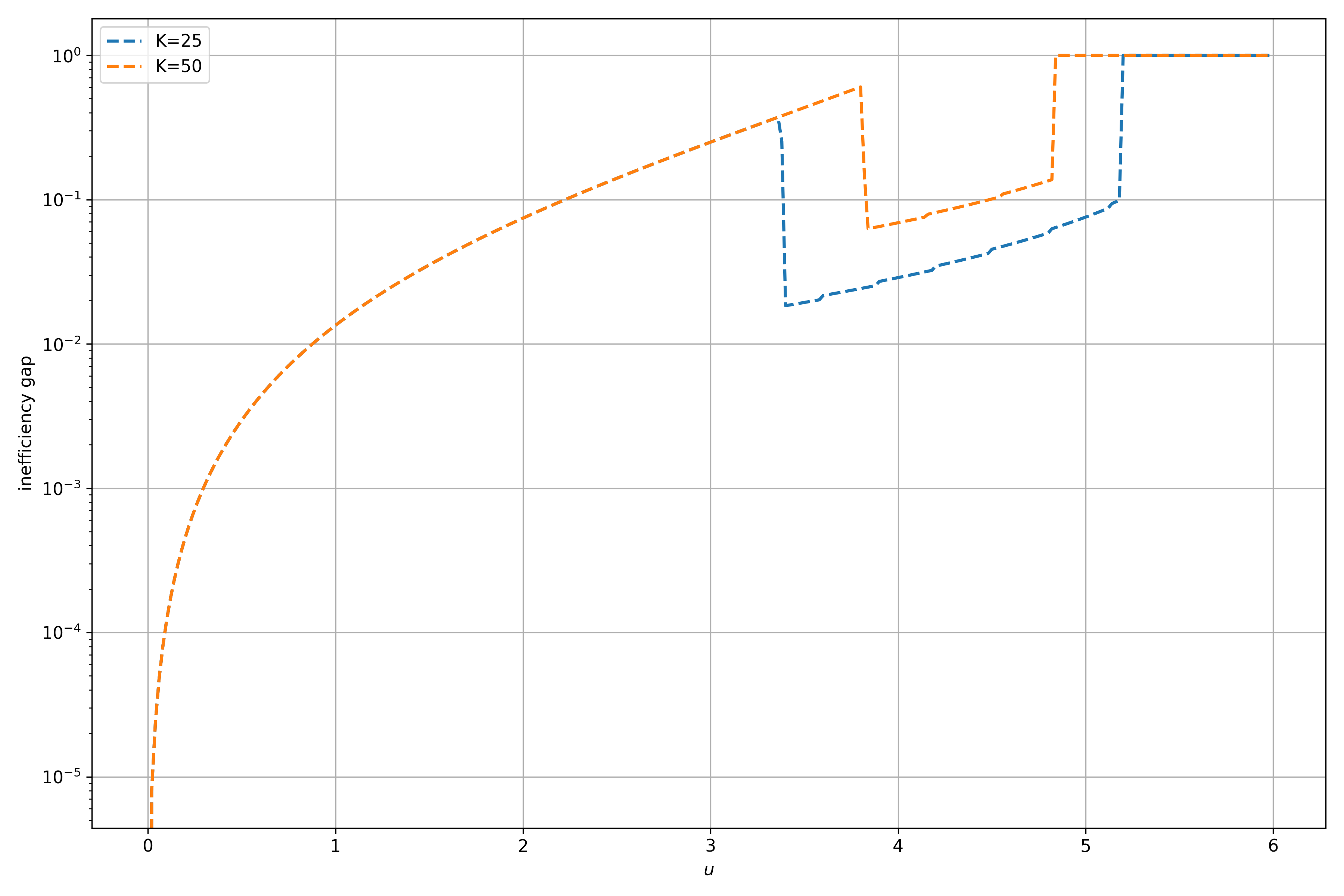}
  \end{subfigure}\hfill
  \begin{subfigure}[t]{0.48\textwidth}
    \centering
    \includegraphics[width=\linewidth]{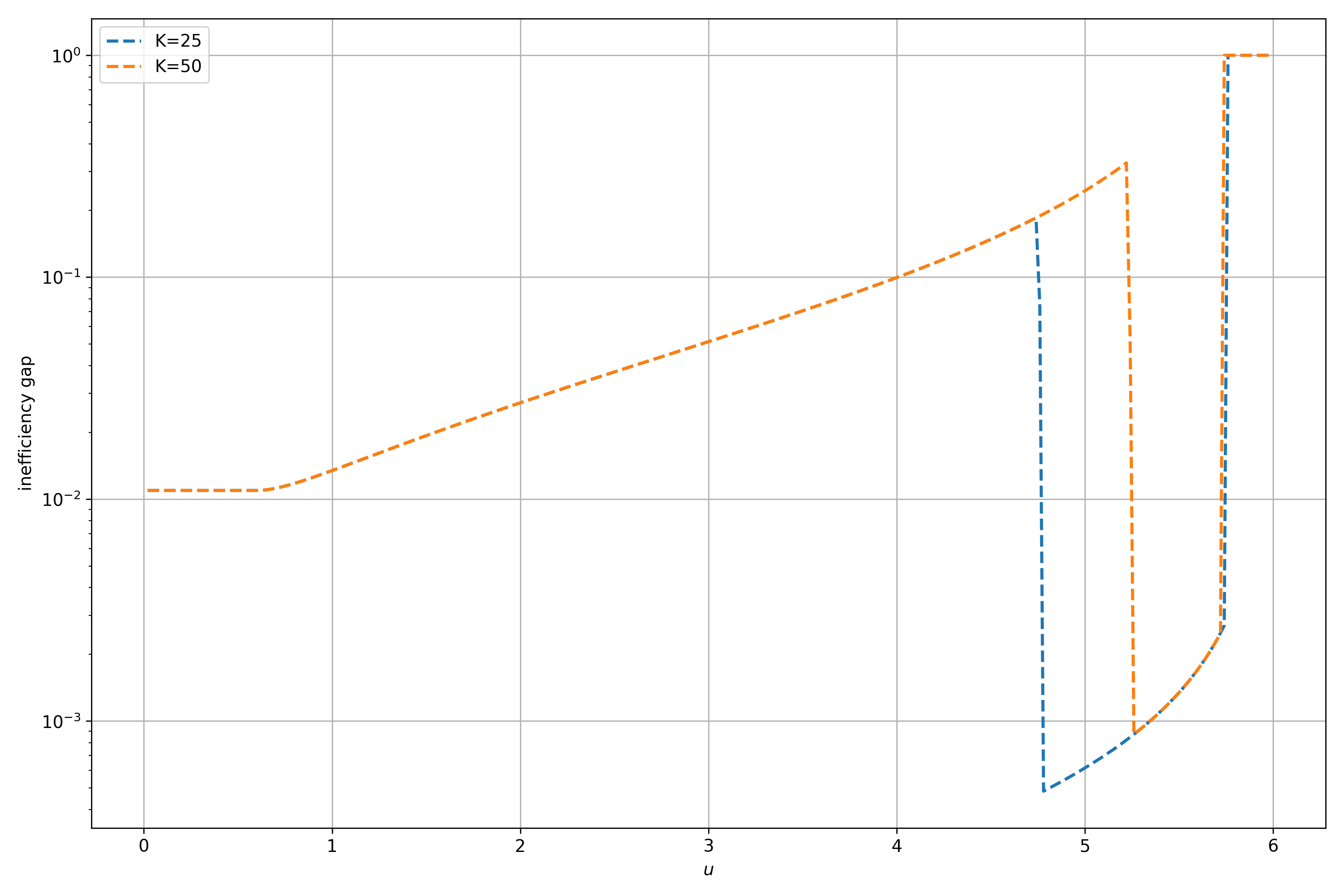}
  \end{subfigure}

  \caption{Relative inefficiency gap for different inspection costs $K$ for independent (left) and power (right) interference.}
  \label{fig:gap_vs_u_K}
\end{figure}

Figure~\ref{fig:gap_vs_u_K} plots this gap as a function of the commercial user’s private valuation $u$ for values of $K$ in $\{25, 50 \}$ for both types of interference. Specifically, we fix $v=6$ and $f$ and $g$ gaussians on $[0,10]$ with parameters $(\mu=5,\sigma=7)$.  Two distinct regions of inefficiency emerge, in line with Theorems~\ref{thm:threshold_mainbody} an~\ref{thm:main_threshold_power}. 

First, when $u\leq u_{\mathrm{bot}}$, inspection is avoided and exclusivity is granted to the incumbent even though sharing might be efficient. This \emph{no-inspection inefficiency} produces a positive gap at low $u$. 

Second, when $u_{\mathrm{bot}}<u<u_{\mathrm{top}}$, the regulator may inspect to discipline high reports of $\alpha$; while this reduces misreporting, the cost of inspection itself lowers both welfare and the area where an inspection can be made, again generating a positive gap. For both the power and independent interference, the elevated cost of inspection may produce three kinds of inefficiencies: 1) increasing the value of $u_{\mathrm{bot}}$, 2) decreasing the value of $u_{\mathrm{top}}$, and 3) a translation of the threshold to the right. 

Outside these two regions of potential misallocation, namely when $u\geq v$, sharing is strictly optimal, and no inefficiency arises (see Lemma~\ref{lem:u_above_v}). Consequently, any welfare loss stems from either avoiding inspection for low $u$ or from balancing inspection costs with uncertain harm in the intermediate range while reducing this same region of inspection.

\begin{figure}[htbp]
  \centering
  \begin{subfigure}[t]{0.48\textwidth}
    \centering
    \includegraphics[width=\linewidth]{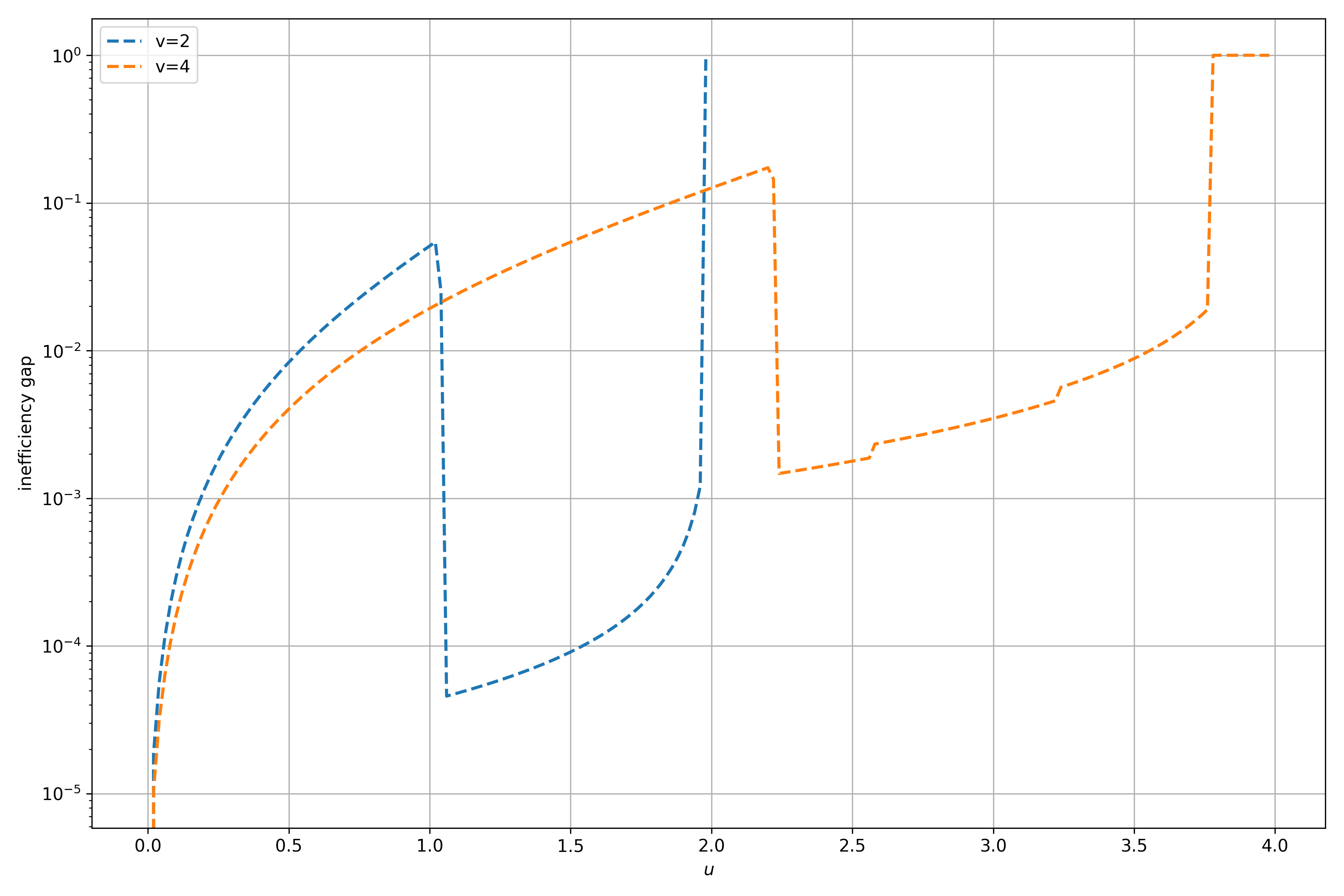}
  \end{subfigure}\hfill
  \begin{subfigure}[t]{0.48\textwidth}
    \centering
    \includegraphics[width=\linewidth]{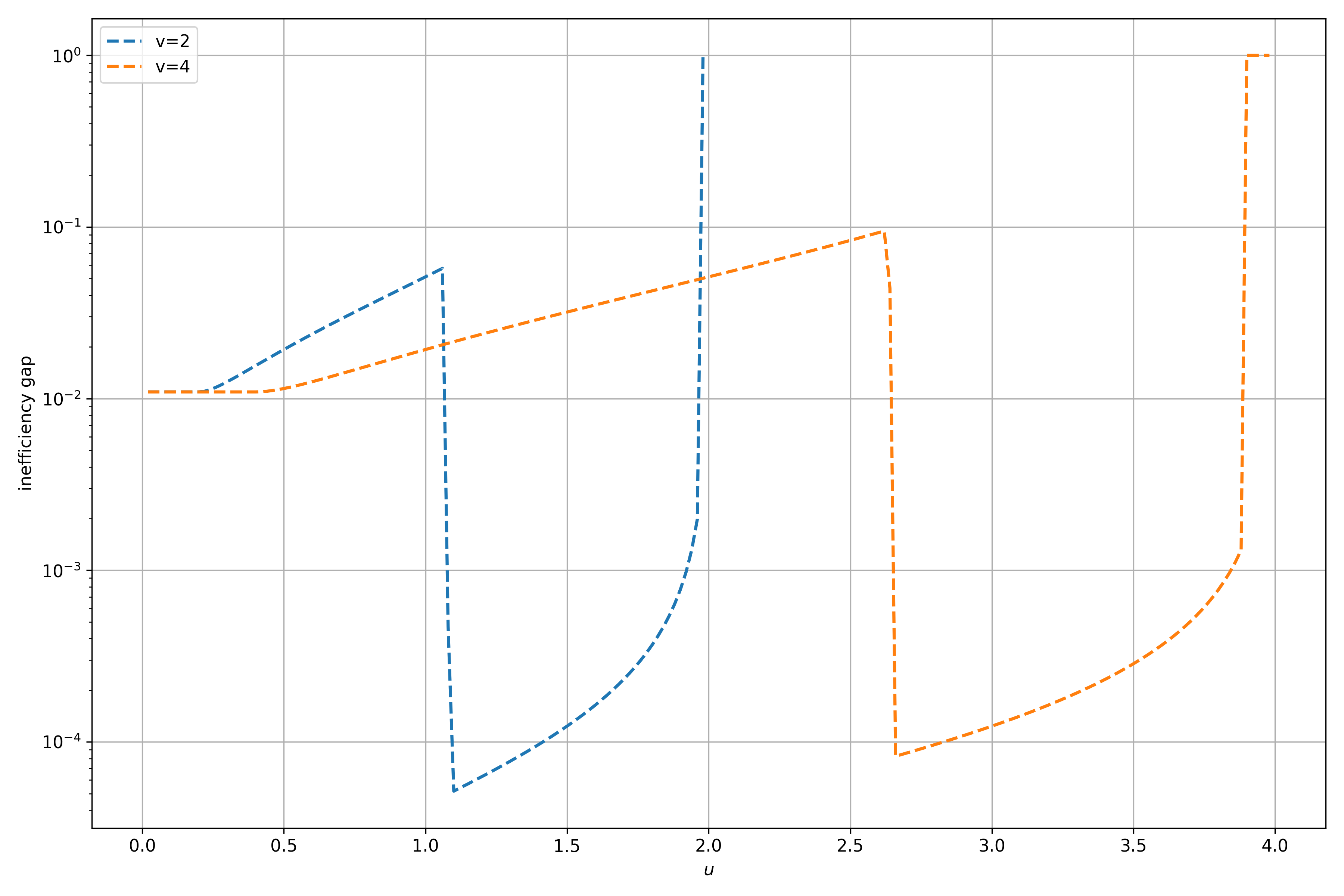}
  \end{subfigure}

  \caption{Relative inefficiency gap for different values $v$ for independent (left) and power (right) interference.}
  \label{fig:gap_vs_u_v}
\end{figure}

In Figure~\ref{fig:gap_vs_u_v} we plot the gap as a function of the commercial user’s private valuation $u$ for values of $v$ in $\{2, 4 \}$ for both types of interference. Specifically, we fix $K=20$ and $f$ and $g$ gaussians on $[0,10]$ with parameters $(\mu=5,\sigma=7)$.  We observe that as the valuation of the resource for the incumbent increases, there is an increased inefficiency loss for the commercial user caused by the increased region of default exclusive access. Such loss is determined by the necessity to maintain inspection for higher values of commercial utility, while the cost of inspection remains the same.

\section{Conclusions \& Future Work}\label{sec:conclusions}



In this paper, we have developed a mechanism design framework for allocating resources in settings where both the commercial utility and the negative externality borne by a non-commercial incumbent remain private, and monetary transfers to or from this incumbent are infeasible. Motivated by challenges such as ensuring “clean” spectrum for scientific use and balancing economic gains with ecological protection, our approach substitutes direct pricing with costly inspections and carefully designed transfer schemes. This framework offers policymakers an alternative to traditional market-based mechanisms, which often fail to capture non-monetizable values.

\textbf{Limited Transfers and Costly Inspections}
Our model diverges from the standard settings in which all agents can exchange monetary transfers or none. Instead, we consider a scenario where the incumbent—such as a radio telescope—is precluded from engaging in monetary transactions, while the commercial user—be it a satellite operator—can. This restriction is critical in settings like spectrum allocation and invasive species control, where the intrinsic value of the resource (or habitat) cannot be monetized directly.  By incorporating costly inspections, our regulator is empowered to verify the incumbent’s claim about the negative externality, thereby disciplining both the overstatement of harm and the inflation of commercial utility. We assume that the money raised to ensure truthful revelation from the commercial user, are fully used to ensure truthful revelation from the incumbent. 

\textbf{Deterministic Allocation \& Inspection}
A key theoretical insight of our work is that the optimal mechanism is deterministic. This simplifies policy implementation. 
 In practical terms, whether in allocating scientific spectrum—where even slight interference can be critical—or in managing invasive species—where ecological balance is paramount—the binary outcome ensures that decisions are unambiguous and directly interpretable in terms of frequency or occurrence. We further provide analytical characterization of the threshold functions for both cases of interference. 

\textbf{Inefficiency}
The optimal mechanism partitions the domain of the commercial user’s private information (denoted by $u$) into three distinct regions. In the first, where $u$ is low, exclusive access is the default; in the intermediate region, the mechanism must compare the commercial user’s reported utility $u$ with the incumbent’s negative externality $t$, often triggering costly inspections; finally, in the third region, where $u$ is sufficiently high, sharing becomes strictly optimal. The potential for an inefficient allocation occurs always in the first two regions arising from the cost and uncertainty of inspections. We explore these trade-offs through detailed experimental evaluations, shedding light on how efficiency losses vary with different levels of externality and reported utility. 

\textbf{Reformulation as a Knapsack Problem for Practical Solvability.} Finally, we show that our mechanism design problem can be reformulated as a variant of the continuous knapsack problem. This reformulation not only underscores the computational tractability of our solution but also enables the application of established algorithmic techniques. Whether the objective is to maximize the efficiency of scientific spectrum usage or to balance economic gains with ecological preservation in the management of invasive species, the knapsack formulation provides a practical and powerful tool for policy design. Indeed, it allows a commercial entrant to know in advance if their expected utility is sufficiently high to make the entry fee a profitable investment.

Overall, our contributions bridge a significant gap in the literature by offering a rigorous and implementable framework for resource reallocation in environments characterized by non-monetizable values and costly inspections. We believe that our results will offer valuable insights for policymakers tasked with balancing economic and intrinsic objectives in complex, real-world scenarios.




\subsection*{Future Work}

Several directions remain open for investigation. 

\paragraph{Ex-Post Budget Balance}
Our mechanism ensures budget balance in \emph{expectation}, but real-world applications may demand \emph{ex-post} budget balance, i.e., at each realization the mechanism must be self-funding. 
%
%
%

\paragraph{Multiple Commercial Users with Identity-Dependent Externalities}
Our approach generalizes readily to multiple commercial users if the externality is not user-specific. However, if the harm depends on the \emph{identity} of each user, the planner might need to inspect \emph{sequentially}, turning the problem into a Pandora-like search \cite{weitzman-optimalSearch} with added complexities (see also \cite{changliu,doval}). Such problems are generally more difficult to solve but highly relevant to many real-world resource allocation questions.

\paragraph{Imperfect or Noisy Inspection}
We assumed the regulator’s inspection perfectly verifies the incumbent’s claim. In practice, inspection may yield only a noisy signal of true harm. Determining optimal threshold policies in the face of imperfect signals presents a compelling avenue for further research, specially given the necessity of maintaining incentive compatibility and budget considerations. 

\paragraph{Precision of Inspection} When the regulator can choose the \emph{precision} of inspection rather than a binary yes/no decision, the optimal inspection and allocation policy itself may no longer be binary. 
This richer setting requires a delicate analysis, since the mechanism must balance both the probability and the intensity of inspection, and the structure of the solution becomes considerably more complex.

\bibliographystyle{plainnat}  
\bibliography{new_bibliography}

\appendix

\section{Appendix: Qualitative Results for Small $K$}\label{sec:appendix:small_K}

We analyze the optimal allocation and inspection policies for small inspection costs under both interference regimes: independent and power. Parameters are fixed as $\bar{\alpha}=\bar{u}=10$, $v=6$, and Gaussian densities $f,g$ with $(\mu=5,\sigma=7)$. We focus on values $K<K_{\text{low}}$ (see Lemma~\ref{lemma:sufficient_small_value_K}), specifically $K\in{0.1,0.5}$.

\begin{figure}[htbp]
  \centering
  \begin{subfigure}[t]{0.48\textwidth}
    \centering
    \includegraphics[width=\linewidth]{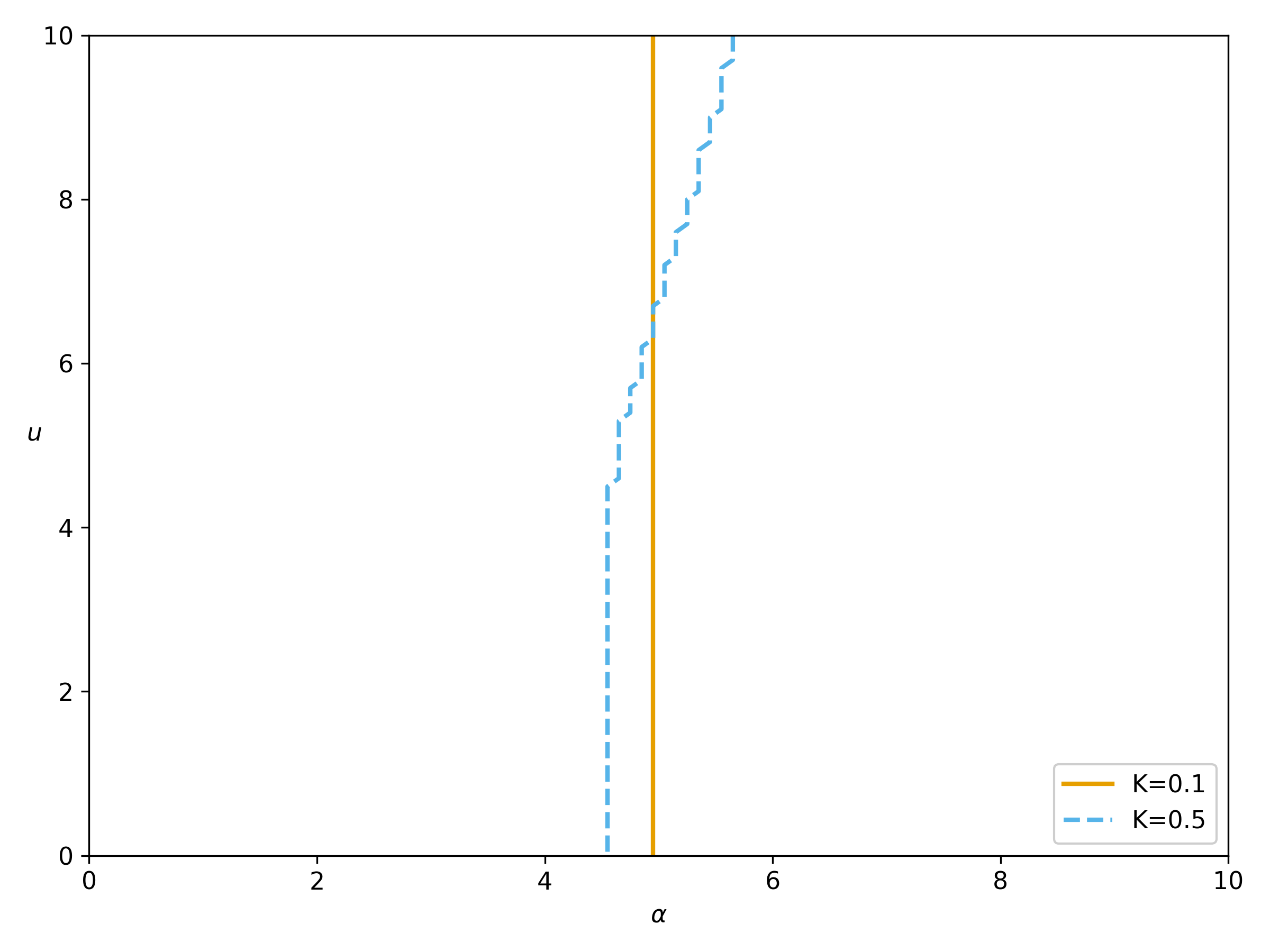}
  \end{subfigure}\hfill
  \begin{subfigure}[t]{0.48\textwidth}
    \centering
    \includegraphics[width=\linewidth]{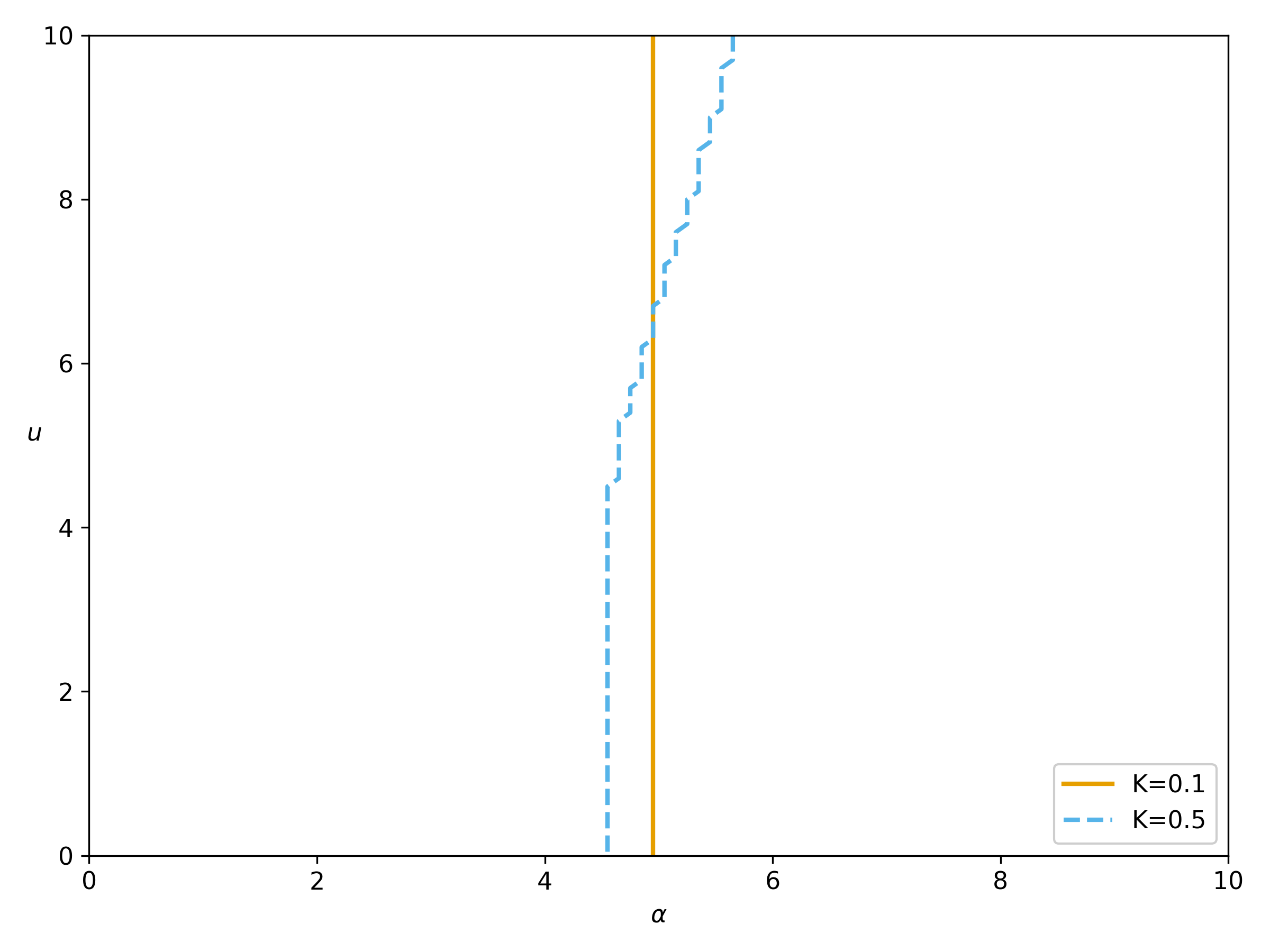}
  \end{subfigure}
\caption{Optimal thresholds for decreasing cost of inspection $K$ under independent interference. The left (right) column shows allocation (inspection) thresholds.}
  \label{fig:independent_low_K}
\end{figure}

For \emph{independent interference}, Figure~\ref{fig:independent_low_K} shows on the left the optimal allocation policy and on the right the corresponding inspection policy. The two coincide, indicating that inspection occurs whenever exclusive access is granted—typical when inspection costs are small. As $K$ decreases, the threshold function rotates counterclockwise and shifts rightward, converging toward a vertical boundary that separates the sharing and exclusive-access regions.

\begin{figure}[htbp]
  \centering
  \begin{subfigure}[t]{0.48\textwidth}
    \centering
    \includegraphics[width=\linewidth]{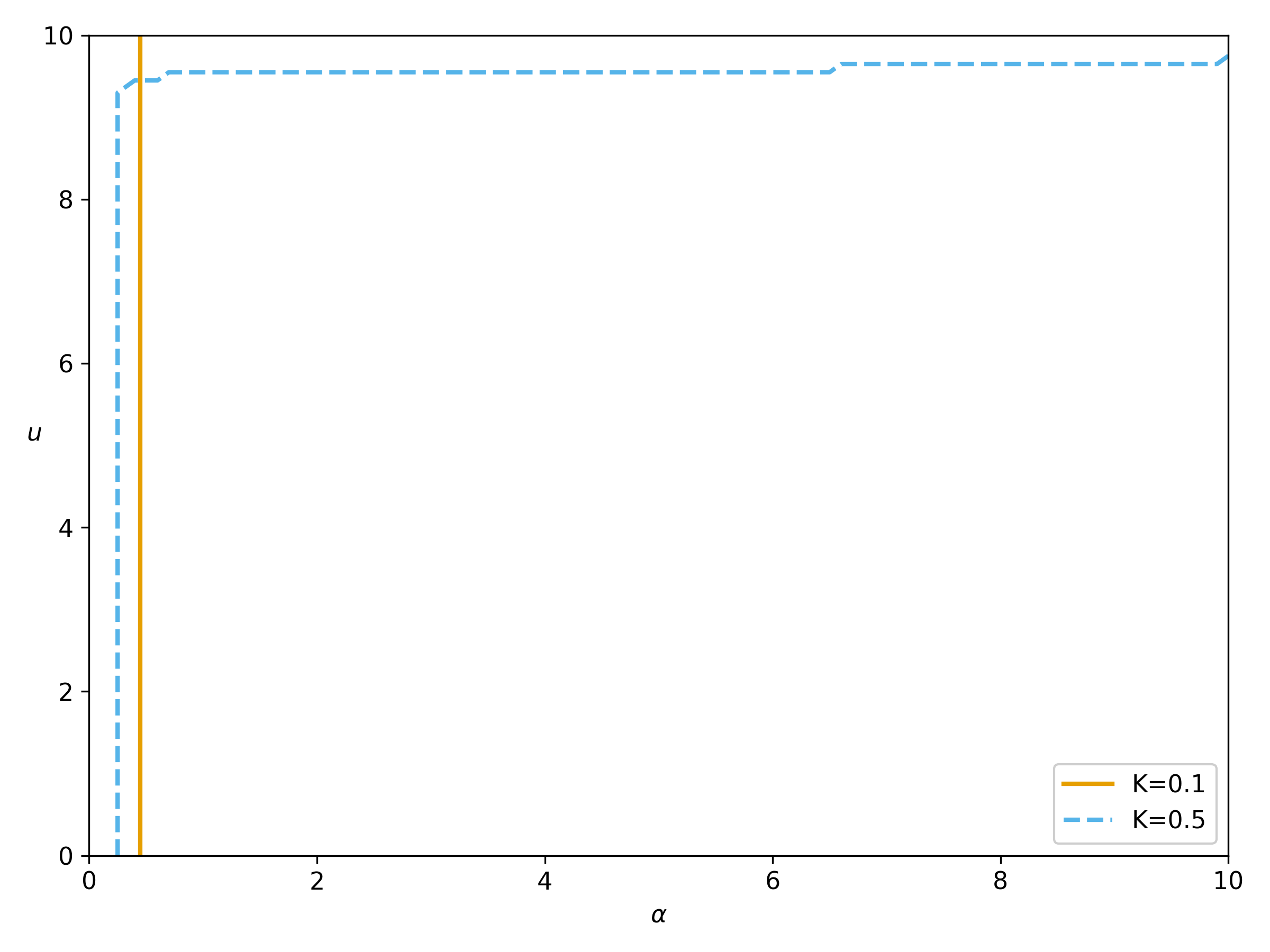}
  \end{subfigure}\hfill
  \begin{subfigure}[t]{0.48\textwidth}
    \centering
    \includegraphics[width=\linewidth]{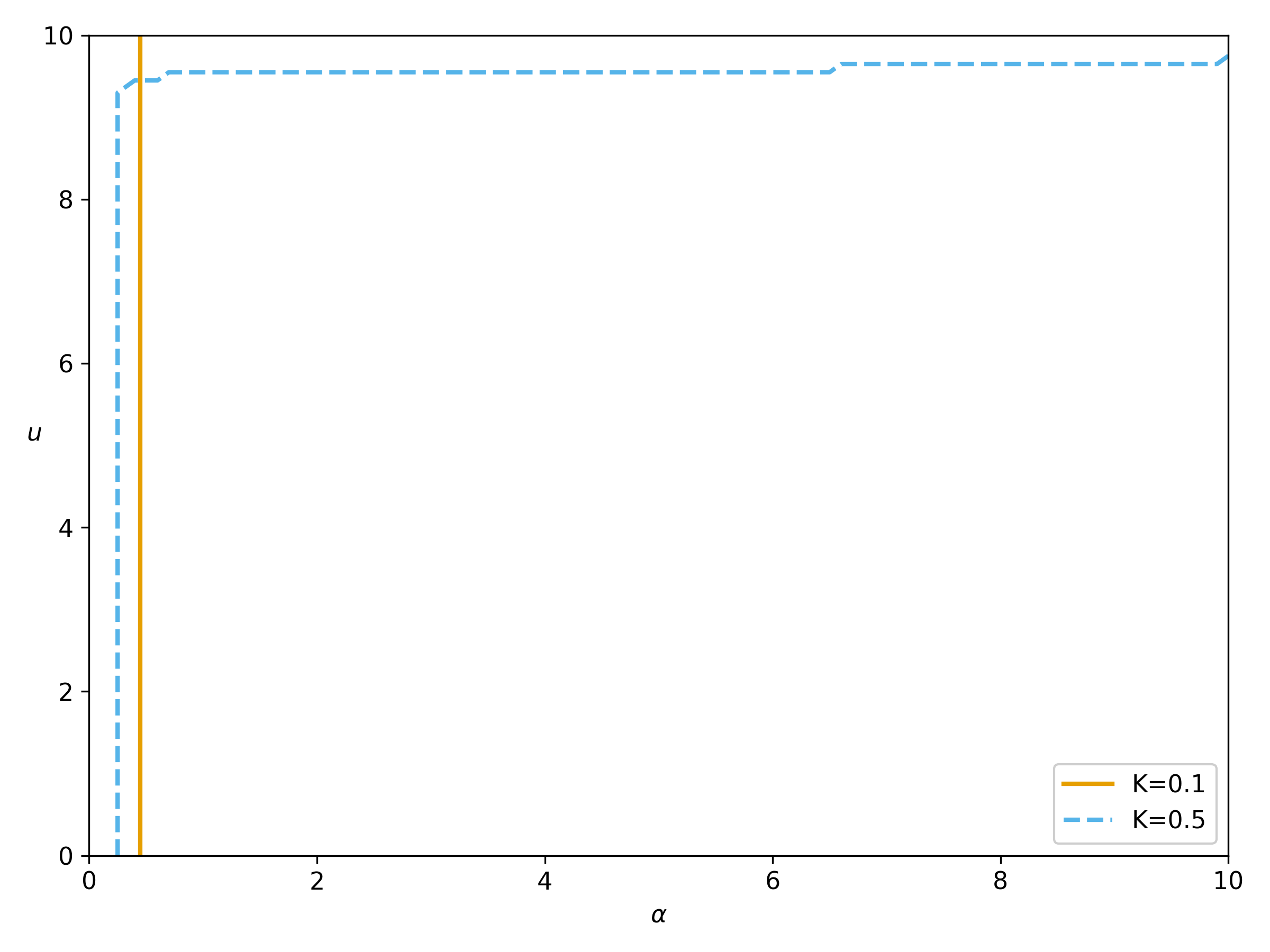}
  \end{subfigure}
\caption{Optimal thresholds for decreasing cost of inspection $K$ under independent interference. The left (right) column shows allocation (inspection) thresholds.}
  \label{fig:power_low_K}
\end{figure}

For \emph{power interference} (Figure~\ref{fig:power_low_K}), the qualitative pattern is analogous. Lower inspection costs translate the threshold upward and to the right, again approaching a vertical boundary in the limit. This shift reflects the regulator’s increasing willingness to enforce exclusive access even for higher commercial utility values when inspection becomes cheap, thus resorting to a simple expected-outcome threshold policy.

\section{Appendix: Technical Preliminaries and Auxiliary Results}\label{sec:appendix}

Throughout the appendix, unless stated otherwise, we assume that $f,\, g$ are continuous and positive with c.d.f. $F,\, G$ respectively. 
When the nature of the interference $t(\alpha,u)$ is not specified, it is understood that the statement holds for both the independent and power interference. 

Before we begin the analysis of the optimal solution it is worth analyzing the behavior of the first best solution, i.e., the optimal solution when there is truthful revelation of the private information.

Consider $u$, we define $\alpha^*(u):=\min\{ \alpha>0\colon \, \min(t(\alpha,u),v) = u\} $, the smallest $\alpha$ such that the interference is equal to the commercial utility. 
Note that given  $t(\alpha,u)=\alpha$ then $\alpha^*(u)=u$, and  if $t(\alpha,u)=\alpha\cdot u$ then $\alpha^*(u)=1$.

\begin{lemma}\label{lemma:first_best_solution}
    The optimal solution $(a^\circ,c^\circ)$ of the unconstrained version of Formulation~\ref{formulation:main_problem} takes the following form. For $u>v$ it holds 

    $$ a^\circ(\alpha,u)=c^\circ(\alpha,u)=0.$$
    
    Otherwise, if $u<v$, then
    
    $$ a^\circ(\alpha,u)=1 , \,\, c^\circ(\alpha,u)=0 \text{ if } \alpha>\alpha^*(u), \text{ and}$$

    $$ a^\circ(\alpha,u)=c^\circ(\alpha,u)= 0 \text{ if } \alpha<\alpha^*(u).$$
    
\end{lemma}
\begin{proof}
For  $u>v$ the objective function is always negative valued therefore the optimal solution must impose $ a^\circ(\alpha,u)=c^\circ(\alpha,u)=0.$ On the other side, if $u<v$ the objective function is positive for values of $\alpha>\alpha^*(u)$.
\end{proof}

Observe that the first best solution satisfies both sets of monotonicity constraints by construction. 
Moreover, note that it also respects Constraints~\eqref{formulation:working_problem_IC} 
and~\eqref{formulation:working_problem_inspection}. 

For technical convenience we focus on a relaxed version of the problem where the budget constraint is relaxed. Specifically, the expected budget amount raised from the \textit{CU} is at least as large as the expected cost of inspection. When $K$ is sufficiently large, and we characterize the threshold for that, the relaxed budget constraint will hold at equality in an optimal solution.

\begin{subequations}\label{formulation:working_problem}
\begin{alignat}{2}
\max_{a,c} \quad & \int_0^{\Bar{u}} g(u) \int_0^{\Bar{\alpha}} f(\alpha) \big( \min(v , t(\alpha,u) ) - u \big) \cdot a(\alpha,u) \, \mathrm{d}\alpha \, \mathrm{d}u  \label{formulation:working_problem_obj}\\
\text{ s.t.   }\quad & \notag\\
&  \int_0^{\Bar{u}} g(u) \int_0^{\Bar{\alpha}} f(\alpha)\left( \int_0^u a(\alpha,x) \, dx-u a(\alpha,u)\right) \mathrm{d}\alpha  \mathrm{d}u \geq K \int_0^{\Bar{u}} g(u) \int_0^{\Bar{\alpha}} f(\alpha) c(\alpha,u)  \mathrm{d}\alpha  \mathrm{d}u \label{formulation:working_problem_budget} \\
&   a(\alpha,u) - a(\alpha,u)c(\alpha,u) \leq a(0,u) \quad\quad\quad \forall \alpha,\, \forall u \label{formulation:working_problem_IC} \\
 &  c(\alpha,u) \leq a(\alpha,u) \quad\quad\,\,\quad\quad\quad\quad\quad\quad\,\,\,\,\,\, \forall \alpha,\, \forall u  \label{formulation:working_problem_inspection}\\
&  a(\alpha,u) \leq a(\alpha,u') \quad\quad\,\,\quad\quad\quad\quad\quad\quad\,\,\, \forall \alpha,\, \forall u \geq u'  \label{formulation:working_problem_umono}\\
&   a(0,u) \leq a(\alpha,u) \quad\quad\,\,\quad\quad\quad\quad\quad\quad\,\,\, \forall u,\, \forall  \alpha.\label{formulation:working_problem_tmono}\\
&  0 \leq a(\alpha,u) \leq 1 \quad\quad\,\,\quad\quad\quad\quad\,\quad\quad\quad \forall \alpha,u, \\
&  0 \leq c(\alpha,u) \leq 1 \quad\quad\,\,\quad\quad\quad\quad\,\,\quad\quad\quad \forall \alpha,u,
\end{alignat}
\end{subequations}

We characterize  the smallest value of $K$ for which constraint~(\ref{formulation:working_problem_budget}) binds at optimality.

\begin{lemma}\label{lemma:sufficient_small_value_K}
Let
$$ K_{\mathrm{low}} := \dfrac{ \int_0^{v}  (1-F(\alpha^*(u)))  \left( 1 - G(u) - u g(u) \right)   \mathrm{d}u}{ \int_0^{v}  (1-F(\alpha^*(u))) g(u)      \mathrm{d}u }. $$
For any $K \geq K_{\mathrm{low}}$,  constraint~(\ref{formulation:working_problem_budget}) binds in an optimal solution to problem (\ref{formulation:working_problem}). 
%
\end{lemma}
\begin{proof}
First, let us rearrange the left-hand side of the budget constraint

\[
L(a)
=\int_{0}^{\bar u} g(u)\int_{0}^{\bar\alpha} f(\alpha)
\Big(\int_{0}^{u} a(\alpha,x)\,dx
-\,u\,a(\alpha,u)\Big)
\,\mathrm{d}\alpha\,\mathrm{d}u.
\]

Swap the order in the first term (Fubini). For fixed $\alpha$:
\[
\int_{0}^{\bar u} g(u)\Big(\int_{0}^{u} a(\alpha,x)\,dx\Big)\,\mathrm{d}u
= \int_{0}^{\bar u}\!\!\int_{0}^{\bar u} g(u)\,a(\alpha,x)\,\mathbf{1}\{x\le u\}\,dx\,\mathrm{d}u
= \int_{0}^{\bar u} \Big(\int_{x}^{\bar u} g(u)\,\mathrm{d}u\Big) a(\alpha,x)\,dx.
\]
Define  $S(x):=1-G(x)=\int_{x}^{\bar u} g(u)\,\mathrm{d}u$.
 Renaming $x$ as $ u$\ gives
\[
\int_{0}^{\bar u} g(u)\Big(\int_{0}^{u} a(\alpha,x)\,dx\Big)\,\mathrm{d}u
= \int_{0}^{\bar u} S(u)\,a(\alpha,u)\,\mathrm{d}u.
\]

Therefore,
\[
L(a)=\int_{0}^{\bar\alpha}\!\!\int_{0}^{\bar u}
f(\alpha)\,a(\alpha,u)\,\Big(S(u)-u\,g(u)\Big)\,\mathrm{d}u\,\mathrm{d}\alpha.
\]

Set
\[
d(u):=S(u)-u\,g(u)=1-G(u)-u\,g(u).
\]

When the budget constraint is not binding the optimal solution of Formulation~\ref{formulation:working_problem} is the first best solution as we noticed earlier. Therefore, at the threshold value $K_{\mathrm{low}}$, the following equation should hold for the first best optimal solution

$$  \int_0^{\Bar{\alpha}} f(\alpha)  \int_0^{\Bar{u}} a(\alpha,u) \left( 1 - G(u) - u g(u) \right) \mathrm{d}u \mathrm{d}\alpha   = K_{\mathrm{low}} \int_0^{\Bar{\alpha}} f(\alpha) \int_0^{\Bar{u}} g(u)  c(\alpha,u)    \mathrm{d}u   \mathrm{d}\alpha  . $$

Implementing the first best optimal solution we find that the budget constraint reduces to 

$$    \int_0^{v} \int_{\alpha^*(u)}^{\Bar{\alpha}} f(\alpha)  \left( 1 - G(u) - u g(u) \right) \mathrm{d}\alpha  \mathrm{d}u   = K_{\mathrm{low}}  \int_0^{v} \int_{\alpha^*(u)}^{\Bar{\alpha}} f(\alpha) g(u)    \mathrm{d}\alpha  \mathrm{d}u , $$

which becomes

$$    \int_0^{v}  (1-F(\alpha^*(u)))  \left( 1 - G(u) - u g(u) \right)   \mathrm{d}u   = K_{\mathrm{low}}  \int_0^{v}  (1-F(\alpha^*(u))) g(u)      \mathrm{d}u  .   $$

Therefore

$$ K_{\mathrm{low}} = \dfrac{ \int_0^{v}  (1-F(\alpha^*(u)))  \left( 1 - G(u) - u g(u) \right)   \mathrm{d}u}{ \int_0^{v}  (1-F(\alpha^*(u))) g(u)      \mathrm{d}u }.$$

\end{proof}

We can prove that for a certain class of $g$ and for every $K \geq K_{\text{low}},$ the optimal solution imposes sharing for all $(\alpha,u)$ with $u>v$. Let us define \(r(u):=\frac{d(u)}{g(u)}:=\frac{1-G(u)}{g(u)}-u\).


\begin{lemma}\label{lem:u_above_v}
Consider $K>K_{\text{low}}$, \(\alpha > 0\) and \(u > v\). 
If $g$ is such that $r(u)$ is decreasing, then every optimal solution 
  \(\bigl(a^\circ, c^\circ\bigr)\) satisfies 
  $$ a^\circ(\alpha,u) = 0 \text{ and }  c^\circ(\alpha,u) = 0 .$$
\end{lemma}
\begin{proof}

Let us focus on the first condition. Fix any feasible $(a,c)$ with the budget equality.
Let $E\subset\{\alpha\ge v,\,u>v\}$ be a set of positive measure where $a>0$.
Since $K>r(u)$ for all $u>v$, we have $d(u)-K g(u)<0$ on $E$.
Turning $a$ and $c$ to zero on $E$ strictly \emph{increases} the objective
(because $v-u<0$ there) and \emph{increases} the residual:
\[
\Delta B_E \;=\; -\!\!\int_E f(\alpha)\,[d(u)-K g(u)]\,\mathrm{d}\alpha\,\mathrm{d}u\;>\;0.
\]

The net effect is a strict objective improvement and strict budget increase.

Let us now consider the second condition.  
%
Given the continuity of $f,g$, the objective function and the constraints, we consider the first best solution as the starting point to analyze in which regions it is suboptimal to increase $a(\alpha,u)$. Denote $(a^\circ,c^\circ)$ the first best solution, which is certainly infeasible for any $K>K_{\text{low}}$. We want to show that the condition $g'(u)> -\frac{2g(u)}{u}$ is sufficient to guarantee that increasing $a(\alpha,u)$ for $u>v$ is suboptimal compared to increasing $a(\alpha',u')$ for some $u'<v$.  In order to prove that, we need to show that the objective/budget ratio of $(\alpha',u')$ is higher than that of $(\alpha,u)$. Specifically, we need to show that 

$$ \dfrac{g(u)f(\alpha) (\min(t(\alpha,u),v)-u)}{f(\alpha) (1-G(u)-u g(u))} < \dfrac{g(u')f(\alpha') (\min(t(\alpha',u'),v)-u')}{f(\alpha') (1-G(u')-u' g(u'))} , $$

where both objective values are negative by the nature of the first best solution. Indeed, our only interest is in increasing the budget as much as possible with as little cost as possible. Note that for every value of $(\alpha,u)$ we can find $(\alpha',u')$ such that $\min(t(\alpha',u'),v)-u' \geq \min(t(\alpha,u),v)-u $. Without loss of generality, we assume that equality holds in the latter expression. Therefore, we find that the comparison of the ratios reduces to 

$$ \dfrac{g(u) }{ (1-G(u)-u g(u))} < \dfrac{g(u')}{ (1-G(u')-u' g(u'))}  ,$$

which turns into, $$ \dfrac{ (1-G(u)-u g(u))}{g(u) } > \dfrac{ (1-G(u')-u' g(u'))}{g(u')}.$$

\end{proof}

The condition that $r(u)$ is decreasing captures the spirit of a monotone hazard rate condition. Observe that another condition that ensures sharing ($a^\circ(\alpha,u)=0$) for every $u>v$ is that $K \;>\; \sup_{u>v} r(u)$, i.e., the cost of inspection is  higher than any budget that could be raised. Conditions of this kind are often imposed in the study of optimal mechanisms.\footnote{See for example, \cite{myerson1981optimal}.} Examples of  $g$ that could satisfy this condition are $g$ uniform or gaussian. 

\paragraph{Uniform density on $[0,1]$.}
Here $g\equiv 1$, $G(u)=u$, $S(u)=1-u$, $d(u)=1-2u$, so
\[
K_{\mathrm{low}}=\frac{v(1-v)}{G(v)}=\frac{v(1-v)}{v}=1-v,
\qquad
r(u)=\frac{d(u)}{g(u)}=1-2u\ \ \text{(decreasing)}.
\]
Hence $K>r(v)=1-2v$ implies $d(u)-K g(u)<0$ for all $u>v$ and both the condition $K \;>\; \sup_{u>v} r(u)$ and Lemma~\ref{lem:u_above_v} apply.
Note also $K_{\mathrm{low}}=1-v\ge 1-2v=r(v)$, so $K\ge K_{\mathrm{low}}$ is \emph{stronger}
than $K>r(v)$ for the uniform case.

\paragraph{Gaussian (half-normal) on $[0,\infty)$.}
Let $g$ be a (truncated) Gaussian density with cdf $G$ and survival $S$.
The Mills ratio $m(u):=S(u)/g(u)$ is strictly decreasing; therefore
\[
r(u)=m(u)-u \quad\text{is strictly decreasing on }[0,\infty).
\]



 For the remainder of the paper we assume that $g$ is chosen within the class of functions that make $r(u)$ decreasing.

\subsection{Dependence on $K$}


In this section we examine how the cost of inspection $K$ influences the choice of the optimal solution. 

As $K$ grows large and eventually tends to infinity, the structure of the optimal solution changes radically. The following lemma characterizes this behavior by showing that, at the limit, the optimal allocation takes the form of a threshold rule.

\begin{lemma}\label{lemma:K_infinite_threshold}
    Assume \(K = \infty\). If $t(\alpha,u)= \alpha$, then there exists a threshold \(u_\infty \le v\) such that 
\[
a^\circ(\alpha,u) = 
\begin{cases}
1 & \text{if } u < u_\infty, \\
0 & \text{if } u \ge u_\infty.
\end{cases}
\]

The threshold is given by
\[
u_\infty 
\;:=\; v - \int_0^{v} f(\alpha)\,\mathrm{d}\alpha
\;=\; v\bigl(1 - F(v)\bigr) + \mathbb{E}[\,t \mid t < v\,]\,F(v).
\]

For $t(\alpha,u)= \alpha\cdot u$, while assuming both $f,g$ are uniform and that $\bar{\alpha}>2$, then 

$$u_\infty=\frac{v}{2} \left( 1 + \sqrt{1- \frac{2}{\bar{\alpha}}} \right).$$
\end{lemma}
\begin{proof}
Set \(K = \infty\). The budget constraint forces \(c(\alpha,u) = 0\) for every \((\alpha,u)\), 
since any positive inspection cost would immediately violate the infinite-cost requirement.  

\paragraph{Claim: \(a(\alpha,u)\) is independent of \(\alpha\).}
From Constraints~\ref{formulation:working_problem_IC}, we have
\[
a(\alpha,u) \;-\; c(\alpha,u)a(\alpha,u) \;\le\; a(0,u)
\quad\Longrightarrow\quad
a(\alpha,u) \;\le\; a(0,u),
\]
because \(c(\alpha,u)=0\). Meanwhile, the monotonicity in \(\alpha\) (Constraints~\ref{formulation:working_problem_tmono}) 
provides us with \(a(0,u) \le a(\alpha,u)\). Combining these, we obtain 
\[
a(\alpha,u) \;\le\; a(0,u) \;\le\; a(\alpha,u)
\quad\Longrightarrow\quad
a(\alpha,u) \;=\; a(0,u).
\]
Hence, for each \(u\), \(a(\alpha,u)\) must be constant in \(\alpha\).

\paragraph{Claim: The solution is a threshold in \(u\).}
Since \(a(\alpha,u)\) does not depend on \(\alpha\), denote it simply by \(a(u)\). 
 By the monotonicity in \(u\), \(a(u)\) must be non-increasing. By Lemma~\ref{lem:u_above_v} we know that for every $u>v$ it must hold $a(\alpha,u)=0$. Therefore, we focus on those $u<v$; in particular, there must be a threshold \(u_\infty\) for which  
\[
a(u) \;=\; 
\begin{cases}
b>0 & u < u_\infty,\\
0 & u \ge u_\infty,
\end{cases}
\]

Note that $b>0$ can only take the value of 1. 
Indeed, for any given $u< u_\infty$ we must have that the contribution to the objective function is positive, i.e., $\int_0^{\Bar{\alpha}} f(\alpha) ( \min(v , t(\alpha,u)) - u ) a(\alpha,u) \, \mathrm{d}\alpha>0$. Therefore, for these $u$, we want to maximise their contribution to the objective function by choosing $a^\circ(\alpha,u)=1$.

\noindent\textbf{\(u_\infty\) for independent interference.} 
Next, we determine \(u_\infty\) by maximizing the objective function for $t(\alpha,u)= \alpha $.
Let us start by simplifying the objective.
With the threshold structure, the objective becomes
\[
I(u_\infty) 
\;=\;\int_{0}^{\Bar{\alpha}} f(\alpha) \,\int_{0}^{u_\infty} g(u)\,\bigl(\min\{t(\alpha,u),v\} - u\bigr)\,\mathrm{d}u\,\mathrm{d}\alpha.
\]
We split the outer integral at \(t=v\):
\[
I(u_\infty) 
\;=\;\underbrace{\int_0^v f(\alpha)\,\int_0^{u_\infty} g(u)\,\bigl(t - u\bigr)\,\mathrm{d}u\,\mathrm{d}\alpha}_{\text{Case }t\le v}
\;+\;\underbrace{\int_v^{\Bar{\alpha}} f(\alpha)\,\int_0^{u_\infty} g(u)\,\bigl(v - u\bigr)\,\mathrm{d}u\,\mathrm{d}\alpha}_{\text{Case }t>v}.
\]

- For \(0 \le t \le v\):
\[
\int_0^{u_\infty} g(u)\,\bigl(t - u\bigr)\,\mathrm{d}u 
\;=\; t\,G(u_\infty) \;-\; \int_0^{u_\infty} u\,g(u)\,\mathrm{d}u 
\;=\; t\,G(u_\infty) \;-\; A(u_\infty),
\]
where \(G(u_\infty) = \int_0^{u_\infty} g(u)\,\mathrm{d}u\) and we define \(A(u_\infty) := \int_0^{u_\infty} u\,g(u)\,\mathrm{d}u.\)

- For \(v < t \le \Bar{\alpha}\):
\[
\int_0^{u_\infty} g(u)\,\bigl(v - u\bigr)\,\mathrm{d}u 
\;=\; v\,G(u_\infty) \;-\; A(u_\infty).
\]

Substitute into \(I(u_\infty)\):
\[
\begin{aligned}
I(u_\infty)
&=\int_0^v f(\alpha)\,\Bigl[t\,G(u_\infty) - A(u_\infty)\Bigr]\,\mathrm{d}\alpha
+\int_v^{\Bar{\alpha}} f(\alpha)\,\Bigl[v\,G(u_\infty) - A(u_\infty)\Bigr]\,\mathrm{d}\alpha\\
&=\;G(u_\infty)\,\Bigl[\underbrace{\int_0^v t\,f(\alpha)\,\mathrm{d}\alpha}_{C}
\;+\;\underbrace{v\!\!\int_v^{\Bar{\alpha}} f(\alpha)\,\mathrm{d}\alpha}_{B}\Bigr]
\;-\;A(u_\infty)\,\underbrace{\Bigl[\int_0^v f(\alpha)\,\mathrm{d}\alpha
\;+\;\int_v^{\Bar{\alpha}} f(\alpha)\,\mathrm{d}\alpha\Bigr]}_{=1}\\
&=G(u_\infty)\,\Bigl[C + B\Bigr]\;-\;A(u_\infty),
\end{aligned}
\]
since \(\int_0^v f(\alpha)\,\mathrm{d}\alpha + \int_v^{\Bar{\alpha}} f(\alpha)\,\mathrm{d}\alpha = 1\). 

Further, observe that by applying integration by parts on the first term we get
\[
C + B \;=\;\int_0^v t\,f(\alpha)\,\mathrm{d}\alpha \;+\; v\!\bigl[1 - F(v)\bigr]
\;=\; v - \int_0^v f(\alpha)\,\mathrm{d}\alpha.
\]
Hence,
\[
I(u_\infty) 
\;=\; G(u_\infty)\,\Bigl[v - \int_0^v f(\alpha)\,\mathrm{d}\alpha\Bigr] \;-\; A(u_\infty).
\]

\paragraph{ Maximize \(I(u_\infty)\).}
Taking the derivative of \(I(u_\infty)\) with respect to \(u_\infty\) and setting it to 0 gives
\[
\frac{dI}{\mathrm{d}u_\infty}
\;=\;\frac{d}{\mathrm{d}u_\infty}\Bigl[G(u_\infty)\,\bigl(v - \!\int_0^v f(\alpha)\,\mathrm{d}\alpha\bigr)\Bigr]
\;-\;\frac{dA(u_\infty)}{\mathrm{d}u_\infty}
\;=\;g(u_\infty)\,\bigl[v - \!\int_0^v f(\alpha)\,\mathrm{d}\alpha\bigr] 
\;-\;u_\infty\,g(u_\infty)
\;=\;0.
\]
Hence, under the condition that $g(u_\infty)\neq 0$, it must hold that 
\[
u_\infty 
\;=\; v - \int_0^v f(\alpha)\,\mathrm{d}\alpha.
\]
A straightforward integral identity shows
\[
v - \int_0^v f(\alpha)\,\mathrm{d}\alpha
\;=\;\int_0^v\bigl[1 - f(\alpha)\bigr]\,\mathrm{d}\alpha
\;=\;v\bigl[1 - F(v)\bigr] + \mathbb{E}\bigl[t\,\mathbf{1}_{t<v}\bigr]
\;=\;v\bigl[1 - F(v)\bigr] + \mathbb{E}[\,t \mid t < v\,]\,F(v).
\]

\paragraph{ Check the second derivative.}
To confirm that this critical point is indeed a maximum, compute
\[
\frac{d^2I}{d(u_\infty)^2}
\;=\;\frac{d}{\mathrm{d}u_\infty}\Bigl[\bigl(v - \!\int_0^v f(\alpha)\,\mathrm{d}\alpha\bigr)g(u_\infty)
\;-\;u_\infty\,g(u_\infty)\Bigr]
\;=\;\Bigl[\bigl(v - \!\int_0^v f(\alpha)\,\mathrm{d}\alpha\bigr)\!-\!u_\infty\Bigr]\,g'(u_\infty)
\;-\;g(u_\infty).
\]
At \(u_\infty = v - \int_0^v f(\alpha)\,\mathrm{d}\alpha\), the bracketed term is zero, so
\[
\frac{d^2I}{d(u_\infty)^2}
\;=\;-\;g(u_\infty)\;<\;0,
\]
which confirms \(u_\infty\) is a maximizer. Thus, the threshold solution with 
\(\displaystyle u_\infty = v - \int_0^v f(\alpha)\,\mathrm{d}\alpha\) 
is optimal under \(K=\infty\). 

\paragraph{ Uniqueness.}
Finally, given the linearity of \(I(u_\infty)\) in each region determined by the threshold,
the constant threshold we have derived is unique. In other words, there cannot be two 
distinct threshold values that both satisfy the first-order and second-derivative conditions.

Hence, the threshold solution with 
\(\displaystyle u_\infty = v - \int_0^v f(\alpha)\,\mathrm{d}\alpha\) 
is the unique optimum under \(K=\infty\).

    
\noindent\textbf{\(u_\infty\) for power interference.} Next, we find the value of $u_\infty$ for $t(\alpha,u)= \alpha \cdot u$ when both $f,g$ are uniform density distributions. The objective function takes the following form, where $u'=\frac{v}{\Bar{\alpha}}$. 

\[
\int_{0}^{u'} g(u)\!\int_{0}^{\bar{\alpha}} f(\alpha)\bigl(u\alpha-u\bigr)\,\mathrm{d}\alpha\,\mathrm{d}u
\;+\;
\int_{u'}^{u_\infty} g(u)\!\left[
      \int_{0}^{\,v/u} f(\alpha)\bigl(u\alpha-u\bigr)\,\mathrm{d}\alpha
      +\int_{\,v/u}^{\bar{\alpha}} f(\alpha)\bigl(v-u\bigr)\,\mathrm{d}\alpha
\right]\mathrm{d}u.
\]

If we do the derivative of the objective function \text{obj} by $u_\infty$, then we obtain

\[
\frac{d\text{obj}}{du_\infty}=g(u_\infty)\,h(u_\infty)
\quad\text{where}\quad
h(u_\infty)=
\int_{0}^{\,v/u_\infty} f(\alpha)\bigl(u_\infty\alpha-u_\infty\bigr)\,\mathrm{d}\alpha
\;+\;
\int_{\,v/u_\infty}^{\bar{\alpha}} f(\alpha)\bigl(v-u_\infty\bigr)\,\mathrm{d}\alpha.
\]

If we solve for the first-order condition \(h(u_\infty)=0\), then we obtain  
\[
v \bar{\alpha}- \frac{1}{2}\cdot \frac{v^2}{u_\infty} - u_\infty \bar{\alpha} \;=\; 0.
\]

If we solve the ensuing quadratic expression (assuming $u_\infty>0)$, then we find that $u_\infty $ may take one of the following two values

\begin{itemize}
    \item $\frac{v}{2} \left( 1 - \sqrt{1- \frac{2}{\bar{\alpha}}} \right)$,
    \item $\frac{v}{2} \left( 1 + \sqrt{1- \frac{2}{\bar{\alpha}}} \right)$,
\end{itemize}

in both cases we need to further assume $\bar{\alpha}>2$ in order to have a real number. Let us study the second derivative to find which of the two values is a minimum. The second derivative of the objective function by  $u_\infty$ is 

\[
 \frac{1}{2}\cdot \frac{v^2}{u_\infty^2} - u_\infty \bar{\alpha} \;=\; 0.
\]

If we substitute the first stationary point into the equation, we obtain

$$ \dfrac{\bar{\alpha}}{\bar{\alpha}-1 - \sqrt{1- \frac{2}{\bar{\alpha}}}} - \bar{\alpha},$$

which has always a positive sign as the denominator is smaller than 1.  

Otherwise, if we substitute the value of the second stationary point for $u_\infty$, then we find that the expression 

$$ \bar{\alpha}-1 - \sqrt{1- \frac{2}{\bar{\alpha}}} >1 $$

is always true, therefore the second derivative of the objective function evaluated in $u_\infty=\frac{v}{2} \left( 1 + \sqrt{1- \frac{2}{\bar{\alpha}}} \right)$ is negative. 


\end{proof}

Observe that for any finite $K$, there is no purely $u$-dependent (constant) threshold. Indeed, for every large finite $K$, one can always identify a small region around the would-be threshold $u_\infty$ in which the incumbent retains exclusive access and is subject to inspection. 

\begin{theorem}\label{cor:finite_K_no_constant_threshold}
    Let $K<\infty$, then there is $u$ and $\alpha>0$ for which $a^\circ(0,u)=0$ and $a^\circ(\alpha,u)>0$.
\end{theorem}
\begin{proof}
    Recall that since, by assumption, $K_{\text{low}}<K$, the budget constraint must be binding. Assume there is no $u$ for which the thesis holds. Note that the negation of the thesis formally goes as: For every $u$ and for every $\alpha$, or (vel) $a^\circ(0,u)>0$ or (vel) $a^\circ(\alpha,u)=0$. Note, in particular, that if for a certain $u$ it holds that $a^\circ(0,u)=0$, then it must hold for every $\alpha>0$ that $a^\circ(\alpha,u)=0$. On the other hand, if $a^\circ(0,u)>0$, then by the $\alpha$ monotonicity in 0, it must hold that $a^\circ(\alpha,u)>0$ for every $\alpha>0$. Therefore, by the monotonicity property in $u$, there is a constant threshold $u^*$ for which for every $u>u^*$ and for every $\alpha\geq 0$ $a^\circ(\alpha,u)=0$, and for every $u<u^*$ then $a^\circ(0,u)>0$. Clearly, it must be that $u^*\leq v $. 

If for every $u'<u^*$ it holds that  $a^\circ(0,u')=1$ (i.e., $a^\circ(\alpha,u')=1$ for every $u'<u^*$ and for every $\alpha$),  then the optimal policy for inspection can be $c^\circ(\alpha,u)=0$ everywhere. Let $\varepsilon>0$ and consider a set $\mathcal{H}(\varepsilon) = \{(\alpha,u)\colon \, u^*-\varepsilon<u<u^* \text{ and } \alpha<\alpha^*(u) \}$. We can choose $\varepsilon>0$ small enough to define a new solution $(a^*(\alpha,u), c^*(\alpha,u))$ in the following way: $a^*(\alpha,u)$ coincides with $a^\circ(\alpha,u)$ everywhere except for the points in $\mathcal{H}(\varepsilon)$; at the points in $\mathcal{H}(\varepsilon)$ we set  $a^*(\alpha,u)=0$. Moreover, we set $c^*(\alpha,u)=1$ for all $u^*-\varepsilon<u<u^*$ and $\alpha>\alpha^*(u)$, otherwise $c^*(\alpha,u)=0$. The choice of $\varepsilon$ is small enough such that the budget constraint holds (which always exists for a finite $K$) for the solution $(a^*(\alpha,u), c^*(\alpha,u))$. Note that the value of the solution $(a^*(\alpha,u), c^*(\alpha,u))$ improves on the one of $(a^\circ(\alpha,u), c^\circ(\alpha,u))$.

    If there is at least one $u'''$ for which $0<a^\circ(\alpha,u''')=a^\circ(0,u''')<1$,  for each $\alpha>0$, then we can find a better solution depending on the sign of $\text{val}^\circ=\int_0^{\Bar{\alpha}} f(\alpha) \big( \min(v , t(\alpha,u''')) - u''' \big) a^\circ(\alpha,u''') \, \mathrm{d}\alpha$. Without loss of generality, we assume $u'''$ is the only commercial utility for which $0<a^\circ(\alpha,u''')=a^\circ(0,u''')<1$ for every $\alpha>0$. If sign of $\text{val}^\circ>0$ then we simply choose as a new solution  $a^\star(0,u^{iv})=1$ for each  $u^{iv}\leq u'''$ and $a^\star(\alpha,u)=a^\circ(\alpha,u)$ everywhere else; otherwise, if sign of $\text{val}^\circ<0$, then we define a new solution $a^\star(0,u^{iv})=a^\star(\alpha,u^{iv})=0$ for each  $u^{iv}\geq u'''$ and for every $\alpha>0$, and $a^\star(\alpha,u)=a^\circ(\alpha,u)$ everywhere else. Note that in both cases, the new solution yields a better objective value. Therefore, from now on, we assume that for each $u'''$ such that $0<a^\circ(0,u''')<1$, there is a $\alpha$ that we denote by $\alpha^\dagger(u''')\geq \alpha^*(u''')$ for which $0<a^\circ(0,u''')<a^\circ(\alpha^\dagger(u'''),u''')$.

    Now we show that for every $u$ such that $0<a^\circ(0,u)<1$, it must hold that $c^\circ(\alpha,u)< a^\circ(\alpha,u)$ whenever $\alpha>\alpha^\dagger(u)$ almost everywhere. Otherwise, assume there is $\alpha>\alpha^\dagger(u)$ for which  $c^\circ(\alpha,u)= a^\circ(\alpha,u)$. From Constraints~\ref{formulation:working_problem_IC} we find that $a^\circ(\alpha,u)-a^\circ(\alpha,u)^2\leq a^\circ(0,u)$. If it holds that  $a^\circ(\alpha,u)-a^\circ(\alpha,u)^2< a^\circ(0,u)$, then we can reduce the value of $c^\circ(\alpha,u)$ and increase it somewhere else to improve the objective function. Otherwise, it must be that $a^\circ(\alpha,u)-a^\circ(\alpha,u)^2= a^\circ(0,u)$ for a non negligible set of  $\alpha>\alpha^*(u)$, but this can hold at most for two values of $a^\circ(\alpha,u)$ once $a^\circ(0,u)$ is fixed (just solve the quadratic expression for $a^\circ(\alpha,u)$), thus proving our statement. 

    Next, we prove that Constraint~\ref{formulation:working_problem_IC} must be tight. Otherwise, assume there is $u'$ (without loss of generality, let $u'$ be the smallest with the following properties) such that $0<a^\circ(0,u')<1$ and there is $\alpha''>\alpha^\dagger(u')$ for which $a^\circ(\alpha'',u')(1-c^\circ(\alpha'',u'))<a^\circ(0,u')$. Let us choose $\alpha''$ to be the largest and increase $a(\alpha'',u)$ until the constraint  is tight. Note that we find a feasible solution with a better objective value. 

    Consider $u'<u^*$ the lowest for which $0<a^\circ(0,u')<1$. Consider the highest $\alpha''$ for which $a^\circ(\alpha'',u')<1$, then, we can increase the value of  $a^\circ(\alpha'',u')$  and $a^\circ(0,u')$ in the following way. From the IC constraint (Constraint~\ref{formulation:working_problem_IC}) we know that $a^\circ(\alpha'',u')(1-c^\circ(\alpha'',u'))=a^\circ(0,u')$. For every $\alpha'\leq \alpha''$ we look for $\gamma, \, \beta>0$ such that $(a^\circ(\alpha',u')+\beta(\alpha',u'))(1-c^\circ(\alpha',u'))=a^\circ(0,u')+\gamma$.  
    %
    %
    %
    For the new solution defined as $a^\circ(\alpha',u')+\beta(\alpha',u')$ to satisfy both $\alpha$-monotonicity it suffices to require $\beta(\alpha',u')\geq   a^\circ(0,u')+\gamma -a^\circ(\alpha',u') $. From the IC constraint, we deduce that it must hold $(\beta(\alpha',u'))(1-c^\circ(\alpha',u'))=\gamma$, which yields $\beta(\alpha',u')\geq \gamma$. Note that $u$-monotonicity follows from the assumption that $u'$ is the smallest.  
    Constraint~\ref{formulation:working_problem_inspection} and~\ref{formulation:working_problem_budget} hold as we do not change the value of any $c^\circ(\alpha,u')$ and increase the values of $a^\circ(\alpha',u')$. The only remaining thing we need to be careful about is that $a^\circ(\alpha',u')+\beta(\alpha',u')\leq 1$, and it suffices to write that $\beta(\alpha'',u')\leq 1-a^\circ(\alpha'',u')$ as $\alpha''$ is the largest $\alpha$ for which we increase the solution. 
    Finally, we need to ensure that the newly found solution is better than the previous one. First, we note that from the assumption that $u'<u^*$ it holds that $ \int_0^{\Bar{\alpha}} (\min(t(\alpha,u'),v) - u' ) a(\alpha,u')\,  \mathrm{d}\alpha>0$ for every positive choice of $a(\alpha,u')$ that satisfies the constraints. If we denote by ObjVal$^\circ$ the optimal value of the solution  $(a^\circ, c^\circ)$, then we have that the value of the newly defined solution is: ObjVal$^\circ$ + $ \int_0^{\alpha^\dagger(u')} (\min(t(\alpha,u'),v) - u' ) \gamma \,  \mathrm{d}\alpha$ + $ \int_{\alpha^\dagger(u')}^{\Bar{\alpha}} (\min(t(\alpha,u'),v) - u' ) \beta(\alpha,u')\,  \mathrm{d}\alpha $.  From the hypothesis that $u'<u^*$, it follows that if $\beta(\alpha,u')=\gamma$, then our thesis holds; since $\beta(\alpha,u')\geq \gamma$ for every $\alpha>\alpha^\dagger(u')$, we have found a better solution.

\end{proof}

In the next section, we  analyze and characterize an optimal solution $(a^\circ, c^\circ)$ of the planner's problem displayed in Formulation~\ref{formulation:working_problem}.

\section{Appendix: Determining the Optimal Policy}\label{sec:solution}
The next result shows that any optimal solution to Formulation~\ref{formulation:working_problem} must, in fact, be binary.


\begin{theorem}\label{G_theo:binary_optimal_solution}
  Let $(a^\circ, c^\circ)$ be an optimal solution of Formulation~\ref{formulation:working_problem}.  For every $(\alpha,u)$ in the domain, the following properties hold: 
  \begin{enumerate}
    \item[(i)] If $a^\circ(\alpha,u)>0$, then $a^\circ(\alpha,u)=1$.
    \item[(ii)]  $a^\circ(0,u)=0$ if and only if $a^\circ(\alpha,u)=0$ for every $\alpha<\alpha^*(u)$. 
  \end{enumerate}
\end{theorem}
\begin{proof}
  \emph{ Preliminary Observations.}  
  By Lemma~\ref{lem:u_above_v}, if $u>v$, then $a^\circ(\alpha,u)=0$ for all $\alpha\ge 0$. Therefore, we only need to consider $u < v$. We distinguish two main cases based on whether $a^\circ(0,u)=0$ or not.

  \emph{ Case $a^\circ(0,u)=0$.}  
  Suppose there exists $\alpha>0$ such that $a^\circ(\alpha,u)>0$. By constraint~\eqref{formulation:working_problem_IC}, having $a^\circ(\alpha,u)>0$ forces $c^\circ(\alpha,u)=1$. Then, by the inspection constraint~\eqref{formulation:working_problem_inspection}, it follows that $a^\circ(\alpha,u)$ must be 1 (since $c(\alpha,u)\le a(\alpha,u)$). Consequently, $a^\circ(\alpha,u)=1$.

  Now we prove that if $a^\circ(0,u)=0$, there can be no $\alpha$ such that $0<\alpha<\alpha^*(u)$ and $a^\circ(\alpha,u)>0$. Suppose by contradiction that it exists such an $\alpha'$ with $\alpha'<\alpha^*(u)$ and, without loss of generality,  let us assume that $\alpha'$ is the smallest and $u'$ is the largest for which $\alpha'<\alpha^*(u')$, $a^\circ(0,u')=0$ and $a^\circ(\alpha',u')>0$. From the previous argument we have that it must hold that $a^\circ(\alpha',u')=1$. Take $u''<u'$ in the neighborhood of $u'$ such that there is an $\alpha''<\alpha'$ such that $a(\alpha'',u'')=0$ (we can choose the highest $u'$ that admits such $u''$).
  %
  %
  %
   We now define a new solution $(a',c')$ where $a'(\alpha',u')=c'(\alpha',u')=0$ and $a'(\alpha'',u'')=c'(\alpha'',u'')=1$, while maintaining the same value of the original solution everywhere else. Note that by continuity we can assume there is no variation in the cost of inspection (rather a slight increase in the raised budget) and all the other constraints continue to hold too. Moreover, the objective function improves.

  Clearly, if $a^\circ(\alpha,u)=0$ for every $\alpha$ such that  $0<\alpha<\alpha^*(u)$, then $a^\circ(0,u)=0$ by monotonicity. 

  \emph{ Case $a^\circ(0,u)>0$.}  We must distinguish the cases of the value of $K$: For $K=\infty$ we have already proved this result inlord va Lemma~\ref{lemma:K_infinite_threshold}, respectively. Therefore, we must prove that if $a^\circ(\alpha,u)>0$, then $a^\circ(\alpha,u)=1$ for $K_{\text{low}}<K<\infty$.
    Since $K_{\text{low}}<K$, then the budget constraint is binding, and for some points $(\alpha,u)$ in the domain we should set $c^\circ(\alpha,u)<1$. By Theorem~\ref{cor:finite_K_no_constant_threshold} there exists $u'$ and $\varepsilon>0$ such that $a^\circ(0,u')=0$ and $a^\circ(u'+\varepsilon,u')>0$. 
  
  By the $u$‐monotonicity constraint~\eqref{formulation:working_problem_umono}, we may assume $u$ is the largest value for which $a^\circ(0,u)>0$. Since $K<\infty$, there must be  $\delta,\varepsilon>0$ for which $a^\circ(0,u+\delta)=0$, and $a^\circ(u+\varepsilon,u+\delta)>0$; from the previous argument, we have that $a^\circ(u+\varepsilon,u+\delta)=1$. Note that by $u$-monotonicity, it must hold that $a^\circ(u+\varepsilon,u)=1$ and $c^\circ(u+\varepsilon,u)=1$. 

Finally, we need to show that if $a^\circ(0,u)>0$, then $a^\circ(0,u)=1$. Assume not. If there is some $\alpha>\alpha^*(u)$ for which $c^\circ(\alpha,u)=1$, then we can decrease $c^\circ(\alpha,u)$ in order to redistribute the inspection mass to improve the objective function somewhere else (as shown in Theorem~\ref{cor:finite_K_no_constant_threshold}). Therefore, $c^\circ(\alpha,u)<1$ almost everywhere, and we can choose the minimal and maximal $u''$ and $u'$, respectively, such that $0<a^\circ(0,u)<1$. 
%
%
%
%
Clearly, for every $\alpha<\alpha^*(u)$ it must hold that $a^\circ(\alpha,u)=a^\circ(0,u)$ (otherwise we could decrease the value of $a^\circ(\alpha,u)$ and improve it locally thus obtaining a better objective value). Additionally, for every $\alpha>\alpha^*(u)$ it must hold that $a^\circ(\alpha,u)=1$ (otherwise we could augment it to 1 and obtain locally a better solution), and from the IC constraint (Constraints~\ref{formulation:working_problem_budget}) it holds that $1-c(\alpha,u)\leq a^\circ(0,u)$; note that it must hold an equality, i.e., $1-c(\alpha,u)= a^\circ(0,u)$, otherwise we could improve the objective value by reducing $a^\circ(0,u)$. Our goal is to reduce the value of $a^\circ(0,u')$ and increase the value of $a^\circ(0,u'')$ while satisfying all the constraints. We will show that by doing so, we find a feasible solution that improves on the initial optimal allocation in which $0<a^\circ(0,u')<1$. 

Consider $\varepsilon, \phi>0$. We define a new solution $(a^*(\alpha,u), c^*(\alpha,u))$ as follows: for $\alpha<\alpha^*(u')$ we choose $a^*(\alpha,u')= a^\circ(\alpha,u')-\varepsilon$; for $\alpha>\alpha^*(u')$ we choose $c^*(\alpha,u')= c^\circ(\alpha,u')+\varepsilon$; for $\alpha<\alpha^*(u'')$ we choose $a^*(\alpha,u'')= a^\circ(\alpha,u'')+\phi$; for $\alpha>\alpha^*(u'')$ we choose $c^*(\alpha,u'')= c^\circ(\alpha,u'')-\phi$; everywhere else we choose  $a^*(\alpha,u)=a^\circ(\alpha,u)$ and $c^*(\alpha,u)=c^\circ(\alpha,u)$. Note that for the feasibility of the new solution we need to impose that $\varepsilon<a^\circ(0,u')$, $\varepsilon<1- c^\circ(\alpha,u')$, $\phi<c^\circ(\alpha,u'')$, $\phi<1- a^\circ(0,u'')$, which also imply that the Constraints~\ref{formulation:working_problem_inspection} are satisfied. Note that the monotonicity constraints are valid by the way we picked $u',u''$ and that Constraints~\ref{formulation:working_problem_IC} hold by construction. We need to verify the budget constraint, and to do that we choose $\beta,\gamma>0$ arbitrarily small; we use the following notation: $[\ldots]:= \int_0^u a^\circ(\alpha,x)\,\mathrm{d}x - u a^\circ(\alpha,u)$.  The left-hand side of the budget constraint takes the following form:

\begin{align*}
& \int_{0}^{u''-\beta} g(u)\,\int_{0}^{\Bar{\alpha}} f(\alpha) [\ldots] \,\mathrm{d}\alpha \,\mathrm{d}u \\ 
& + \int_{u''-\beta}^{u''} g(u)\, \Bigg[ \int_{0}^{\alpha^*(u)} f(\alpha) \Bigg( \int_0^{u''-\beta} (a^\circ(\alpha,x) ) \,\mathrm{d}x  +\int_{u''-\beta}^{u} (a^\circ(\alpha,x) +\phi) \,\mathrm{d}x - u (a^\circ(\alpha,u)+\phi) \Bigg) \,\mathrm{d}\alpha  \\ 
& + \int_{\alpha^*(u)}^{\Bar{\alpha}} f(\alpha) [\ldots] \,\mathrm{d}\alpha  \Bigg] \,\mathrm{d}u  \\ 
& + \int^{u'-\gamma}_{u''} g(u)\, \Bigg[ \int_{0}^{\alpha^*(u)} f(\alpha) \Bigg( \int_0^{u''-\beta} a^\circ(\alpha,x) \,\mathrm{d}x  +\int_{u''-\beta}^{u''} (a^\circ(\alpha,x) +\phi) \,\mathrm{d}x \\ 
& + \int_{u''}^{u} a^\circ(\alpha,x)  \,\mathrm{d}x - u a^\circ(\alpha,u) \Bigg) \,\mathrm{d}\alpha    + \int_{\alpha^*(u)}^{\Bar{\alpha}} f(\alpha) [\ldots] \,\mathrm{d}\alpha  \Bigg] \,\mathrm{d}u  \\
& + \int_{u'-\gamma}^{u'} g(u)\, \Bigg[ \int_{0}^{\alpha^*(u)} f(\alpha) \Bigg( \int_0^{u''-\beta} a^\circ(\alpha,x) \,\mathrm{d}x  +\int_{u''-\beta}^{u''} (a^\circ(\alpha,x) +\phi) \,\mathrm{d}x \\ 
& + \int_{u''}^{u'-\gamma} a^\circ(\alpha,x)  \,\mathrm{d}x + \int^{u}_{u'-\gamma} (a^\circ(\alpha,x) -\varepsilon) \,\mathrm{d}x - u (a^\circ(\alpha,u)-\varepsilon) \Bigg) \,\mathrm{d}\alpha    + \int_{\alpha^*(u)}^{\Bar{\alpha}} f(\alpha) [\ldots] \,\mathrm{d}\alpha  \Bigg] \,\mathrm{d}u  \\
& + \int_{u'}^{\Bar{u}} g(u)\, \Bigg[ \int_{0}^{\alpha^*(u)} f(\alpha) \Bigg( \int_0^{u''-\beta} a^\circ(\alpha,x) \,\mathrm{d}x  +\int_{u''-\beta}^{u''} (a^\circ(\alpha,x) +\phi) \,\mathrm{d}x \\ 
& + \int_{u''}^{u'-\gamma} a^\circ(\alpha,x)  \,\mathrm{d}x + \int^{u'}_{u'-\gamma} (a^\circ(\alpha,x) -\varepsilon) \,\mathrm{d}x + \int^{u}_{u'} a^\circ(\alpha,x)  \,\mathrm{d}x - u a^\circ(\alpha,u) \Bigg) \,\mathrm{d}\alpha    + \int_{\alpha^*(u)}^{\Bar{\alpha}} f(\alpha) [\ldots] \,\mathrm{d}\alpha  \Bigg] \,\mathrm{d}u  \\
\end{align*}

Note that since the budget constraint is binding for the solution $(a^\circ, c^\circ)$ we can focus simply on the terms involving $\phi$ and $\epsilon$. Therefore, for small $\beta,\gamma$ the left-hand side of the budget constraint becomes

$$ g(u'') \int_{0}^{\alpha^*(u'')} f(\alpha) (-\phi \cdot u'')\, \mathrm{d}\alpha + g(u') \int_{0}^{\alpha^*(u')} f(\alpha) (\varepsilon \cdot u') \, \mathrm{d}\alpha. $$

Following a similar reasoning, we find that the right-hand side of the budget constraint takes the following form:

$$ K \Bigg( g(u'') \int^{\Bar{\alpha}}_{\alpha^*(u'')} f(\alpha) (-\phi )\, \mathrm{d}\alpha + g(u') \int^{\Bar{\alpha}}_{\alpha^*(u')} f(\alpha) (\varepsilon ) \, \mathrm{d}\alpha \Bigg) . $$

In order to satisfy the budget constraint for the new solution $(a^*, c^*)$ we need to satisfy the following inequality,

$$  \dfrac{g(u'')\Bigg( K\int^{\Bar{\alpha}}_{\alpha^*(u'')} f(\alpha) \, \mathrm{d}\alpha -u''\int_{0}^{\alpha^*(u'')} f(\alpha) \, \mathrm{d}\alpha \Bigg)}{g(u')\Bigg( K\int^{\Bar{\alpha}}_{\alpha^*(u')} f(\alpha) \, \mathrm{d}\alpha -u'\int_{0}^{\alpha^*(u')} f(\alpha) \, \mathrm{d}\alpha \Bigg)} \geq \frac{\varepsilon}{\phi} >0.$$

Now we want to verify that the variation in the objective function caused by $\varepsilon$ and $\phi$ yields a better objective value. Specifically, we want the decrease incurred by $\varepsilon $ to be smaller than the increase provided by $\phi$. 

$$- g(u') \varepsilon  \int_{0}^{\alpha^*(u')} f(\alpha) (\min (t(\alpha,u),v)-u') \, \mathrm{d}\alpha + g(u'') \phi  \int_{0}^{\alpha^*(u'')} f(\alpha) (\min (t(\alpha,u),v)-u'') \, \mathrm{d}\alpha <0 ,$$

and by rearranging the terms, we find

$$ \frac{\varepsilon }{\phi } > \dfrac{g(u'')   \int_{0}^{\alpha^*(u'')} f(\alpha) (\min (t(\alpha,u),v)-u'') \, \mathrm{d}\alpha}{g(u')   \int_{0}^{\alpha^*(u')} f(\alpha) (\min (t(\alpha,u),v)-u') \, \mathrm{d}\alpha}.$$

If we find the right hand side to be greater than 0, we have verified the condition on the budget constraint and found a better feasible solution than $(a^\circ, c^\circ)$. This must be true, as both the numerator and the denominator are negative for the choice of $\alpha^*(u)$ (recall that $\alpha^*(u)$ is the largest $\alpha$ for which $\min (t(\alpha,u),v)<u''$).

  %

  \emph{Conclusion.}  
  The above arguments establishes that if $a^\circ(\alpha,u)>0$, it must be $1$, proving part (i). 
\end{proof}

By Constraints~\ref{formulation:working_problem_tmono}, we immediately obtain the following corollary,

\begin{corollary}\label{cor:a(0,u)}
  Let $u$ be such that $a^\circ(0,u)=1$ in an optimal solution. Then $a^\circ(\alpha,u)=1$ for every $\alpha \ge 0$.
\end{corollary}




In addition to restricting $a^\circ(\alpha,u)$ to be binary, the optimal solution also imposes specific bounds on the probability of inspection, $c^\circ(\alpha,u)$. 

\begin{corollary}\label{cor:choice_of_c}
  Suppose $u$ is such that $a^\circ(0,u)=0$ in an optimal solution. Then, for every $\alpha \ge 0$, we have $c^\circ(\alpha,u)=a^\circ(\alpha,u)$.
\end{corollary}
\begin{proof}
  By constraint~\ref{formulation:working_problem_IC}, we know 
  \[
    a^\circ(\alpha,u) - c^\circ(\alpha,u)\,a^\circ(\alpha,u) \;\le\; 0.
  \]
  Hence, if $a^\circ(\alpha,u)$ is positive, then it must be $a^\circ(\alpha,u)=1$ and it forces $c^\circ(\alpha,u)=1$. 
  Meanwhile, if $a^\circ(\alpha,u)=0$ then from Constraints~\ref{formulation:working_problem_inspection} it must hold that $c^\circ(\alpha,u)=0$. 

  Thus, $c^\circ(\alpha,u)=a^\circ(\alpha,u)$ for all $\alpha\ge 0$.
\end{proof}

Corollary~\ref{cor:choice_of_c} implies that whenever sharing occurs at a point with no negative externality (i.e., $a^\circ(0,u)=0$ and $\alpha>\alpha^*(u)$), an inspection will happen exactly if the good is allocated exclusively to the incumbent.

When the planner chooses to grant the incumbent exclusive access at a specific utility value 
\(u\) (i.e., \(a^\circ(0,u)=1\)), we want to determine how inspection is assigned at all higher 
interference value \(\alpha\ge 0\). Intuitively, if the cost of inspection \(K\) is sufficiently large, 
the planner may prefer not to inspect at all in the region where the commercial user declares a low utility. The following corollary formalizes this point. 

\begin{corollary}\label{cor:inspection-zero}
  Suppose \(a^\circ(0,u)=1\) for some \(u\). Then, in any optimal solution, 
  \(c^\circ(\alpha,u)=0\) for all \(\alpha\ge 0\). 
\end{corollary}
\begin{proof}
  Assume \(a^\circ(0,u) = 1\). By Corollary~\ref{cor:a(0,u)},
  whenever \(a^\circ(0,u)=1\), it follows that \(a^\circ(\alpha,u)=1\) for all \(\alpha < \alpha^*(u)\). 
  In other words, the incumbent is granted exclusive access in that region of the \((\alpha,u)\) space.

  Now, if \(K\) was sufficiently small, one might attempt to set \(a^\circ(\alpha,u)=c^\circ(\alpha,u)=0\) for \(\alpha < \alpha^*(u)\)
  to avoid the negative contribution caused by exclusive access at small \(\alpha\). However, doing so 
  contradicts our hypothesis that \(a^\circ(\alpha,u)=1\) for \(\alpha < \alpha^*(u)\). Consequently, $K$ must be large, and in order to satisfy Constraint~\ref{formulation:working_problem_budget}, for some points of the domain the probability must be set to zero. We argue that the probability of inspection is set to zero in correspondence of those $u$ for which \(a^\circ(0,u) = 1\). 

  If the probability of inspection would be set to zero in correspondence to some $(\alpha,u)$ for which \(a^\circ(\alpha,u) = 1\) and \(a^\circ(0,u) = 0\), then we would violate Constraints~\ref{formulation:working_problem_IC}: $1-0\leq 0$. Therefore, we could only set \(c^\circ(\alpha,u) = 0\) in correspondence of those $(\alpha,u)$ for which \(a^\circ(0,u) = 1\). In order to minimize the negative impact on the objective function, we would set  \(c^\circ(\alpha,u) = 0\) for every $t$ and every $u$ for which \(a^\circ(0,u) = 1\).  
\end{proof}

Corollary~\ref{cor:inspection-zero} shows that if the incumbent is granted exclusive access at 
\((0,u)\), then this would happen only under high inspection costs, and the planner finds it optimal to forgo inspection 
throughout all negative externality levels \(\alpha \ge 0\). This aligns with the intuition that costly 
inspections (combined with incumbent exclusivity) lead to strictly positive allocations 
(\(a^\circ\)=1) and zero inspection (\(c^\circ\)=0) over that region.

Next, we show that Constraints~\ref{formulation:working_problem_umono} is sufficient to guarantee that every optimal allocation is weakly decreasing in $\alpha$.

\begin{lemma}\label{lemma:alpha_full_monotonicity}
    Every optimal allocation $a^\circ(\alpha,u)$ is non-decreasing monotone in $\alpha$.
\end{lemma}
\begin{proof}
    Fix $u'\leq v$. From Constraints~\ref{formulation:working_problem_tmono} we deduce that if $a(0, u')=1$, then $a(\alpha, u')=1$ for every $\alpha>0$. Let us then consider $a(0, u')=0$. By reorganizing the left-hand side of the budget constraint, we find that for a fixed $(\alpha,u')$ we have that $f(\alpha) \left( \int_u^{\Bar{u}} g(y) \, dy - u \cdot g(u) \right)$. Therefore, the ratio between the value objective function and the budget constraint at $(\alpha,u')$ takes the following form:

    $$ \dfrac{ f(\alpha)g(u')\left( \min(t(\alpha,u'),v)-u' \right) a(\alpha,u') }{f(\alpha) \left( \int_u^{\Bar{u}} g(y) \, dy - u \cdot g(u) \right)},$$

where $t(\alpha,u')$ increases in $\alpha$. Therefore, for higher values of $\alpha$ we have a better ratio, thus our weak-increasing $\alpha$ monotonicity. 
\end{proof}

Given that $a^\circ(\alpha,u)$ is  \emph{weakly  non-decreasing} in $\alpha$ and \emph{weakly  non-increasing} in $u$, the domain over which the optimal policy imposes sharing must be connected. Specifically, for each fixed $\alpha$, the function $u \mapsto a^\circ(\alpha,u)$ takes on binary values (either $>0$ or $0$) and is monotonic in $u$. Consequently, it can only \emph{switch} from $>0$ to $0$ at most once. Hence, for every $\alpha$, there is a unique threshold in $u$ that determines whether sharing is imposed. 

We denote this threshold by $\phi(\alpha,K,v)$, where $K$ and $v$ are common knowledge parameters,  
and $t$ is known too. 
Formally, the allocation decision is written as 

\[
  a^\circ(\alpha,u) \;=\;
  \begin{cases}
    1, & \text{if } u < \phi(\alpha,K,v),\\[6pt]
    0,  & \text{if } u > \phi(\alpha,K,v).
  \end{cases}
\]

Because $a^\circ(\alpha,u)$ is also non-decreasing in $\alpha$, the threshold function $\phi(\alpha,K,v)$ itself inherits a weakly monotonic dependence on $\alpha$. This ensures a single continuous boundary in the $(\alpha,u)$-plane between the region of sharing and the region of no sharing.

Therefore, our task now reduces to finding an explicit expression for the threshold function $\phi(\alpha,K,v)$ that characterizes the allocation decision. In particular, when $K,v$ are treated as fixed parameters, we may focus on the dependence $\phi(\alpha)$ in $\alpha$, and express the threshold piecewise in terms of $\alpha$. In the next section, we study the optimal thresholds, 
for the two kinds of interference $t(\alpha,u)$.

\subsection{Threshold Solution for $t(\alpha,u)=\alpha$}

In our first result, we characterize the structure of the threshold function for $t(\alpha,u)=\alpha$, which depends on two parameters $(u_{\mathrm{bot}}, u_{\mathrm{top}})$; these parameters divide the three regions of the domain identified by the optimal allocation-inspection policy. Later, in Theorem~\ref{thm:knapsack_independent}, we provide an efficient method for computing $(u_{\mathrm{bot}}, u_{\mathrm{top}})$.

\begin{theorem}\label{thm:threshold_independent}
  Consider $K,v \geq 0$, then there exist two constants $u_{\mathrm{bot}}$ and $u_{\mathrm{top}}$ satisfying
  \[
    0 \;\le\; u_{\mathrm{bot}} 
          \;\le\; u_{\mathrm{top}}
          \;\le\; v,
  \]
  
  such that the threshold function $\phi(\alpha)$ takes the following piecewise form:
  \[
    \phi(\alpha) \;=\;
    \begin{cases}
      u_{\mathrm{top}}, & \text{if } \alpha \,\ge\, v,\\[6pt]
      \alpha- (v-u_{\mathrm{top}}),                & \text{if } u_{\mathrm{bot}} + v-u_{\mathrm{top}} \,<\, \alpha \,<\, v,\\[6pt]
      u_{\mathrm{bot}}, & \text{if } \alpha \,\le\, u_{\mathrm{bot}}+v-u_{\mathrm{top}}.
    \end{cases}
  \]
  In other words, the threshold for $K>K_{\text{low}}$ is an horizontal translation of the threshold for $K_{\text{low}}$. Specifically, the threshold is constant at $u_{\mathrm{bot}}$ for small $\alpha$, then coincides with $\alpha$ minus an inefficiency in the intermediate range, and finally becomes constant at $u_{\mathrm{top}}$ for large $\alpha$.
\end{theorem}
\begin{proof}  
    Consider an optimal solution $\bigl(a^\circ(\alpha,u),\,c^\circ(\alpha,u)\bigr)$. By 
Theorem~\ref{G_theo:binary_optimal_solution}, every optimal allocation $a^\circ$ 
takes only binary values (either zero or positive). 

\emph{Definition of $u_{\mathrm{bot}}$.} 
Suppose there exists some $u' > 0$ such that $a^\circ(0,u') = 1$. Then, by 
Corollary~\ref{cor:a(0,u)}, it follows that $a^\circ(\alpha,u') = 1$ for every 
$\alpha \ge 0$. Because $a^\circ$ is non-increasing in $u$, we can define
\[
  u_{\mathrm{bot}}
  \;=\;
  \sup\!\bigl\{\,u \ge 0 : a^\circ(0,u)>0  \bigr\}.
\]

By Lemma~\ref{lem:u_above_v}, this supremum is actually a maximum, and, moreover 
$u_{\mathrm{bot}} \le v$. If no such $u' > 0$ exists (i.e., $a^\circ(0,u)=0$ for 
all $u>0$), then we define $u_{\mathrm{bot}}=0$.

\emph{ Definition of $u_{\mathrm{top}}$.} 
\begin{itemize}
\item \emph{Case 1:} If there is no $u > 0$ and no $\alpha \ge 0$ for 
which $a^\circ(\alpha,u)>0$ and $a^\circ(0,u)=0$, then we define  $u_{\mathrm{top}}=u_{\mathrm{bot}}$.

\item \emph{Case 2:} Otherwise, define
\[
  u_{\mathrm{top}}
  \;=\;
  \sup\!\bigl\{\,u \ge 0 : \exists\,\alpha>0 
    \text{ with } a^\circ(\alpha,u)>0 \bigr\}.
\]

Again, by the same monotonicity arguments and Lemma~\ref{lem:u_above_v}, this 
supremum is a maximum, and $u_{\mathrm{top}} \le v$.
\end{itemize}

\textbf{Behavior for $u \in \bigl(u_{\mathrm{bot}},\,u_{\mathrm{top}}\bigr)$.} 
Fix $u$ such that $u_{\mathrm{bot}} < u \le u_{\mathrm{top}}$. By 
$\alpha$-monotonicity, if $a^\circ(\alpha,u)=1$ for some $\alpha$, then 
$a^\circ(\alpha',u)=1$ for all $\alpha' \ge \alpha$. Conversely, by 
Theorem~\ref{G_theo:binary_optimal_solution} for any $\alpha' < \alpha^*(u)$, we must have 
$a^\circ(\alpha',u)=0$. Hence, the minimal $\alpha>0$ for which $a^\circ(\alpha,u)$ can be 
positive must exceed $\alpha^*(u)$.
Next, we show that (1) for every $\alpha, u$ for which  $u<\alpha<u+ v-u_{\mathrm{top}}$ it must hold that $a(\alpha,u)=0$, and (2) for every $\alpha>u+ v-u_{\mathrm{top}}$ it must hold that $a(\alpha, u)=1$. 

Let us first show (1). Assume $u_{\mathrm{top}}<v$, and consider $u'$ the smallest $u$ for which $a^\circ(\Bar{\alpha},u)=0$. Per absurd, we assume the existence of $(u,\alpha)$ such that $a^\circ(0,u)=0$ and $a^\circ(\alpha,u)=1$ where $\alpha$ is such that $\alpha^\star(u)<\alpha<\alpha^\star(u) + v-u_{\mathrm{top}}$; we assume that both $u''$ and $\alpha^\dagger(u'')$ are the smallest for which $a^\circ(0,u'')=0$ and $a^\circ(\alpha^\dagger(u''),u'')=1$ such that $\alpha^\star(u)<\alpha^\dagger(u'')<\alpha^\star(u) + v-u_{\mathrm{top}}$. 

Next, we show that we can find a feasible solution with a better objective value. Consider $(a^*,c^*)$ defined as follows: Let $\varepsilon, \beta>0$ sufficiently small, for every $u$ such that $u''\leq u\leq u'' + \beta$, and for every $\varepsilon'\leq \varepsilon - (\alpha^\dagger(u)-\alpha^\dagger(u''))  $, we define $a^*(\alpha^\dagger(u)+\varepsilon',u )=0$ and $c^*(\alpha^\dagger(u)+\varepsilon',u )=0$; now consider $\delta, \gamma>0$ sufficiently small, for every $u$ such that $u'\leq u\leq u' + \gamma$,  and for every $\delta'\leq \delta $, we define $a^*(\Bar{\alpha}-\delta',u )=1$ and $c^*(\Bar{\alpha}-\delta',u )=1$; finally, everywhere else, $(a^*,c^*)$ coincides with $(a^\circ,c^\circ)$. It is immediate to verify that $(a^*,c^*)$ satisfies all the constraints except for the Budget, that we analyze next. Considering very small values of $\beta$ and $\gamma$, we can reduce the budget constraint to the following expression (we are assuming $(a^\circ,c^\circ)$ has a binding budget constraint). 

$$ - \int_{\alpha^\dagger(u'')}^{\alpha^\dagger(u'')+\varepsilon} f(\alpha) \cdot [1- G(u'') - u''g(u'')] \, \mathrm{d}\alpha +  \int_{\Bar{\alpha}-\delta}^{\Bar{\alpha}} f(\alpha) \cdot  [1- G(u') - u'g(u')] \, \mathrm{d}\alpha \geq $$ 
$$ K \Bigg( g(u'') \int_{\alpha^\dagger(u'')}^{\alpha^\dagger(u'')+\varepsilon} - f(\alpha)  \, \mathrm{d}\alpha + g(u') \int_{\Bar{\alpha}-\delta}^{\Bar{\alpha}} f(\alpha)  \, \mathrm{d}\alpha \Bigg)$$

which can be rewritten as 

$$ g(u'') \int_{\alpha^\dagger(u'')}^{\alpha^\dagger(u'')+\varepsilon} f(\alpha) \cdot (r(u'')-K) \, \mathrm{d}\alpha \leq g(u') \int_{\Bar{\alpha}-\delta}^{\Bar{\alpha}} f(\alpha) \cdot ( r(u') -K) \, \mathrm{d}\alpha $$

note that for each $\varepsilon$ small enough, we can always find a $\delta$ to satisfy the previous expression. Now let us check whether the objective function yields a positive increment. The objective function can be reduced to 

$$ -g(u'') \int_{\alpha^\dagger(u'')}^{\alpha^\dagger(u'')+\varepsilon} f(\alpha) \cdot( \alpha -u'') \, \mathrm{d}\alpha + g(u') \int_{\Bar{\alpha}-\delta}^{\Bar{\alpha}} f(\alpha) \cdot (v- u') \, \mathrm{d}\alpha $$ 

and considering binding the inequality in the budget constraint, we can transform the objective function to the following expression, 

$$ g(u'') \int_{\alpha^\dagger(u'')}^{\alpha^\dagger(u'')+\varepsilon} f(\alpha) \cdot\Bigg[-( \alpha -u'') + (v- u')\cdot \frac{\lvert r(u'')-K \rvert}{\lvert r(u')-K \rvert }\Bigg]\, \mathrm{d}\alpha.  $$ 

Note that the ratio $\frac{\lvert r(u'')-K \rvert}{\lvert r(u')-K \rvert }>1$ by the working hypothesis of decreasing monotonicity on $r(u)$. Moreover, we can assume that $\alpha-u''= u'' +v - u_{\text{top}} -\iota - u''=v - u_{\text{top}} -\iota$ for a certain $\iota>0$. On the other side, we can assume that $v-u'= v-u_{\text{top}} -\eta$ for a certain $\eta>0$. Without loss of generality we can assume that $\iota\geq \eta$, since we chose $u'$ as the smallest $u>u_{\text{top}}$. Therefore, for every  $\varepsilon$ we can find a solution with a better objective value.

Finally, let us prove (2). Specifically, we want to prove that given $u$ such that $u_{\mathrm{top}}>u>u_{\mathrm{bot}}$, and given $\alpha$ such that $t(\alpha,u)-v+ u_{\mathrm{top}} >u$, it holds  $a^\circ(\alpha,u)=1 $.  Assume $u_{\mathrm{top}}<v$ and consider the highest $u'$ for which $a^\circ(\Bar{\alpha},u')=1$. Per absurd, we assume the existence of $u$ such that $a^\circ(0,u)=0$ and $a^\circ(\alpha^*(u)+v- u_{\mathrm{top}}+\varepsilon,u)=0$ for a given $\varepsilon>0$; let  $u''$ be the smallest for which $a^\circ(0,u'')=0$ and $a^\circ(\alpha^*(u)+v- u_{\mathrm{top}}+\varepsilon,u'')=0$. If $r(u'')-K\geq 0$, then turning to positive the points at $a^\circ(\alpha^*(u)+v- u_{\mathrm{top}}+\varepsilon,u'')=0$ would increase both the budget and the objective function; hence, let us assume that $r(u'')-K< 0$, note that by the monotonicity assumption on $g$ it holds that for every $u'>u''$ it must also hold $r(u')-K< 0$.

We show we can find a feasible solution with a better objective value. Consider $(a^*,c^*)$ defined as follows: Let $\varepsilon, \beta>0$ sufficiently small, for every $u$ such that $u''\leq u\leq u'' + \beta$, and for every $\varepsilon'\leq \varepsilon - (\alpha^*(u)-\alpha^*(u''))  $, we define $a^*(\alpha^*(u)+v-u_{\mathrm{top}}+\varepsilon',u )=1$ and $c^*(\alpha^*(u)+\varepsilon',u )=1$; now consider $ \gamma>0$ sufficiently small, for every $u$ such that $u'- \gamma\leq u\leq u' $, and for every $\alpha>\alpha^*(u)$,   we define $a^*(\alpha,u)=0$ and $c^*(\alpha,u )=0$; finally, everywhere else, $(a^*,c^*)$ coincides with $(a^\circ,c^\circ)$. It is immediate to verify that $(a^*,c^*)$ satisfies all the constraints except for the Budget, that we analyze next. Considering very small values of $\beta$ and $\gamma$, we can reduce the budget constraint to the following expression (we are assuming that $(a^\circ,c^\circ)$ has a binding budget constraint). 

$$ g(u'') \int_{\alpha^*(u'')+v-u_{\mathrm{top}}}^{\alpha^*(u'')+v-u_{\mathrm{top}}+\varepsilon} f(\alpha) \cdot r( u'') \, \mathrm{d}\alpha - g(u') \int_{\alpha^*(u')}^{\Bar{\alpha}} f(\alpha) \cdot r( u') \, \mathrm{d}\alpha \geq $$ 
$$ K \Bigg( g(u'') \int_{\alpha^*(u'')+v-u_{\mathrm{top}}}^{\alpha^*(u'')+v-u_{\mathrm{top}}+\varepsilon}  f(\alpha)  \, \mathrm{d}\alpha - g(u') \int_{\alpha^*(u')}^{\Bar{\alpha}}  f(\alpha)  \, \mathrm{d}\alpha \Bigg)$$

which can be rewritten as 

$$ g(u'') \int_{\alpha^*(u'')+v-u_{\mathrm{top}}}^{\alpha^*(u'')+v-u_{\mathrm{top}}+\varepsilon} f(\alpha) \cdot (K-r(u'')) \, \mathrm{d}\alpha \leq g(u') \int_{\alpha^*(u')}^{\Bar{\alpha}} f(\alpha) \cdot (K-r(u')) \, \mathrm{d}\alpha $$

note that we can always find $\varepsilon$ small enough to satisfy the previous expression. Now let us check whether the objective function yields a positive increment. The relevant part of the objective function can be written as 

$$ g(u'') \int_{\alpha^*(u'')+v-u_{\mathrm{top}}}^{\alpha^*(u'')+v-u_{\mathrm{top}}+\varepsilon} f(\alpha) \cdot( \min(v,\alpha) -u'') \, \mathrm{d}\alpha - g(u') \int_{\alpha^*(u')}^{\Bar{\alpha}} f(\alpha) \cdot (\min(v,\alpha)- u') \, \mathrm{d}\alpha \geq $$ 

$$ g(u'') \int_{\alpha^*(u'')+v-u_{\mathrm{top}}}^{\alpha^*(u'')+v-u_{\mathrm{top}}+\varepsilon} f(\alpha) \cdot( \alpha -u'') \, \mathrm{d}\alpha - g(u') \int_{\alpha^*(u')}^{\Bar{\alpha}} f(\alpha) \cdot (v- u') \, \mathrm{d}\alpha  $$ 

Note that by construction, $t(\alpha,u'') -u''\geq v- u'$, if we choose $u'=u_{\mathrm{top}}$; considering $u'=u_{\mathrm{top}}$ is  a reasonable  assumption as we chose $u'$ as the highest $u$ with non-zero $a^\circ(\Bar{\alpha}, u)$. Without loss of generality, we can consider as binding the inequality in the Budget constraint, thus we can transform the objective function to the following expression, 

$$ g(u'') \int_{\alpha^*(u'')+v-u_{\mathrm{top}}}^{\alpha^*(u'')+v-u_{\mathrm{top}}+\varepsilon} f(\alpha) \cdot\Bigg[( \alpha -u'') - (v- u')\cdot \frac{(K-r(u''))}{(K-r(u'))}\Bigg]\, \mathrm{d}\alpha.  $$ 

If we prove that the integrand is always positive, we have found a feasible solution with a better objective value. Hence, we must prove that  $\alpha -u'' - (v- u')\cdot \frac{(K-r(u''))}{(K-r(u'))}>0$, which can be rewritten as $\frac{\alpha -u''}{(v- u')} > \frac{(K-r(u''))}{(K-r(u''))}$. Indeed, by construction we have that $\frac{\alpha -u''}{(v- u')} >1$, and since $r(u')<r(u'')$ we also have $1> \frac{(K-r(u''))}{(K-r(u'))}$, thus completing the proof.

Combining these observations shows that $u_{\mathrm{bot}}$ and $u_{\mathrm{top}}$ 
indeed define the unique transition points in $u$-space below and above which the 
allocation is fully determined by monotonicity and the binary constraint. 
\end{proof}

Next, we define a threshold function for the probability of inspection. We denote this threshold by $\psi(\alpha, K,v, u_{\mathrm{bot}}, u_{\mathrm{top}})$, shortly as $\psi(\alpha)$ when the parameters are fixed. Formally, the inspection policy decision is written as 

\[
  c^\circ(\alpha,u) \;=\;
  \begin{cases}
     1, & \text{if } u_{\mathrm{bot}}< u < \psi(\alpha),\\[6pt]
    0,  & \text{otherwise}. 
  \end{cases}
\]

\begin{corollary}
    Given  $K$, v, $u_{\mathrm{bot}}$, and $u_{\mathrm{top}}$  the threshold function $\psi(\alpha)$ takes the following piecewise form:
  \[
    \psi(\alpha) \;=\;
    \begin{cases}
      u_{\mathrm{top}}, & \text{if } \alpha \,\ge\, v,\\[6pt]
      \alpha - (v-u_{\mathrm{top}}),                & \text{if } u_{\mathrm{bot}} + v-u_{\mathrm{top}} \,<\, \alpha \,<\, v,\\[6pt]
      0, & \text{if } \alpha \,\le\, u_{\mathrm{bot}} + v-u_{\mathrm{top}}.
    \end{cases}
  \]
  In other words, the threshold is zero for $\alpha$ smaller than $u_{\mathrm{bot}}+ v-u_{\mathrm{top}}$, then coincides with $\alpha$ minus an inefficiency in the intermediate range, and finally becomes zero at $u_{\mathrm{top}}$ for large $\alpha$. 
\end{corollary}

In the previous corollary, we established the threshold structure of the optimal solution, characterizing the values of \(\alpha\) and \(u\) for which the allocation \(\bigl(a^\circ(\alpha,u), c^\circ(\alpha,u)\bigr)\) is positive.

Building on those results, we now show that determining the specific values of \(u_{\mathrm{top}}\) and \(u_{\mathrm{bot}}\) reduces to a classical \emph{Knapsack problem}. 
The following theorem and its proof illustrate how the original problem~\eqref{formulation:working_problem} can be transformed into the simpler formulation~\eqref{formulation:knapsack} for \(\bigl(u_{\mathrm{top}}, u_{\mathrm{bot}}\bigr)\).

\begin{theorem}\label{thm:knapsack_independent} 
  Let \(K,v \geq 0\), the solution to the \emph{Knapsack problem} given in Formulation~\ref{formulation:knapsack} characterizes a threshold function $\phi,$ for the allocation rule. 
  \begin{subequations}\label{formulation:knapsack}
  \begin{alignat}{1}
  \max_{u_{\mathrm{top}},\,u_{\mathrm{bot}}} 
  \,  \quad & \int_{0}^{u_{\mathrm{bot}}} g(u) \int_{0}^{\Bar{\alpha}} f(\alpha) (\min(v,\alpha) -u ) \mathrm{d}\alpha  \mathrm{d}u  \, + \notag \\
     &    \quad\quad\quad\quad\quad\quad\quad\quad\quad\quad\quad\quad\quad + 
     \int_{u_{\mathrm{bot}}}^{u_{\mathrm{top}}} g(u) \int_{u+ v - u_{\mathrm{top}}}^{\Bar{\alpha}} f(\alpha)  (\min(v,\alpha) -u )  \mathrm{d}\alpha        
             \mathrm{d}u
         \label{formulation:knapsack_obj}\\
  \text{s.t.}\quad 
& \int^{u_{\mathrm{bot}}}_{0} \int_0^{\Bar{\alpha}} f(\alpha)  \left( 1-G(u)-ug(u) \right) \mathrm{d}\alpha \mathrm{d}u \, + 
   \notag \\
      &   
 + \int_{u_{\mathrm{bot}}}^{u_{\mathrm{top}}} \int^{\Bar{\alpha}}_{u+v
-u_{\mathrm{top}} } f(\alpha)\left( 1-G(u)-ug(u) \right) \mathrm{d}\alpha \mathrm{d} u
       \geq   K \int_{u_{\mathrm{bot}}}^{u_{\mathrm{top}}} g(u) \int_{u+v-u_{\mathrm{top}} }^{\Bar{\alpha}} f(\alpha) \mathrm{d}\alpha \mathrm{d}u, 
      \label{formulation:knapsack_budget}\\
  & 0 \;\le\; u_{\mathrm{bot}} \;\le\;  u_{\mathrm{top}} \;\le\; v.
  \end{alignat}
  \end{subequations}
\end{theorem}

\begin{proof}
We show that maximizing \eqref{formulation:knapsack_obj} subject to \eqref{formulation:knapsack_budget} is equivalent to solving the original problem \eqref{formulation:working_problem} under a threshold-based policy. 

\paragraph{Threshold structure.}
By Theorem~\ref{G_theo:binary_optimal_solution} and the monotonicity constraints \eqref{formulation:working_problem_umono}--\eqref{formulation:working_problem_tmono} and Lemma~\ref{lemma:alpha_full_monotonicity}, any optimal solution must be deterministic (0–1) and satisfy
\[
   a^\circ(\alpha,u) \text{ is non-decreasing in }\alpha 
   \quad\text{and}\quad
   a^\circ(\alpha,u) \text{ is non-increasing in }u.
\]
Consequently, for each $\alpha$, the map $u \mapsto a^\circ(\alpha,u)$ can “switch” from 1 to 0 at most once.  Therefore, there exist two constants
\[
  0 \;\le\; u_{\mathrm{bot}} \;\le\; u_{\mathrm{top}} \;\le\; v,
\]
such that the $(\alpha,u)$-plane is partitioned into three regions: $(0 \le u < u_{\mathrm{bot}})$, $(u_{\mathrm{bot}} \le u \le u_{\mathrm{top}})$, and $(u > u_{\mathrm{top}})$. We can describe the behavior of the optimal allocation through a threshold function separating these three regions

\[
  a^\circ(\alpha,u) \;=\;
  \begin{cases}
    1, & \text{if } u \;<\; \phi(\alpha),\\[6pt]
    0, & \text{if } u \;>\; \phi(\alpha),
  \end{cases}
\]
where $\phi(\cdot)$ is a non-decreasing function in $\alpha$ and non-increasing in $u$. 

\paragraph{Rewriting the objective.}
The original objective from \eqref{formulation:working_problem_obj} is
\[
  \int_0^{\bar{u}} g(u) \int_0^{\Bar{\alpha}} f(\alpha)\,\bigl(\min(v,\alpha) - u\bigr)\,
  a^\circ(\alpha,u)\,\mathrm{d}\alpha\,\mathrm{d}u.
\]
Under the threshold structure, $a^\circ(\alpha,u)$ is either $1$ or $0$ depending on which of the three intervals $u$ falls into.  

\begin{enumerate}
\item For $0 \le u < u_{\mathrm{bot}}$, the policy sets $a^\circ(\alpha,u)=1$ for all $\alpha \in [0,\Bar{\alpha}]$. 
  Hence,
  \[
    \int_0^{u_{\mathrm{bot}}} g(u)\,\int_0^{\Bar{\alpha}} f(\alpha)\,\bigl(\min(v,\alpha) - u\bigr)\,
    \underbrace{1}_{a^\circ(\alpha,u)}\,\mathrm{d}\alpha\,\mathrm{d}u
    \;=\; \] 
    \[
   \;=\;  \int_0^{u_{\mathrm{bot}}} g(u)\,\int_0^{\Bar{\alpha}} f(\alpha)\,(\min(v,\alpha) - u)\,\mathrm{d}\alpha\,\mathrm{d}u.
  \]
\item For $u_{\mathrm{bot}} \le u \le u_{\mathrm{top}}$, the policy sets $a^\circ(\alpha,u)=1$ if and only if $\alpha-v + u_{\mathrm{top}} \ge u$ (i.e., we switch from 0 to 1 at $\alpha$ such that $\alpha-v + u_{\mathrm{top}} = u$.  This splits the integral in $\alpha$ from [$\alpha^*(u)+ v - u_{\mathrm{top}}$] to $\Bar{\alpha}$, where $\alpha^*(u)=u$.
  \[
    \int_{u_{\mathrm{bot}}}^{u_{\mathrm{top}}} g(u)\,\int_{\alpha^*(u)+ v - u_{\mathrm{top}}}^{\Bar{\alpha}} f(\alpha)\,\bigl(\min(v,\alpha) - u\bigr)\,
    \underbrace{1}_{a^\circ(\alpha,u)}\,\mathrm{d}\alpha\,\mathrm{d}u.
  \]
\item For $u > u_{\mathrm{top}}$, the policy sets $a^\circ(\alpha,u)=0$ for all $\alpha$, contributing
  \[
    \int_{u_{\mathrm{top}}}^{\bar{u}} g(u)\,\int_{0}^{\Bar{\alpha}} f(\alpha)\,\bigl(\min(v,\alpha) - u\bigr)\,
    \underbrace{0}_{a^\circ(\alpha,u)}\,\mathrm{d}\alpha\,\mathrm{d}u
    \;=\; 0.
  \]
\end{enumerate}
Summing these three pieces collapses into the two positive terms in \eqref{formulation:knapsack_obj}, since the region $u>u_{\mathrm{top}}$ contributes zero.  This algebraic partitioning directly yields the objective in \eqref{formulation:knapsack_obj}:
\[
  \int_{0}^{u_{\mathrm{bot}}} g(u)\int_{0}^{\Bar{\alpha}} f(\alpha) \bigl(\min(v,\alpha) - u\bigr) \mathrm{d}\alpha\mathrm{d}u
  +
  \int_{u_{\mathrm{bot}}}^{u_{\mathrm{top}}} g(u)\int_{u+ v - u_{\mathrm{top}}}^{\Bar{\alpha}} f(\alpha)\,\bigl(\min(v,\alpha) - u\bigr)\mathrm{d}\alpha\mathrm{d}u.
\]

\paragraph{Knapsack (budget) constraint.}
A similar partition applies to the budget constraint \eqref{formulation:working_problem_budget}.  Define the transfer cost and inspection decisions according to the same three intervals:
\begin{itemize}
\item For $0 \le u < u_{\mathrm{bot}}$, no inspection occurs. 
\item For $u_{\mathrm{bot}} \le u \le u_{\mathrm{top}}$, inspection happens if the incumbent’s report $\alpha$ exceeds $\alpha^*(u)$ plus the inefficiency, incurring cost $K$ whenever exclusive access is ultimately granted.
\item For $u > u_{\mathrm{top}}$, the mechanism defaults to sharing without inspection.
\end{itemize}

Thus the right-hand side of the budget constraint takes the following form:

\[ K\,\int_{u_{\mathrm{bot}}}^{u_{\mathrm{top}}} g(u)\!\int_{\alpha^*(u)+v-u_{\mathrm{top}} }^{\Bar{\alpha}} f(\alpha)\,\mathrm{d}\alpha\,\mathrm{d}u.\]

Next, we want to analyze how the term
\[
\int_{0}^{\bar{u}} g(u) \int_{0}^{\Bar{\alpha}} f(\alpha)\Bigl(\int_{0}^{u} a(\alpha,x)\,\mathrm{d}x \;-\; u\,a(\alpha,u) \Bigr)\,\mathrm{d}\alpha\,\mathrm{d}u.
\]

\noindent
\textbf{Partition the domain of integration in $u$.}\\[4pt]
We split the outer integral in $u$ at the points $u_{\mathrm{bot}}$ and $u_{\mathrm{top}}$:
\[
\int_{0}^{\bar{u}} 
\;\;=\;\;
\int_{0}^{u_{\mathrm{bot}}} 
\;+\;
\int_{u_{\mathrm{bot}}}^{u_{\mathrm{top}}}
\;+\;
\int_{u_{\mathrm{top}}}^{\bar{u}}.
\]
Hence,
\[
\int_{0}^{\bar{u}} g(u) \int_{0}^{\Bar{\alpha}} f(\alpha)\,\bigl(\ldots\bigr)\,\mathrm{d}\alpha\,\mathrm{d}u
\;\;=\;\;\] \[ =
\underbrace{\int_{0}^{u_{\mathrm{bot}}} g(u)\int_{0}^{\Bar{\alpha}} f(\alpha)\bigl(\ldots\bigr)\mathrm{d}\alpha\mathrm{d}u}_{\mathrm{(I)}}
+
\underbrace{\int_{u_{\mathrm{bot}}}^{u_{\mathrm{top}}} g(u)\!\int_{0}^{\Bar{\alpha}} f(\alpha)\bigl(\ldots\bigr)\mathrm{d}\alpha\mathrm{d}u}_{\mathrm{(II)}}
+
\underbrace{\int_{u_{\mathrm{top}}}^{\bar{u}} g(u)\int_{0}^{\Bar{\alpha}} f(\alpha)\bigl(\ldots\bigr)\mathrm{d}\alpha\mathrm{d}u}_{\mathrm{(III)}}.
\]
We next identify what $(\ldots)$ (the integrand) equals in each sub‐interval, based on the \emph{threshold policy} for $a(\alpha,u)$.

\medskip

\noindent
\textbf{Inner expression}: \(g(u)\left(\int_{0}^{u} a(\alpha,x)\,\mathrm{d}x - u\,a(\alpha,u)\right)= (1-G(u)-ug(u))a(\alpha,u)  \).\\[4pt]





\begin{itemize}
\item \(\mathrm{(I)}\): For \(0 \le u < u_{\mathrm{bot}}\), $a(\alpha,u)=1$ for $\alpha\geq0$. Therefore we have

 $$ \int^{u_{\mathrm{bot}}}_{0} \int_0^{\Bar{\alpha}} f(\alpha)  \left( 1-G(u)-ug(u) \right) \mathrm{d}\alpha \mathrm{d}u  .  $$

\item \text{(II) For \(u_{\mathrm{bot}} \le u \le u_{\mathrm{top}}\).}
The contribution to the left-hand side reduces to 

$$\int_{u_{\mathrm{bot}}}^{u_{\mathrm{top}}} \int^{\Bar{\alpha}}_{\alpha^*(u)+v
-u_{\mathrm{top}} } f(\alpha)\,\left( 1-G(u)-ug(u) \right) \mathrm{d}\alpha \mathrm{d} u . $$

\item \text{(III) For \(u > u_{\mathrm{top}}\) we have that $ a(\alpha,u)=0$. Therefore the contribution of the left-hand side is 0 too.}
%





\end{itemize}
Hence, summing (II) and (III) leads to

\[
\int^{u_{\mathrm{bot}}}_{0} \int_0^{\Bar{\alpha}} f(\alpha)  \left( 1-G(u)-ug(u) \right) \mathrm{d}\alpha \mathrm{d}u + \int_{u_{\mathrm{bot}}}^{u_{\mathrm{top}}} \int^{\Bar{\alpha}}_{u+v
-u_{\mathrm{top}} } f(\alpha)\,\left( 1-G(u)-ug(u) \right) \mathrm{d}\alpha \mathrm{d} u .
\]


This completes the derivation.
\end{proof}


\subsection{Threshold Solution for $t(\alpha,u)=\alpha\cdot u$}

In this section we provide the optimal solution when the interference is $t(\alpha,u)=\alpha\cdot u$.

Fix constants $0\le u_{\mathrm{top}}\le u_{\mathrm{top}}\le v$ and define
\[
B:=\int_{u_{\mathrm{top}}}^{u_{\mathrm{top}}} g(u)\,\bigl(r(u)-K\bigr)\,du,\qquad
C_0:=\Bigl(\int_{0}^{\bar{\alpha}} f(\alpha)\,d\alpha\Bigr)\Bigl(\int_{0}^{u_{\mathrm{top}}} g(u)\,r(u)\,du\Bigr).
\]
and
\[
C(1)=C_0+\int_{u_{\mathrm{top}}}^{u_{\mathrm{top}}} g(u)\!\int_{1}^{\bar{\alpha}} f(\alpha)\,\bigl(r(u)-K\bigr)\,d\alpha\,du.
\]


Next, we show that the threshold is flat at $u_{\mathrm{bot}}$ for small signals $t=\alpha u$, scales linearly as $t/\alpha^{\mathrm{opt}}$ in the intermediate region, and saturates at $u_{\mathrm{top}}$ for large signals. The cutoff $\alpha^{\mathrm{opt}}$ equals $1$ when the constraint is slack at $x=1$, and otherwise is the unique quantile solving $F(\alpha^{\mathrm{opt}})=F(\bar{\alpha})+C_0/B$ (binding-constraint case).

\begin{theorem}\label{thm:threshold_power}
  Let $K,v \geq 0$. There exist two constants $u_{\mathrm{bot}}$ and $u_{\mathrm{top}}$ satisfying
  \[
    0 \;\le\; u_{\mathrm{bot}} 
          \;\le\; u_\infty
          \;\le\; u_{\mathrm{top}}
          \;\le\; v,
  \]
  such that the threshold function $\phi(\alpha,u)$ takes the following piecewise form:
  \[
\phi(\alpha,u)=
\begin{cases}
u_{\mathrm{bot}}, & t(\alpha,u) \le \alpha^{\mathrm{opt}}\,u_{\mathrm{bot}},\\[6pt]
\dfrac{t(\alpha,u)}{\alpha^{\mathrm{opt}}}, & \alpha^{\mathrm{opt}}\,u_{\mathrm{top}}< t(\alpha,u)< \alpha^{\mathrm{opt}}\,u_{\mathrm{top}},\\[10pt]
u_{\mathrm{top}}, & t(\alpha,u)\ge \alpha^{\mathrm{opt}}\,u_{\mathrm{top}}.
\end{cases}
\]

  where \[
\alpha^{\mathrm{opt}}=
\begin{cases}
1, & \text{if } C(1)\ge 0,\\[6pt]
F^{-1}\!\Bigl(F(\bar{\alpha})+\dfrac{C_0}{B}\Bigr), & \text{if } C(1)<0.
\end{cases}
\]
\end{theorem}

\begin{proof}  
    Consider an optimal solution $\bigl(a^\circ(\alpha,u),\,c^\circ(\alpha,u)\bigr)$. By 
Theorem~\ref{G_theo:binary_optimal_solution}, every optimal allocation $a^\circ$ 
takes only binary values (either zero or positive). 

The definition of $u_{\mathrm{bot}}$ and $u_{\mathrm{top}}$ follows a similar reasoning as the one for Theorem~\ref{thm:threshold_independent}. We focus only on the region of those $u$ for which $u_{\mathrm{bot}} \,<\, t(\alpha,u) \,<\, u_{\mathrm{top}}$.

By $\alpha$-monotonicity, if $a^\circ(\alpha,u)=1$ for some $\alpha$, then 
$a^\circ(\alpha',u)=1$ for all $\alpha' \ge \alpha$. Conversely, by 
Theorem~\ref{G_theo:binary_optimal_solution} for any $\alpha' < \alpha^*(u)$, we must have 
$a^\circ(\alpha',u)=0$. Hence, the minimal $\alpha>0$ for which $a^\circ(\alpha,u)$ can be 
positive must exceed $\alpha^*(u)$.
Next, we show that for every $\alpha, u$ for which  $t(\alpha, u)>u$ it must hold that $a(\alpha,u)=1$. We prove this in two steps: first we show that the curve (segment) connecting the threshold's border between $u_{\mathrm{bot}}$ $u_{\mathrm{top}}$ is vertical; second, given $u_{\mathrm{bot}}<u<u_{\mathrm{top}}$, we find the analytical expression for the coordinate of the threshold segment.

Consider $u_{\mathrm{bot}}<u<u_{\mathrm{top}}$. The threshold value in the $(\alpha,u)$ plane must have positive slope to connect the two horizontal threshold contours at $u_{\mathrm{bot}}$ and $u_{\mathrm{top}}$, as the threshold curve starts at the point $(0,u_{\mathrm{bot}})$ and ends at the point $(\Bar{\alpha},u_{\mathrm{top}})$. By contradiction, assume there is at least one point of the threshold curve where the slope of the tangent is not infinite. Let $\delta>0$ be small and consider a point  $(\alpha,u)$ such that $u+\delta$ is the highest point on the threshold border for which the slope of the tangent is positive, yet not infinite.

Clearly, it must hold  $a^\circ(0,u)=0$, as we are assuming $K<\infty$ and $u_{\mathrm{bot}}<u<u_{\mathrm{top}}$. We want to show that we can find a better solution defined as follows.

Let $\iota,\theta>0$, and consider $u$ be the largest for which $a^\circ(0,u)=0$ and   $a^\circ(\alpha+\iota,u+\theta)=1$ with $\alpha>\alpha^*(u)$, i.e., the largest $u$ for which there can still be a vertical increment. Our goal is to take mass from the bottom-left side of the neighborhood of $(\alpha,u)$, and \lq\lq bring\rq\rq\, it to the top-right side of the neighborhood. Note that the distance between the bottom-left \emph{active mass} and the top-right \emph{inactive mass} may be large. Without loss of generality, and considering $\delta,\gamma>0$ to be sufficiently small, we consider the set \emph{Plus}= $\{(\alpha',u')=(\alpha+\gamma',u-\delta')\colon\, \gamma'<\gamma \, \wedge \, \delta'<\delta  \}$ where the optimal solution yields $a^\circ(\alpha',u')= c^\circ(\alpha',u')=0 $. W.l.o.g., considering $\beta,  \varepsilon>0$ sufficiently small, we can also assume that in the set \emph{Minus} = $\{(\alpha'',u'')=(\alpha-\varepsilon',u-\beta')\colon\, \varepsilon'<\varepsilon \, \wedge \, \beta'<\beta  \}$  the optimal solution yields $a^\circ(\alpha'',u'')= c^\circ(\alpha'',u'')=1 $. If for every $u'$ in \emph{Plus} it holds that $K-r(u')<0$, then we can simply increase to 1 the allocation and the inspection for the points in  \emph{Plus} and we obtain a better feasible solution, an absurd. Hence, we assume that 
$K-r(u''')>0$ for every $u'''>u-\beta$. 

We define a new solution $(a^*,c^*)$ defined as follows: for every $(\alpha'',u'')$ in \emph{Minus} we set $a^*(\alpha'',u'')=c^*(\alpha'',u'')=0$, for every $(\alpha',u')$ in \emph{Plus} we set $a^*(\alpha',u')=c^*(\alpha',u')=1$, and everywhere else $(a^*,c^*)=(a^\circ,c^\circ)$. Note that all the constraints except for the budget constraint are satisfied. We need to show that the budget constraint is valid and that we have found a solution with a better objective value. 

Since the budget constraint in the solution $(a^\circ,c^\circ)$ is tight, the constraint in the new solution reduces to

$$ - \int_{u-\beta}^{u} g(u) \int_{\alpha-\varepsilon}^{\alpha} f(\alpha) r(u) \, \mathrm{d}\alpha \mathrm{d}u + \int_{u}^{u+\delta} g(u) \int_{\alpha}^{\alpha+\gamma} f(\alpha) \cdot  r(u) \, \mathrm{d}\alpha \mathrm{d} u \geq $$ 
$$ K \Bigg( - \int_{u-\beta}^{u} g(u) \int_{\alpha-\varepsilon}^{\alpha} f(\alpha)  \, \mathrm{d}\alpha \mathrm{d}u  +\int_{u}^{u+\delta} g(u) \int_{\alpha}^{\alpha+\gamma} f(\alpha)  \, \mathrm{d}\alpha \mathrm{d}u \Bigg)$$

and if we choose $\beta, \gamma, \delta, \varepsilon$ sufficiently small, we can assume in each neighborhood $f,g$ have the same value; since the two regions \emph{Minus} and \emph{Plus} are taken within a neighborhood of $\alpha$, we can assume that $f$ takes the same value in both regions.  Hence, we obtain

$$\int_{u}^{u+\delta} g(u)  \int_{\alpha}^{\alpha+\gamma}    (K-r(u)) \, \mathrm{d}\alpha \mathrm{d} u \leq \int_{u-\beta}^{u} g(u) \int_{\alpha-\varepsilon}^{\alpha}  (K-r(u)) \, \mathrm{d}\alpha \mathrm{d}u  .$$ 

From the assumption that $r(u)$ is decreasing, we can always find parameters for which the inequality holds.

Let us now analyze the variation in the objective function. Without loss of generality, assume that $\alpha\cdot u<v$, hence we have that the variation can be reduced to

$$   -\int_{u-\beta}^{u} g(u) \int_{\alpha-\varepsilon}^{\alpha}  u(\alpha-1)  \, \mathrm{d}\alpha \mathrm{d}u + \int_{u}^{u+\delta}  \int_{\alpha}^{\alpha+\gamma}  g(u)  u(\alpha-1) \, \mathrm{d}\alpha \mathrm{d} u.  $$

Let us assume that $\beta,\gamma$ are sufficiently large so that we can reduce \emph{Minus} to an horizontal neighborhood of $u''$ and we  can reduce \emph{Plus} to an horizontal neighborhood of $u'$. Moreover, without loss of generality, we can pick $\delta=\beta$. From the budget constraint, we can write 

$$ \gamma \cdot \dfrac{g(u')(K-r(u'))}{g(u'')(K-r(u''))}\leq \varepsilon ,$$

where $r(u''), r(u')$ is the value of $r$ in \emph{Minus}, \emph{Plus}, respectively; moreover, we can choose $\varepsilon $ satisfying the equality.

On the other side, the objective function can be reduced to 

$$     - g(u'') \cdot u''\cdot (\varepsilon \alpha-\frac{\varepsilon^2}{2}-\varepsilon)    +  g(u') \cdot u'\cdot( \gamma \alpha+\frac{\gamma^2}{2}-\gamma)    .$$

Substituting the value of $\varepsilon$ and focusing only on the meaningful term involving $\alpha$, the objective function reduces to 

$$  \gamma \cdot\alpha\cdot g(u') \cdot\left(  u'  - u'' \cdot \dfrac{K-r(u')}{K-r(u'')} \right)  .$$

Thus, it remains to prove that 

\[
1>\frac{u''}{u'}\cdot\frac{K-r(u')}{K-r(u'')}.
\]

Since $r$ is decreasing and $u'>u''$, we have $r(u')\le r(u'')<K$, hence
$K-r(u')>0$ and $K-r(u'')>0$. Therefore all factors below are positive and we can
cross-multiply without changing the inequality sign:
\[
\frac{u''}{u'}\cdot\frac{K-r(u')}{K-r(u'')}<1
\;\Longleftrightarrow\;
u''\bigl(K-r(u')\bigr)<u'\bigl(K-r(u'')\bigr).
\]
Rearranging gives
\[
u'r(u'')-u''r(u')<K\,(u'-u'').
\]
Using $r(u')\le r(u'')$,
\[
u'r(u'')-u''r(u')\;\le\;u'r(u'')-u''r(u'')
=(u'-u'')\,r(u'')
\;<\;(u'-u'')\,K,
\]
where the last strict inequality follows from $r(u'')<K$ and $u'-u''>0$.


Next, we find the optimal choice of the coordinate for the segment connecting $u_{\mathrm{bot}}$ and $u_{\mathrm{top}}$, that we denote as $\alpha^{\text{opt}}$. 
In order to find the optimal coordinate $\alpha^{\text{opt}}=x$ for given pair of $u_{\mathrm{bot}}$ and $u_{\mathrm{top}}$, we need to solve the following optimization problem.

\begin{subequations}
  \begin{alignat}{2}
  \max_{x} 
  \quad & O(x) := \int_{0}^{u_{\mathrm{bot}}} g(u) \int_{0}^{\Bar{\alpha}} f(\alpha) (\min(v,\alpha u) -u ) \mathrm{d}\alpha  \mathrm{d}u  + \int_{u_{\mathrm{bot}}}^{u_{\mathrm{top}}} g(u) 
              \int_{x}^{\Bar{\alpha}} f(\alpha)  (\min(v,\alpha u) -u )  \mathrm{d}\alpha        
             \mathrm{d}u
         \notag\\
  \text{s.t.}\quad 
  & C(x) := \int^{u_{\mathrm{bot}}}_{0} g(u)\int_{0}^{\Bar{\alpha} } f(\alpha)\left( r(u) \right)\mathrm{d}\alpha \mathrm{d}u
  +
  \int^{u_{\mathrm{top}}}_{u_{\mathrm{bot}}} g(u)\int_{x}^{\Bar{\alpha}} f(\alpha)\bigl(r(u)-K\bigr)\mathrm{d}\alpha \mathrm{d}u 
      \;\geq  0 \notag \\
  & 1 \;\le\; x \;\le\; \Bar{\alpha}, \notag 
  \end{alignat}
  \end{subequations}

Let
\[
A(x)=\int_{u_{\mathrm{bot}}}^{u_{\mathrm{top}}} g(u)\Bigl(\min\{v,xu\}-u\Bigr)\,du,\qquad
B=\int_{u_{\mathrm{bot}}}^{u_{\mathrm{top}}} g(u)\bigl(r(u)-K\bigr)\,du\ (<0),
\]
\[
T(x)=\int_{x}^{\bar{\alpha}} f(\alpha)\,d\alpha,\qquad
C_0=\Bigl(\int_{0}^{\bar{\alpha}} f(\alpha)\,d\alpha\Bigr)\Bigl(\int_{0}^{u_{\mathrm{bot}}} g(u)\,r(u)\,du\Bigr)\ (>0).
\]

Note that only the second inner $\alpha$-integral in both $O(x)$ and $C(x)$ depends on $x$, hence by Leibniz:
\[
O'(x)=-f(x)\,A(x),\qquad
C(x)=C_0+B\,T(x),\qquad
C'(x)=-f(x)\,B>0.
\]

Since $u_{\mathrm{top}}<v$ and $x\ge 1$, we have $\min\{v,xu\}\ge u$ for all $u\in[u_{\mathrm{bot}},u_{\mathrm{top}}]$,
with strict inequality on a set of positive measure when $x>1$. Thus
\[
A(1)=0,\qquad A(x)>0\ \text{for }x>1
\ \Rightarrow\ 
O'(x)=-f(x)A(x)\le 0,\ \ \text{and}\ \ O'(x)<0\ \text{for }x>1.
\]
Therefore $O$ is strictly decreasing on $[1,\bar{\alpha}]$. Over any feasible interval,
the maximizer is its \emph{leftmost} point.

Because $B<0$ we have $C'(x)>0$, so $C$ is strictly increasing on
\(
\{x\in[1,\bar{\alpha}]:\ C(x)\ge 0\}=[x_c,\bar{\alpha}]
\)
for some threshold $x_c$ (possibly $x_c\le 1$). In particular,
if $C(1)\ge 0$ then $x=1$ is feasible; otherwise $x_c>1$ is uniquely defined by
$T(x_c)=-C_0/B$.

Let us now check the KKT conditions with a single multiplier on the domain $x\in[1,\bar{\alpha}]$. 
We use the Lagrangian with one multiplier:
\[
\mathcal L(x,\delta)=O(x)+\delta\,C(x),\qquad \delta\ge 0.
\]
And the KKT conditions take the following form as we are considering as a domain for $x$ the interval $[1,\bar{\alpha}] $.
\begin{itemize}
\item Interior $x\in(1,\bar{\alpha})$: \quad $A(x)+\delta B=0$, \ \ $C(x)\ge 0$, \ \ $\delta\,C(x)=0$.
\item Left boundary $x=1$ (only right moves allowed): \quad $O'_+(1)+\delta\,C'_+(1)\le 0$.
\end{itemize}
At $x=1$, $O'_+(1)=0$ (since $A(1)=0$) and $C'_+(1)=-f(1)B>0$, hence
\[
0+\delta\,(-f(1)B)\le 0\ \Rightarrow\ \delta=0.
\]
With $\delta=0$, complementary slackness is automatic; only primal feasibility
$C(1)\ge 0$ remains.

To conclude, we have two different cases: 
\begin{itemize}
\item If $C(1)\ge 0$, then $x=1$ is feasible, satisfies the boundary KKT with $\delta=0$,
and since $O$ strictly decreases for $x>1$, it is the unique maximizer:
\[
\alpha^{\text{opt}}=1.
\]
\item If $C(1)<0$, the smallest feasible point is $x_c>1$ with $C(x_c)=0$; since
$O$ decreases, the maximizer is $\alpha^{\text{opt}}=x_c$, where $T(x_c)=-C_0/B$, i.e., $$ \alpha^{\text{opt}} = F^{-1}\left(F(\Bar{\alpha}) + \frac{C_0}{B} \right).$$
\end{itemize}

Finally, given $u_{\mathrm{bot}}$, $u_{\mathrm{top}}$, and $\alpha^{\mathrm{opt}}$, we derive the threshold structure.
 For each $(\alpha,u)$, consider the pointwise choice of $\phi=\phi(\alpha,u)$ over 
\(
u_{\mathrm{bot}} \;\le\; \phi \;\le\; u_{\mathrm{top}}
\). Since $t(\alpha,u)=\alpha u$ and $u>0$, the condition $\alpha>\alpha^{\mathrm{opt}}$ is equivalent to
\[
t(\alpha,u)>\alpha^{\mathrm{opt}}\,u \ \Longleftrightarrow\ u<\frac{t(\alpha,u)}{\alpha^{\mathrm{opt}}},
\]
so in the interior region the threshold must satisfy $\phi(\alpha,u)=t(\alpha,u)/\alpha^{\mathrm{opt}}$.
\end{proof}

The threshold function is expressed as a function of $u_{\mathrm{top}},\,u_{\mathrm{bot}}$, next we show there is a polynomial time way to retrieve the values to these parameters.

Recall that

\[  
C_0 \;:=\; \Bigl(\int_{0}^{\bar{\alpha}} f(\alpha)\,d\alpha\Bigr)\Bigl(\int_{0}^{u_{\mathrm{bot}}} g(u)\,r(u)\,du\Bigr),\qquad
B \;:=\; \int_{u_{\mathrm{bot}}}^{u_{\mathrm{top}}} g(u)\,\bigl(r(u)-K\bigr)\,du,
\]

where $B<0$ for $K>K_{\text{low}}$ by the decreasing monotonicity condition on $r(u)$. Moreover, note that writing the budget at $x$ as
\[
C(x)=C_0+\int_{u_{\mathrm{bot}}}^{u_{\mathrm{top}}}\!g(u)\!\int_{x}^{\bar{\alpha}} f(\alpha)\,\bigl(r(u)-K\bigr)\,d\alpha\,du
\;=\; C_0 + B\bigl(F(\bar{\alpha})-F(x)\bigr),
\]
we have that: if $C(1)\ge 0$ then $F(\bar{\alpha})+\tfrac{C_0}{B}\le F(1)$ and the projection enforces
$\alpha^{\mathrm{opt}}=1$; if $C(1)<0$ the argument lies in $(F(1),F(\bar{\alpha}))$ and
$\alpha^{\mathrm{opt}}=F^{-1}\!\big(F(\bar{\alpha})+\tfrac{C_0}{B}\big)\in(1,\bar{\alpha})$.

\begin{theorem}\label{thm:knapsack_power_interference}
Let $K,v\ge 0$ and $t(\alpha,u)=\alpha u$. The solution to
Formulation~\ref{formulation:knapsack_power} is characterized by a
threshold allocation rule $a(\alpha,u)=\mathbf{1}\{\,u<\phi(\alpha,u)\,\}$
with two cutoffs $0\le u_{\mathrm{bot}}\le u_{\mathrm{top}}\le v$ and an
$\alpha$–cutoff $\alpha^{\mathrm{opt}}\in[1,\bar{\alpha}]$ solving
\begin{subequations}\label{formulation:knapsack_power}
\begin{alignat}{2}
\max_{\,u_{\mathrm{top}},\,u_{\mathrm{bot}}}\quad
& \int_{0}^{u_{\mathrm{bot}}} g(u)\!\int_{0}^{\bar{\alpha}} f(\alpha)\,\bigl(\min\{v,\alpha u\}-u\bigr)\,d\alpha\,du
\;+\;
\int_{u_{\mathrm{bot}}}^{u_{\mathrm{top}}} g(u)\!\int_{\alpha^{\mathrm{opt}}}^{\bar{\alpha}} f(\alpha)\,\bigl(\min\{v,\alpha u\}-u\bigr)\,d\alpha\,du
\label{formulation:knapsack_obj_power}\\[2pt]
\text{s.t.}\quad
& \int_{0}^{u_{\mathrm{bot}}} g(u)\!\int_{0}^{\bar{\alpha}} f(\alpha)\,r(u)\,d\alpha\,du
\;+\;
\int_{u_{\mathrm{bot}}}^{u_{\mathrm{top}}} g(u)\!\int_{\alpha^{\mathrm{opt}}}^{\bar{\alpha}} f(\alpha)\,\bigl(r(u)-K\bigr)\,d\alpha\,du
\;\ge\;0,
\label{formulation:knapsack_budget_power}\\[4pt]
& \alpha^{\mathrm{opt}} \;=\; \max\left(1,
F^{-1}\!\left(1+\frac{C_0(u_{\mathrm{bot}})}{B(u_{\mathrm{bot}},u_{\mathrm{top}})}\right)
\right),
\label{eq:alpha_opt_operational}\\[2pt]
& 0\le u_{\mathrm{bot}}\le u_{\mathrm{top}}\le v,
\end{alignat}
\end{subequations}
The induced threshold for $u$ is
\[
\phi(\alpha,u)=\min\Bigl\{\,u_{\mathrm{top}},\ \max\Bigl\{\,u_{\mathrm{bot}},\ \frac{t(\alpha,u)}{\alpha^{\mathrm{opt}}}\Bigr\}\Bigr\}.
\]
\end{theorem}

\begin{proof}[Proof of Theorem~\ref{thm:knapsack_power_interference}]

By Theorem~\ref{G_theo:binary_optimal_solution} together with the monotonicity
constraints in Formulation~\ref{formulation:working_problem}, any optimal policy
is deterministic and satisfies: $a(\alpha,u)$ is nondecreasing in $\alpha$ and
nonincreasing in $u$. Hence, for each $\alpha$, the map $u\mapsto a(\alpha,u)$
can switch from $1$ to $0$ at most once. There exist $0\le u_{\mathrm{bot}}\le
u_{\mathrm{top}}\le v$ and an $\alpha$–cutoff $\alpha^{\mathrm{opt}}$ such that
\[
a(\alpha,u)=
\begin{cases}
1, & u<u_{\mathrm{bot}},\\
\mathbf{1}\{\alpha\ge \alpha^{\mathrm{opt}}\}, & u_{\mathrm{bot}}\le u\le u_{\mathrm{top}},\\
0, & u>u_{\mathrm{top}}.
\end{cases}
\]
Since $t(\alpha,u)=\alpha u$, the intermediate-region threshold line is
$t=\alpha^{\mathrm{opt}}u$, which is equivalent to $u=t/\alpha^{\mathrm{opt}}$.
Thus the allocation is $a(\alpha,u)=\mathbf{1}\{\,u<\phi(\alpha,u)\,\}$ with
\[
\phi(\alpha,u)=\min\{u_{\mathrm{top}},\max\{u_{\mathrm{bot}},\,t(\alpha,u)/\alpha^{\mathrm{opt}}\}\}.
\]

Using the three $u$-regions above, the objective equals
\[
\int_{0}^{u_{\mathrm{bot}}} g(u)\!\int_{0}^{\bar{\alpha}} f(\alpha)\,\bigl(\min\{v,\alpha u\}-u\bigr)\,d\alpha\,du
\;+\;
\int_{u_{\mathrm{bot}}}^{u_{\mathrm{top}}} g(u)\!\int_{\alpha^{\mathrm{opt}}}^{\bar{\alpha}} f(\alpha)\,\bigl(\min\{v,\alpha u\}-u\bigr)\,d\alpha\,du,
\]
because the region $u>u_{\mathrm{top}}$ contributes zero (there $a\equiv 0$).
This is exactly \eqref{formulation:knapsack_obj_power}.

Let us now consider the budget constraint under the threshold condition. 
Applying the same partition to the budget yields
\[
\int_{0}^{u_{\mathrm{bot}}} g(u)\!\int_{0}^{\bar{\alpha}} f(\alpha)\,r(u)\,d\alpha\,du
\;+\;
\int_{u_{\mathrm{bot}}}^{u_{\mathrm{top}}} g(u)\!\int_{\alpha^{\mathrm{opt}}}^{\bar{\alpha}} f(\alpha)\,\bigl(r(u)-K\bigr)\,d\alpha\,du
\;\ge\;0,
\]
which is \eqref{formulation:knapsack_budget_power}. Aggregating constants gives
\[
C_0:=\Bigl(\int_{0}^{\bar{\alpha}} f\Bigr)\Bigl(\int_{0}^{u_{\mathrm{bot}}} g\,r\Bigr),\qquad
B:=\int_{u_{\mathrm{bot}}}^{u_{\mathrm{top}}} g\,(r-K),
\]
and the “budget at cutoff $x$” can be written as
\[
C(x)=C_0+\int_{u_{\mathrm{bot}}}^{u_{\mathrm{top}}} g(u)\!\int_{x}^{\bar{\alpha}} f(\alpha)\,(r(u)-K)\,d\alpha\,du
\;=\; C_0 + B\bigl(F(\bar{\alpha})-F(x)\bigr).
\]

Choosing $\alpha^{\mathrm{opt}}$ to guarantee feasibility requires $C(\alpha^{\mathrm{opt}})\ge 0$. If $B<0$, then $C(x)$ is
strictly increasing in $x$; the smallest feasible $x$ is therefore
$\alpha^{\mathrm{opt}}=F^{-1}\!\big(F(\bar{\alpha})+C_0/B\big)$, projected onto
$[1,\bar{\alpha}]$ if necessary. If $B\ge 0$, then $C(x)$ is weakly decreasing
in $x$ and its minimum over $[1,\bar{\alpha}]$ is $C(\bar{\alpha})=C_0$; when
$C_0\ge 0$ the constraint is slack for all $x$, and (since the objective is
nonincreasing in the cutoff under $u_{\mathrm{top}}\le v$) the optimal choice is
$\alpha^{\mathrm{opt}}=1$. Both cases are compactly encoded by the projection
formula \eqref{eq:alpha_opt_operational}.

The policy is a threshold in $(\alpha,u)$ with threshold function
$\phi(\alpha,u)=\min\{u_{\mathrm{top}},\max\{u_{\mathrm{bot}},\,t(\alpha,u)/\alpha^{\mathrm{opt}}\}\}$,
and the triplet $(u_{\mathrm{bot}},u_{\mathrm{top}},\alpha^{\mathrm{opt}})$
solves \eqref{formulation:knapsack_power}.
\end{proof}

\end{document}